\documentclass[useAMS,usenatbib]{mn2e}
\usepackage{graphicx}
\usepackage{epsfig}
\usepackage{amssymb}
\usepackage{lscape}
\usepackage{ulem}
\usepackage{txfonts}
\usepackage{lipsum}
\usepackage{float}

\def\mhi{M$_{\rm HI}$}
\def\fgas{M$_{\rm HI}/\rm{M_*}$}
\def\cfgas{C$_{\rm{fgas}}$}
\def\pr{PR}
\def\sigf{$\sigma_{\rm{fit}}$}
\def\sigfn{$\sigma_{\rm{fitN}}$}
\def\s21{S$_{21}$}

\def\sigt{$\sigma_{\rm{fgas}}$}

\def\kms{km~s$^{-1}$}

\title[HI gas fraction estimations] {Pattern recognition in the ALFALFA.70 and Sloan Digital Sky Surveys: A catalog of $\sim$ 500,000 HI gas  fraction estimates based on artificial neural networks.}

\author[Teimoorinia, Ellison \& Patton] {Hossen Teimoorinia$^1$, 
Sara L. Ellison$^1$ \&
David R. Patton$^2$\\
$^1$ Department of Physics \& Astronomy, University
of Victoria, Finnerty Road, Victoria, British Columbia, V8P 1A1,
Canada.\\
$^{2}$Department of Physics and Astronomy, Trent University, 1600 West Bank Drive, Peterborough, ON K9L 0G2, Canada.\\
}

\def\LaTeX{L\kern-.36em\raise.3ex\hbox{a}\kern-.15em
    T\kern-.1667em\lower.7ex\hbox{E}\kern-.125emX}

\begin{document}

\label{firstpage}

\maketitle

\begin{abstract}

The application of artificial neural networks (ANNs) for the estimation of HI gas mass fraction 
(\fgas) is investigated, based on a sample of 13,674 galaxies in the
Sloan Digital Sky Survey (SDSS) with HI detections or upper limits from the Arecibo Legacy Fast 
Arecibo L-band Feed Array (ALFALFA).  We show that, for an example set of fixed input 
parameters ($g-r$ colour and $i$-band surface brightness), a multidimensional 
quadratic model  yields \fgas\ scaling relations with a  smaller scatter (0.22 dex) than 
traditional linear fits (0.32 dex), demonstrating that non-linear methods can lead to
an improved performance over traditional approaches.  A more extensive ANN analysis
is performed using 15 galaxy parameters that capture variation in stellar mass, internal
structure, environment and star formation.  Of the 15 parameters investigated, we
find that $g-r$ colour, followed by stellar mass surface density, bulge fraction and specific
star formation rate have the best connection with \fgas.
By combining two control parameters, that indicate
how well a given galaxy in SDSS is represented by the ALFALFA training set (\pr) and the scatter in the
training procedure (\sigf), we develop a strategy for quantifying which SDSS galaxies our ANN
can be adequately applied to, and the associated errors in the \fgas\ estimation.  In contrast to previous works,
our \fgas\ estimation has no systematic trend with galactic parameters such as M$_{\star}$, $g-r$ and SFR.  
We present a catalog of \fgas\ estimates for more than half a million
galaxies in the SDSS, of which $\sim$ 150,000 galaxies have a secure selection parameter 
with average scatter in the \fgas\ estimation of 0.22 dex. 

\end{abstract}


\section{introduction}

Neutral hydrogen plays an important role in the fuelling pipeline for
star formation activity in galaxies.  The study of HI can therefore
provide insights into some of the main physical processes that drive
galaxy evolution. To this end, numerous surveys (see Giovanelli \&
Haynes 2015 for a recent review) have
been conducted with the aim of determining the HI mass (\mhi) of large numbers of
galaxies, such as  the Arecibo Dual Beam Survey (ADBS, Rosenberg \& Schneider
2000), the HI Parkes All-Sky Survey (HIPASS, Barnes et al. 2001) and the HI
Jodrell All-Sky Survey (HIJASS, Lang et al. 2003). Most recently, the
Arecibo Legacy Fast Arecibo L-band Feed Array (ALFALFA) survey has provided
a wide area, moderate depth (log \mhi/M$_{\odot} >$8.5), blind survey of
HI in the local  ($z<0.06$) universe (Giovanelli
et al. 2005).   Due to the broad inverse correlation
between gas fraction and stellar mass, relatively few massive
galaxies are included in the ALFALFA detections. To complement the
strategy of ALFALFA, the GALEX Arecibo SDSS Survey (GASS) has therefore performed
deep 21~cm observations of $\sim$ 1000 massive (log M$_{\star}$/M$_{\odot} >
10$) galaxies down to a fixed gas fraction limit (Catinella et al. 2010, 2013). 	
Coupled with multi-wavelength data from other large surveys, these large HI
surveys have provided unprecedented
insight into the connections between global gas properties and
environment, nuclear activity, star formation and chemical enrichment (Cortese et al. 2011; Fabello et al. 2011; Bothwell et al. 2013; Lara-Lopez et
al. 2013; Hughes et al. 2013).

Despite the tremendous efforts of HI surveys, the measurement of
HI masses continues to lag behind optical surveys.  For example, in contrast to the
$\sim$ 5000 detections  available for the 40 per cent ALFALFA
data release (Haynes et al. 2011), 
the main galaxy sample of the Sloan
Digital Sky Survey (SDSS) contains more than two orders of magnitude
more galaxies out to $z \sim 0.2$ (e.g. Strauss et al. 2002).  For this reason, there is a
long history of efforts that attempt to infer \mhi\ from more readily
available optical properties (Roberts 1969; Bothun 1984; Roberts \&
Haynes 1994).  In the last decade, such efforts have been able to
capitalize on large, homogeneous datasets, often in combination with
multi-wavelength data in the UV and IR regimes (Kannappan 2004; Zhang
et al. 2009; Catinella et al 2010, 2012; Toribio et al. 2011; Wang et al. 2011; Li et al.
2012; Denes, Kilborn, \& Koribalski 2014; Eckert et al. 2015).
These works have used different linear combinations of
physical parameters to reduce scatter between the estimated and observed
HI gas mass fraction (defined as \fgas\ throughout this paper). 
However, despite the availability of extensive physical parameters, and the
identification of the `best' parameters for \fgas\ estimation, the
improvement in the accuracy of scaling laws has been small.
For example, Zhang et al. (2009) used a linear
combination of $i$-band surface brightness and $g-r$ colour that was
motivated by the Kennicutt-Schmidt law to achieve
a scatter in their \fgas\ estimator of $\sim$ 0.31 dex. Catinella et al.
(2010) found that the near ultraviolet (NUV) NUV$-r$ colour was the single best estimator of
HI mass, and determined a `gas fraction plane' by combining the stellar mass surface density ($\mu_{\star}$), NUV$-r$ and HI gas mass fraction.  However, the scatter in this
relation remained at 0.3 dex (see also Catinella et al. 2012).  Motivated by the
finding that gas fractions are linked to outer disk colours (Wang
et al. 2011), Li et al. (2012) add the colour gradient to their
linear estimator, applied to ALFALFA and GASS samples, but again
the scatter in the HI gas fraction was not improved beyond $\sim0.3$ dex.
Moreover, many of these linear scaling relationships still have significant 
systematics relative to stellar mass, age, colour or concentration indices (e.g., Zhang et al. 2009).

Despite their current limitations, linear scaling relations and the 
gas fraction plane are useful as indicators of `HI normalcy',
which can be used to identify outliers (within a given sample) that are either particularly
gas-rich or gas-poor (e.g. Catinella et al. 2012). Similarly, Cortese et al. (2011) demonstrate
that the distance from the gas fraction plane is a good alternative
to the classic `HI deficiency' parameter (Haynes \& Giovanelli 1984).
However, extreme caution must be applied in using scaling relations
to \textit{estimate} \fgas, due to the
inherently biased nature of the samples from which they are (currently) constructed.
For example, the application of scaling relations derived from the
blind, but shallow, ALFALFA data tends to over-estimate \fgas\ in deeper, or
targeted surveys (e.g., Huang et al. 2012).
Problems may persist even when precautions are taken to identify galaxies
with broadly similar properties as the calibration sample.  For example, Zhang et al. (2009) identify
a sample of star-forming galaxies from which they determine their \fgas\
scaling relation.  It might therefore be expected that the distinction between star-forming and quiescent
galaxies may help to identify samples of galaxies appropriate for the
application of the Zhang et al. (2009) parametrization.  However,
Huang et al. (2012) show that the Zhang et al. (2009) relation is not
a good predictor of \fgas\ in ALFALFA, despite the domination of
star-forming galaxies in the latter.
Therefore, whilst linear scalings reveal the typical relationship
between \fgas\ and other optical/NUV/infrared (IR) properties \textit{for a given sample}, 
they can not be used as a universal predictor of HI gas mass fraction, 
particularly in circumstances when galaxies are expected to deviate from 
the `normal' relation (e.g. Cortese et al. 2011).  

In this paper, we explore the application of non-linear methods, such as artificial
neural networks (ANNs), to the challenge of gas fraction estimation. 
We have previously demonstrated the application of ANN to various astronomical
problems, including pattern recognition for the spectral classification of
galaxies (Teimoorinia 2012), fitting applications in order to estimate
line fluxes which can be used for Active Galactic Nucleus (AGN) classifications, metal abundances etc.
(Teimoorinia \& Ellison 2014) and a large public catalog of estimated IR
 luminosities (Ellison et al. 2016a) that can be used to determined star formation
rates for galaxies dominated by AGN (Ellison et al. 2016b).  Most  recently,
we have presented a method for ranking the parameters involved in galaxy quenching, which
effectively disentangles the complex multi-variate nature of this physical process 
(Teimoorinia et al. 2016).    We refer the interested reader to those
works for details of the general ANN method. The non-linear ANN
techniques used in these previous works are readily  applied to
the estimation of the gas fraction from large input datasets. In this work, we
will train and validate the ANN using galaxies drawn from the ALFALFA.70
data release, matched to galaxies in the SDSS DR7.
The trained network can then be applied to all galaxies in the DR7
(not covered by ALFALFA.70) in order to predict the gas fraction based on their
optical properties.

The layout of the paper is as follows. In Section \ref{sec-sample} we describe
the samples that are used for training the ANN and the method used for rejecting
galaxies that may have their 21~cm fluxes contaminated by the presence of
a close companion within the Arecibo beam. The results in the following sections are
then presented with respect to three main objectives.
First, we will explore whether non-linear methods are able to
reduce the scatter in \fgas\ scaling relations below what has been previously
achieved for linear calibrations 
for a given set of input parameters (Section \ref{sec-example}).  Specifically, 
we use the same optical input parameters as Zhang et al. (2009) in order to compare the scatter of
their linear calibration with that of the non-linear methods. Although this
example uses a limited number of variables for training, it serves as a direct comparison
of the performance of linear and non-linear techniques.
 In Section \ref{sec-example} we demonstrate how
differences in HI surveys can influence the application of scaling relations.
In Section \ref{sec-performance} the input data for the 
ANN  are described and also a performance function, R, is introduced for ranking the weights of
input parameters in the fitting procedure.  A second performance metric is then used to
determined which galaxy variables show the most important physical connection with \fgas.
The third objective is then to develop the non-linear 
relationships between \fgas\ and galactic properties into a procedure that can be
robustly used to estimate \fgas\ for galaxies in the SDSS, the results of which
are presented in Section \ref{sec-fitting}.   In Section \ref{sec-pr} we describe
a pattern recognition procedure that determines how similar a given galaxy is to those
that have been used in the training set (i.e. detections in ALFALFA).  We hence are able to 
assign an ALFALFA detection probability, \pr , to any SDSS galaxy; a robust gas fraction can be 
best determined for galaxies with high detection probabilities (i.e. the galaxy is similar to
the ALFALFA detected galaxies on which the ANN was trained).   The final strategy for
defining a robust sample for applying the ANN is
presented in Section \ref{sec-comb}.  In section \ref{sec-sum} a short summary is presented. 
A public catalog of the  HI gas mass fraction, the control parameters  and errors accompanies 
this paper in electronic format.

\section{Samples}
\label{sec-sample}

\subsection{Training Set}

At the heart of our ANN approach is a training set based on matches
between the SDSS DR7 galaxy catalog and the ALFALFA.70 data release
which contains 6837 galaxies with 21~cm detections at a S/N$>$ 6.5.
Following Haynes et al. (2011), we allocate
a code = 1 to these objects (hereafter AC1). Haynes et al. (2011)
also identify lower S/N detections (code = 2, hereafter, AC2).  Although AC2
galaxies have been matched to galaxies in the SDSS and are therefore likely
detections, we take the conservative measure of not including them as detections
in our training sample.  Our analysis also makes use of galaxies not detected
by ALFALFA.  Non-detections of SDSS galaxies are not explicitly reported in
the ALFALFA data releases, so we make use of the non-detections for
the ALFALFA.40 footprint as computed by Ellison et al. (2015).  The 23,652
non-detections are allocated a code = 3 (AC3).  ) It should be noted that the training sample  is  sufficiently
low redshift (and also validation sets which are discussed in this paper) that it is safe to assume no evolution with redshift
in the quantities of interest.

While each of the galaxies in AC1 consists of an HI source with 
a single optical counterpart, the relatively large diameter of the Arecibo beam 
(3.3$\times$3.8 arcminutes) means that more than one galaxy may contribute to the
21 cm flux of a given source (Haynes et al. 2011; Jones et al. 2015), 
leading to overestimates of the HI fluxes of some galaxies.
We therefore undertake a three-step process to remove from our AC1 sample galaxies whose 
HI flux may be contaminated by neighbouring galaxies within the beam. 

First, for every galaxy in our initial training set, we search for all known spectroscopic 
companions (from Simard et al. 2011) which may contribute significant additional flux to a 
given galaxy's HI measurement. 
Given the paucity of gas in red sequence galaxies, we consider only companions which are blue.  
Following Patton et al. (2011), we classify galaxies as blue if they have Simard et al. (2011) 
global rest-frame colours of $g-r < -0.01*(M_r+21) + 0.65$.
If a galaxy has one or more blue companions which lie within one beam radius (1.9 arcminutes)  
and within a relative velocity of 250 km s$^{-1}$, we conclude that there is a risk of significant HI contamination,
and remove the galaxy from our training set.  
A total of 431 galaxies were removed from our training set for this reason. 

Second, we remove galaxies which lie within one beam radius and 250 km s$^{-1}$ of  
a known ALFALFA source.  While approximately half of these sources have already been removed 
in the previous step, 
a comparable number remain due to spectroscopic incompleteness in SDSS.  In particular, fibre collisions 
lead to a high rate of spectroscopic incompleteness at angular separations less than 
55 arcsec (Blanton et al. 2003, Patton \& Atfield 2008).  
A total of 27 additional galaxies are removed from our training set for this reason.

Finally, we address the possibility of contamination by companions whose centres lie 
outside the ALFALFA beam and yet whose HI flux is likely to overlap the beam.
We use ALFALFA HI gas fractions and the Mendel et al. (2014) stellar masses to 
compute the HI mass of all companions which lie outside the beam but within 250 km s$^{-1}$.
We then estimate the HI radius of each companion, using the relationship between HI mass 
and diameter reported in Wang et al. (2016)\footnote{This equation is in excellent agreement 
with Broeils \& Rhee (1997).}  If a companion's estimated HI radius 
overlaps the beam, we remove the given host galaxy from our training set. 
An additional 50 galaxies are removed from our training set by this criterion, 
leaving a training set of 6329 galaxies whose HI fluxes are unlikely to be 
contaminated by neighbouring galaxies.  Visual inspection of the SDSS 
images of a random subset of these galaxies confirms this interpretation, 
as no obvious blue galaxies are seen within one beam radius (with the 
exception of those with relative velocities greater than 250 km s$^{-1}$).

\subsection{Additional Samples and Ancillary Data}

In order to test our \fgas\ estimator on independent datasets, we will also make use of
several other samples that have been matched with the SDSS. The largest of
these is the GASS sample\footnote{Data publically available at 
http://wwwmpa.mpa-garching.mpg.de/GASS/data.php.}, for which we use the 
342 detections from the data release 3 (Catinella et al. 2013).  
Additionally, we use the Cornell catalog
($\sim 30$ galaxies, Giovanelli et al 2007) as incorporated into the GASS representative sample
and a sample of $\sim40$ post-mergers (PM) presented by Ellison et al. (2015).We have also used 279 galaxies (of 2839, matched with our ANN's input parameter space) from the  Nancay Interstellar Baryons Legacy Extragalactic Survey (NIBLES) sample, presented by van Driel et al. (2016).  They use a sample in the redshift range z$<$0.04 selected on z-band magnitude ($-24 < \rm{M_z} < -13.5$) as a proxy for stellar mass. Galaxies in the NIBLES sample are at the bright end of the SDSS distribution. All of the above samples have large amounts of ancillary information
available, which may be used as input variables for the \fgas\ estimation,
from the following sources:

\begin{itemize}

\item Photometry is taken  from SDSS DR7.  Structural parameters, such as
bulge fractions and galaxy sizes are taken from  the re-processed SDSS images by
Simard et al. (2011) and Mendel et al. (2014).  
The fluxes in different bands ($u, g, r, i, z$)  
are all corrected for  Galactic extinction.

\item  Stellar masses are taken from Mendel et al. (2014) based on
their re-assessment of SDSS photometry.

\item  Total star formation rates were taken from the MPA/JHU catalogs,
which applied a colour dependent aperture correction to account for the
light outside of the SDSS fibre (Brinchmann et al. 2004; Salim et al. 2007).
Star formation rates are only used for those galaxies classified
as `star-forming' by the definition of Kauffmann et al. (2003).
Specific SFRs are determined by combining the SFRs described above
with the stellar masses from Mendel et al. (2014).

\item The halo masses come from the group catalogue of Yang et al. (2007, 2009).

\item  Local environmental densities are computed as $\Sigma_n = \frac{n}{\pi d_n^2}$,
where $d_n$ is the projected distance in Mpc to the $n^{th}$ nearest
neighbour within $\pm$1000 \kms.  Normalized densities, $\delta_n$,
are computed relative to the median $\Sigma_n$ within a redshift slice
$\pm$ 0.01.  In this study we adopt $n=5$.


\item Stellar mass density is defined as log($\mu_{*i})= \log(M_*) -\log(2 \pi R_{50i}^2$) in which M$_*$ is the stellar mass and R$_{50i}$  is the radius (in kpc) enclosing 50 per  cent of the total Petrosian $i$-band flux.

\end{itemize}

\section{A simple example of HI mass estimation from non-linear methods}
\label{sec-example}

 Before launching into a complex many-parameter ANN application for
\fgas\ estimation, we show a simple example of comparing the linear estimation 
of Zhang et al (2009) to a non-linear method.  Zhang et al (2009)
used a sample of $\sim$ 800 galaxies with HI masses in the Hyperleda database that
are matched to galaxies in the SDSS to determine a calibration between \fgas\
and  the $g-$ and $r$-band Petrosian apparent magnitudes
and $i$-band  surface brightness, defined as $\mu_i = m_i +2.5 \log(2 \pi R_{50i}^2$), where 
R$_{50i}$ is the radius (in units of arcsecond) enclosing 50 per cent of the total 
Petrosian $i$-band flux. The calibration presented by Zhang et al. (2009) is:

\begin{equation}
\rm{log~M_{HI}/M_{*}}=-1.73\it{(g-r)}+\rm{0.22\mu_{i}-4.08}
\label{eq-zhnag}
 \end{equation}

\begin{figure}
\centering
\includegraphics[width=6.7cm,height=5.2cm,angle=0]{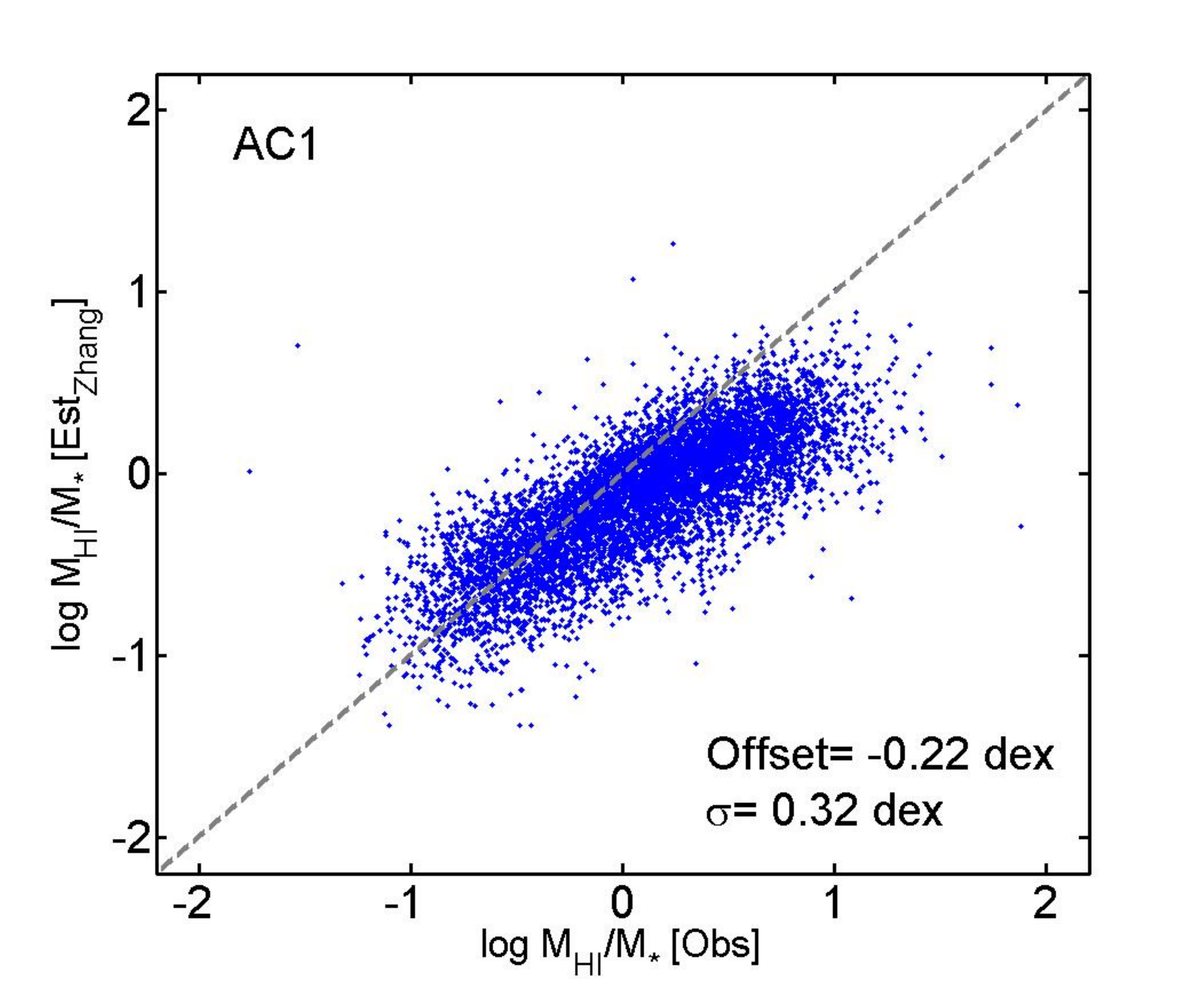}
\includegraphics[width=6.7cm,height=5.2cm,angle=0]{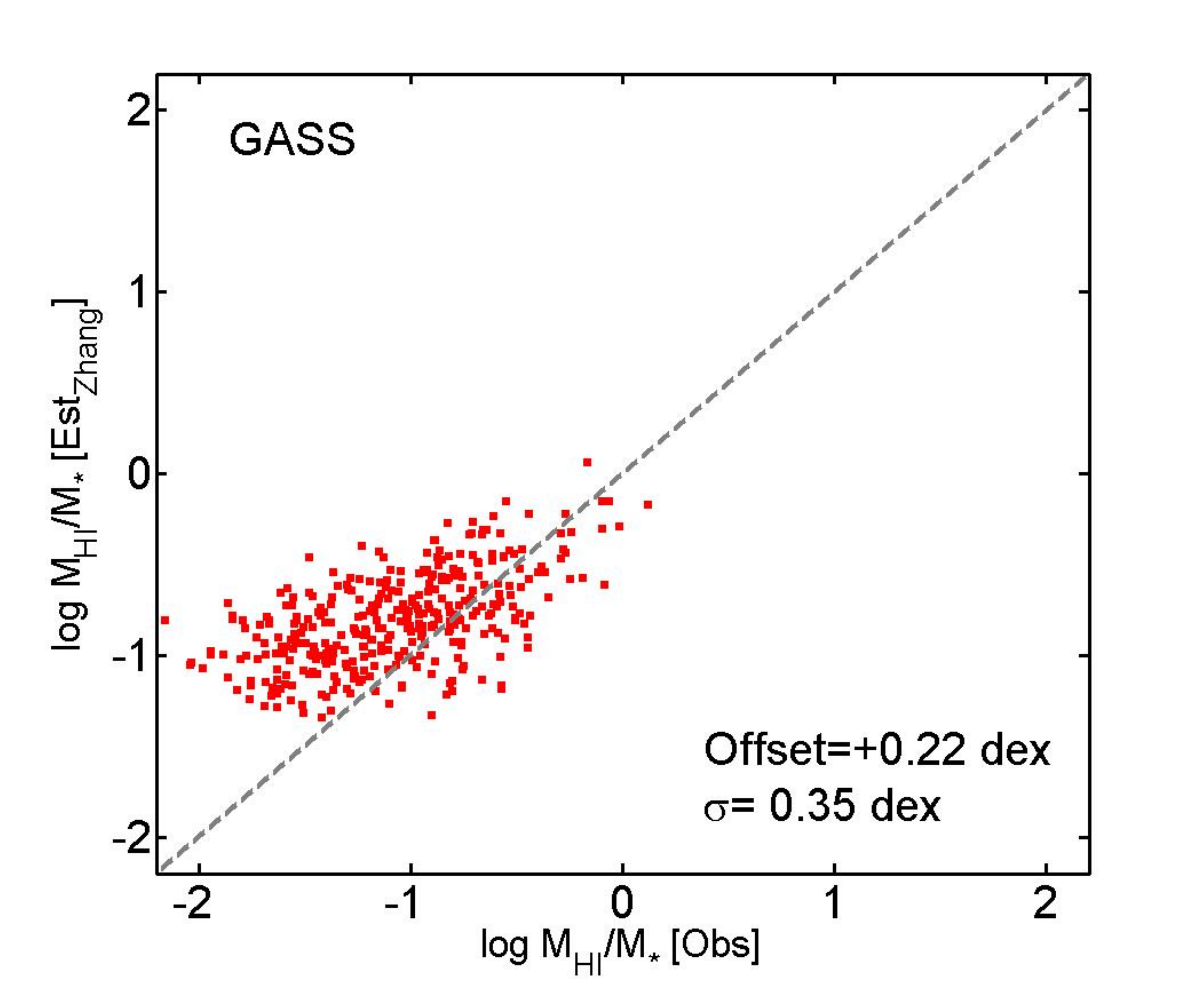}
\includegraphics[width=6.7cm,height=5.2cm,angle=0]{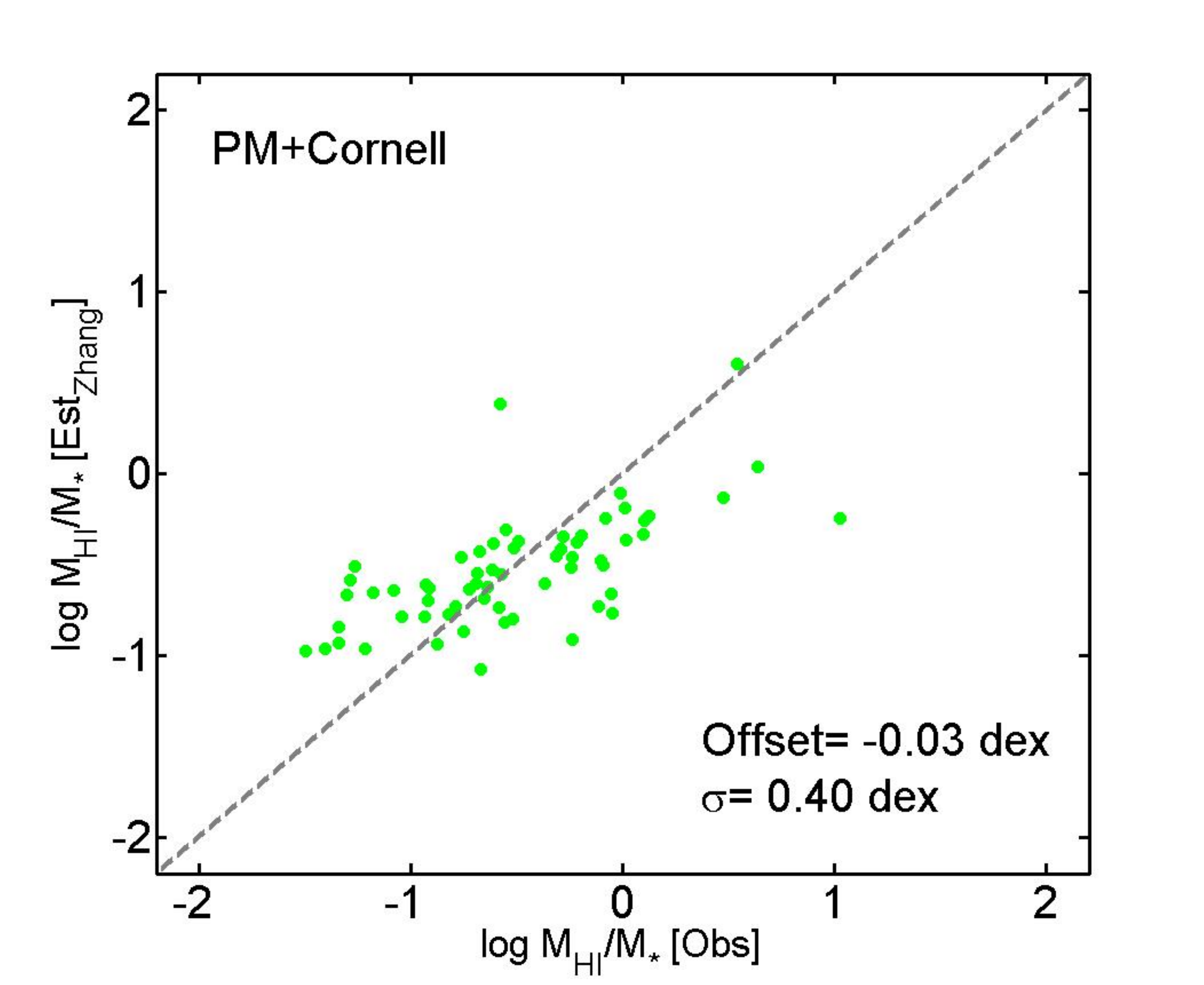}
\includegraphics[width=6.7cm,height=5.2cm,angle=0]{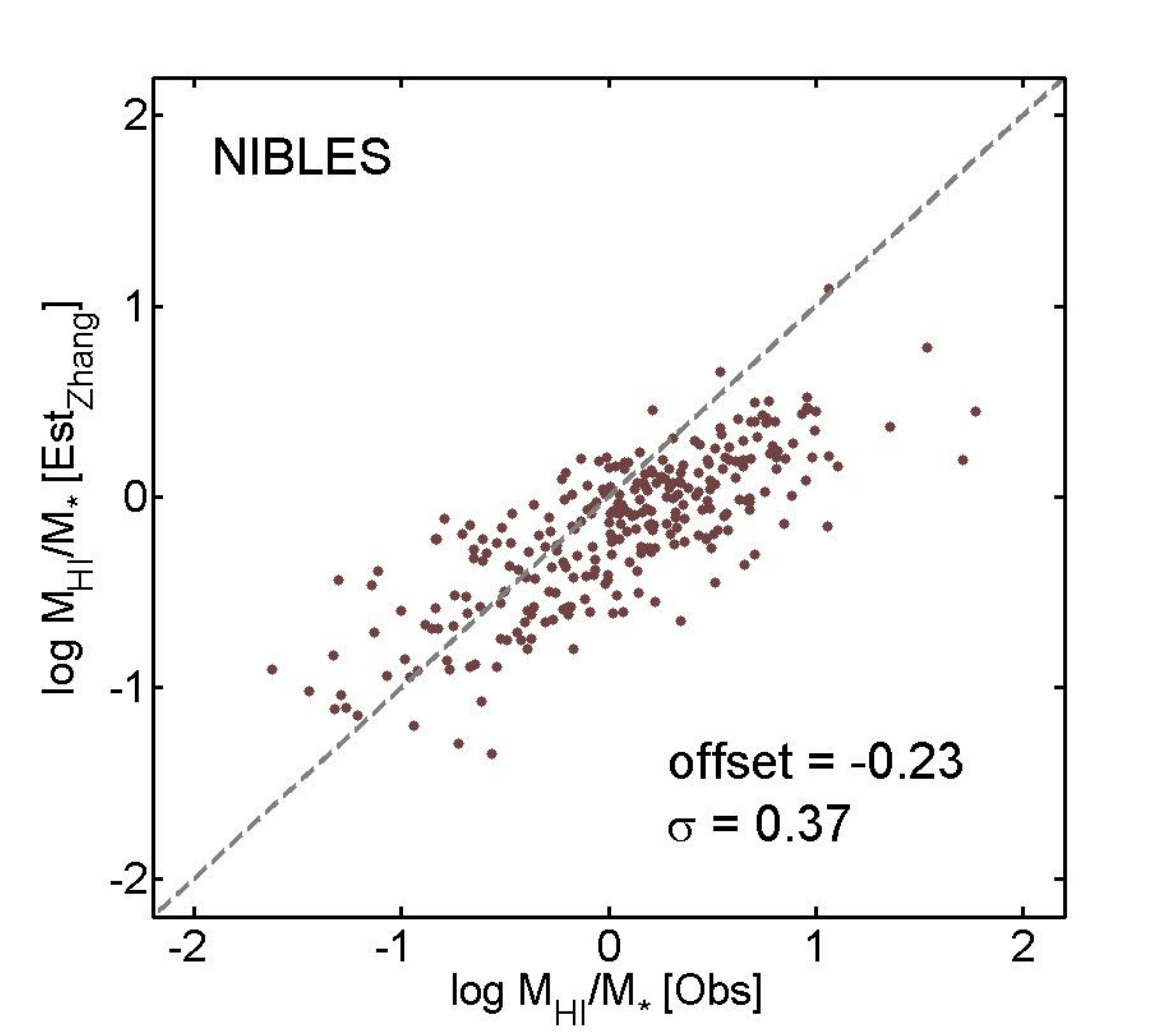}

\caption{Comparison of observed and estimated \fgas\ based on the \fgas\ calibration
of Zhang et al. (2009), given by Eq.\ref{eq-zhnag}.  The mean offset between the
estimated and observed \fgas\ and the scatter ($\sigma$) therein are shown in each plot. The
gray dashed line shows the one-to-one relation of \fgas\ ($_{est}) =$\fgas ($_{obs})$.}
\label{fig-zhang-valid}
\end{figure}

Using Eq. \ref{eq-zhnag} we estimate \fgas\ for the samples described in Section \ref{sec-sample}:
AC1, GASS, PM and Cornell, where we group the latter two samples together for plotting
purposes due to their small size.  In Fig. \ref{fig-zhang-valid} we compare the \fgas\
estimated by Zhang et al. (2009)'s calibration (Eq. \ref{eq-zhnag}) and the observed
values.   In each panel, the values in the lower right
corner give the mean difference between the estimated and observed \fgas,
and the scatter ($\sigma$) in these differences.  There are systematic offsets 
between the observed and estimated \fgas\ values, that vary between the samples.  For example,
the top panel of Fig.  \ref{fig-zhang-valid} shows
that \fgas\ in AC1 (ALFALFA detections) is underestimated, on average,
by 0.22 dex by Zhang's formula. The same seems to be true in the NIBLES data, with the offset occurring in a similar regime (i.e. log \fgas [Obs]$>0$).
The offset between AC1 and the Zhang et al. formulation has been previously reported by 
Huang et al. (2012) who suggest that
differences in the methods of calculating the stellar masses may account for a systematic 
deviation of $\sim0.2$ dex.  The top panel of  Fig. \ref{fig-zhang-valid} and
explanation by Huang et al. (2012)
demonstrate an important caveat for the application
of any calibration method -- if parameters are not uniformly derived, even `perfect'
calibrations will perform poorly.  It is therefore vital to apply calibrations to
datasets whose parameters have been derived as closely as possible to the original
data.

The second plot from the top of  Fig.  \ref{fig-zhang-valid} shows that, in contrast to what
was seen in AC1, the \fgas\ estimated for the GASS sample using Eq. \ref{eq-zhnag}
has a tendency to be
mildly over-estimated.  Taken together, the top and middle panels therefore
imply a total difference between AC1 and GASS estimations $\sim$ 0.45 dex.
As described above, Huang et al. (2012) have suggested that this may be
due, at least in part, to differences in stellar mass calibrations.  In
order to test this suggestion, we can re-derive the best fit linear
coefficients of  ($g-r$) and $\mu_i$ in Eq.\ref{eq-zhnag} using AC1 and test this calibration
on the GASS data.  Since both samples have consistent sources of input
parameters and stellar mass measurements, any systematic trend caused by
inconsistencies therein should be removed. The AC1-calibrated version of
 Eq.\ref{eq-zhnag} is then

\begin{equation}
\rm{log~M_{HI}/M_{*}}=-2.332\it{(g-r)}+\rm{0.168\mu_{i}-2.528}
\label{eq-zhnag_new}
 \end{equation}

\begin{figure}
\centering
\includegraphics[width=8cm,height=6.5cm,angle=0]{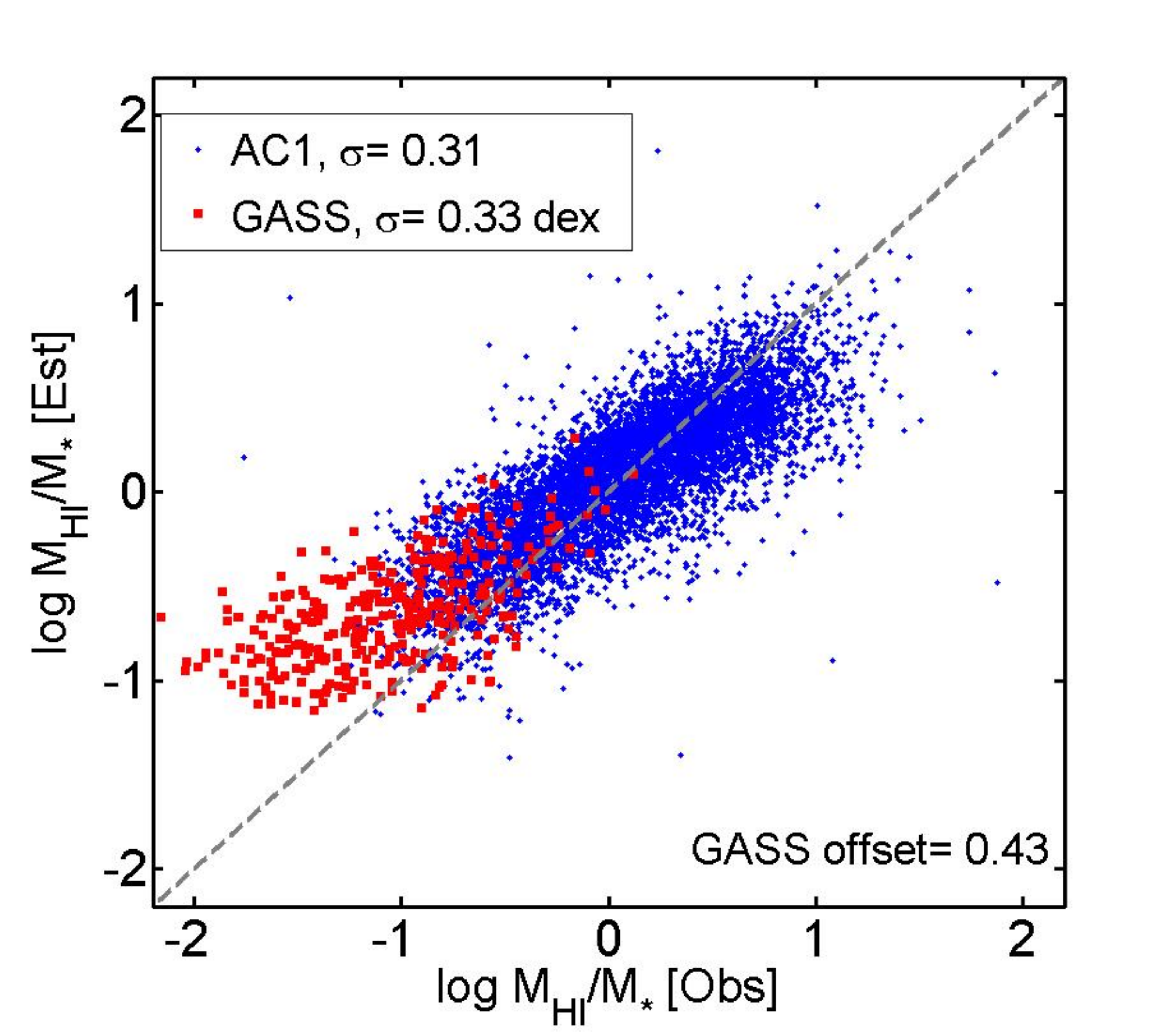}
\caption{A re-calibration of  Eq.\ref{eq-zhnag} based on AC1 data yield the new
coefficients given in  Eq.\ref{eq-zhnag_new}.  This new linear relationship is
applied to AC1 (blue points) and GASS (red points). The
gray dashed line shows the one-to-one relation of \fgas\ ($_{est}) =$ \fgas ($_{obs}$). 
The deviations of the estimated
from the observed \fgas\ in AC1 indicate that a linear combination of $g-r$ and $\mu_i$ 
is insufficient for  estimating the HI gas fraction.   A systematic error of $\sim0.45$ dex is 
found when the AC1 linear calibration  in  Eq.\ref{eq-zhnag_new} is applied to the GASS sample. 
This offset can no longer be due to inconsistencies in the input data.  }
\label{fig-alfa-gass-two-par}
\end{figure}

In Fig. \ref{fig-alfa-gass-two-par} we now compare
the observed and estimated \fgas\ for the newly derived coefficients in Eq.\ref{eq-zhnag_new}
in the AC1 and GASS samples (blue and red points respectively). The first important result
demonstrated by  Fig. \ref{fig-alfa-gass-two-par}, is that  Eq.\ref{eq-zhnag_new} does not
perform well at estimating the data on which it is calibrated (AC1), indicating that AC1
can not be well fit with the parametrization in  Eqs. \ref{eq-zhnag} and \ref{eq-zhnag_new}.
Application to the GASS sample also results in a strong systematic offset.  Importantly,
the mean offset between the estimated (using Eq. \ref{eq-zhnag_new})
and the observed GASS data is still $\sim0.45$ dex, the same as was inferred between AC1 and GASS
using the original coefficients in  Eq. \ref{eq-zhnag}.  This difference can no longer 
be attributed to differences in inhomogeneous stellar mass estimates or photometry, 
which were derived identically for the two samples.  As we will demonstrate below,
the offset is due to the fundamentally different nature of the galaxies in
the two samples.

To compare the performance of the linear (Zhang et al. 2009) and
non-linear approaches, we follow the methods outlined in Teimoorinia
\& Ellison (2014), in which non-linear methods were used to determine
Balmer decrements and emission line fluxes.  Here,  we use the AC1 sample
of HI detections from ALFALFA.70, combined with the same parameters used by Zhang et al. 
(2009) and tested in the above linear fit tests ($g-r$ and $\mu_i$).  Therefore,
the only change we are making is in the methodology of the fitting, not in the
parameters used.  We use the Levenberg-Marquardt algorithm (Marquardt  1963)
to find the coefficients in the following equation:

  \begin{equation}
f(C,X)=\sum\limits_{i,j=1}^N C_{ij}X_iX_j+\sum\limits_{i=1}^N C^{'}_iX_i+C^{''}.
\label{eq-basis-function}
 \end{equation}

In the above equation, $f(x)$= log \fgas\  and X is:

\[ X= \left( \begin{array}{c}
m_g\\
m_r\\
m_i\\
R_{50i}\\
\end{array} \right)\]

Here, N=4 and X$_i$ (i=1 to 4) in which  $m_{g}$  , $m_{r}$, $m_i$ are  apparent magnitudes. These
are the individual variables that comprise $g-r$ and $\mu_i$ in the parametrization
of Zhang et al. (2009) in   Eqs. \ref{eq-zhnag} and \ref{eq-zhnag_new}. We use 
AC1 as the training set  and determine the 15 coefficients required for the above parametrization. 

\begin{figure}
\centering
\includegraphics[width=8cm,height=6.5cm,angle=0]{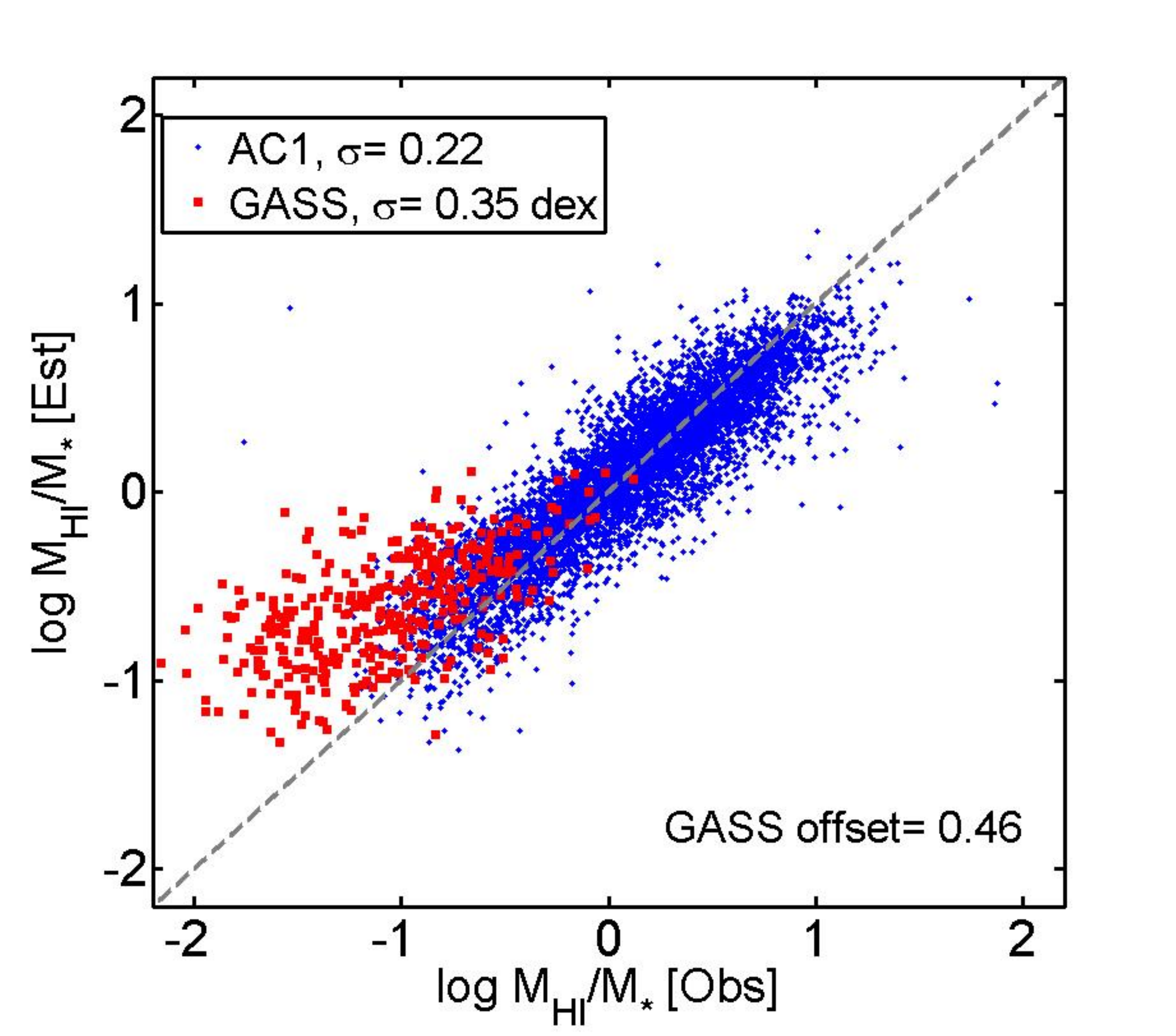}
\caption{The same as Figure \ref{fig-alfa-gass-two-par}, using the same input variables,
but now using the matrix representation in Eq. \ref{eq-basis-function} with 15 independent 
coefficients rather than a linear fit. The systematic trend in AC1 (blue points) is removed and the
scatter is reduced from 0.32 to 0.22 dex. This demonstrates the improvement that is possible with
the application of non-linear methods.   However, a systematic offset remains in the
GASS sample (red points) and there is no improvement in the relative systematic error.}
\label{fig-alfa-gass-4-par}
\end{figure}

Fig. \ref{fig-alfa-gass-4-par} shows the estimated vs. observed gas fractions derived from 
Eqn. \ref{eq-basis-function}, which is a representation of the matrix-based solution. 
The residual tilt in  Fig. \ref{fig-alfa-gass-two-par} in the AC1 data (blue points) is now completely
removed and scatter is reduced from 0.31 dex in the linear calibration (Eq. \ref{eq-zhnag_new}) 
to 0.22 dex. This demonstrates that  although $g-r$ and $\mu_i$ \textit{can} be used for \fgas\ calibration, a more complex representation, such as  the matrix form, can provide a better match (due to  more complex connections between the parameters). Moreover, this matrix form can also remove the deviation and skew seen in Fig. \ref{fig-alfa-gass-two-par} for AC1, which is obtained using only three (different) coefficients.

\begin{figure}
\centering
\includegraphics[width=8.cm,height=6.5cm,angle=0]{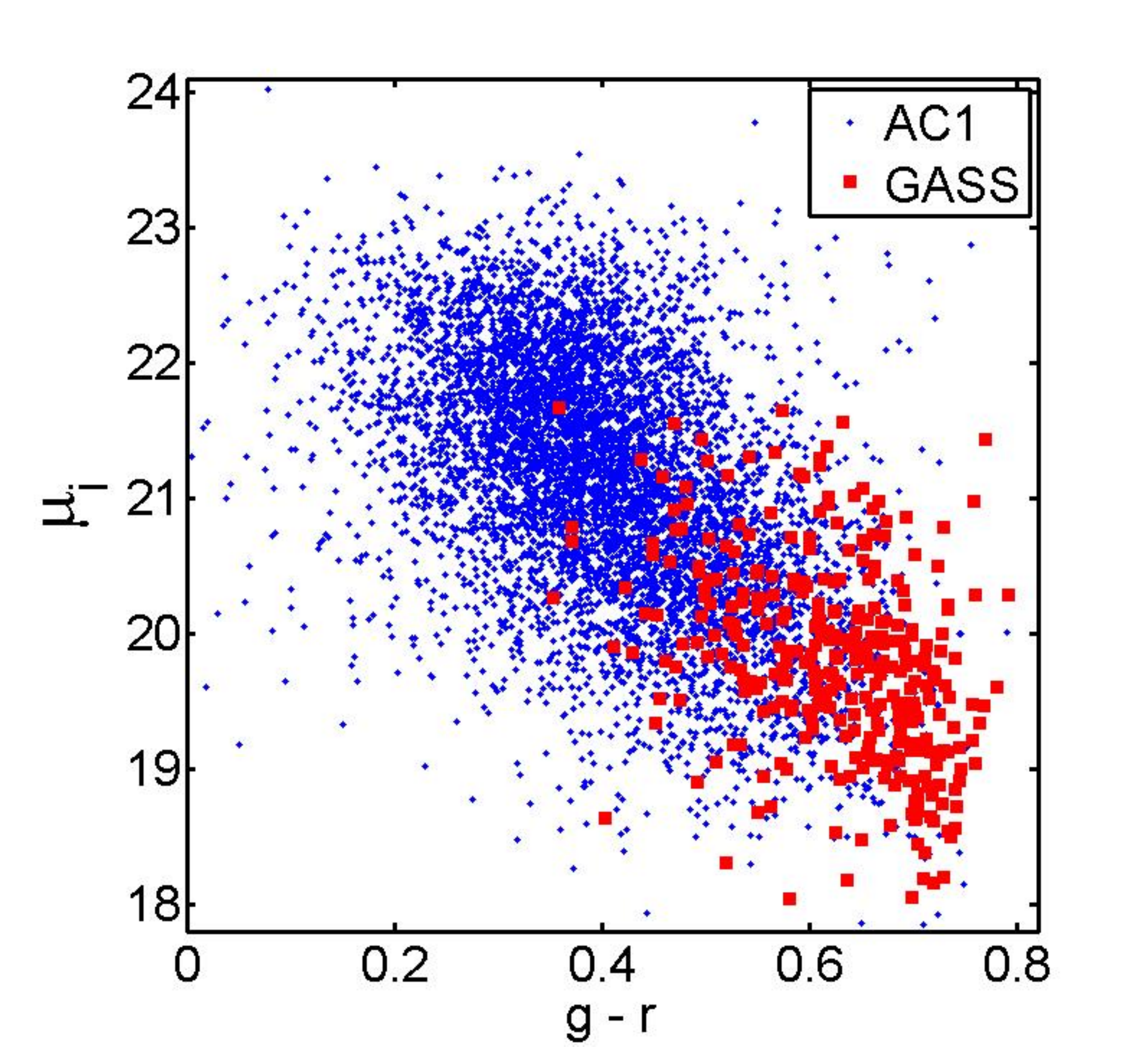}
\includegraphics[width=8cm,height=6.5cm,angle=0]{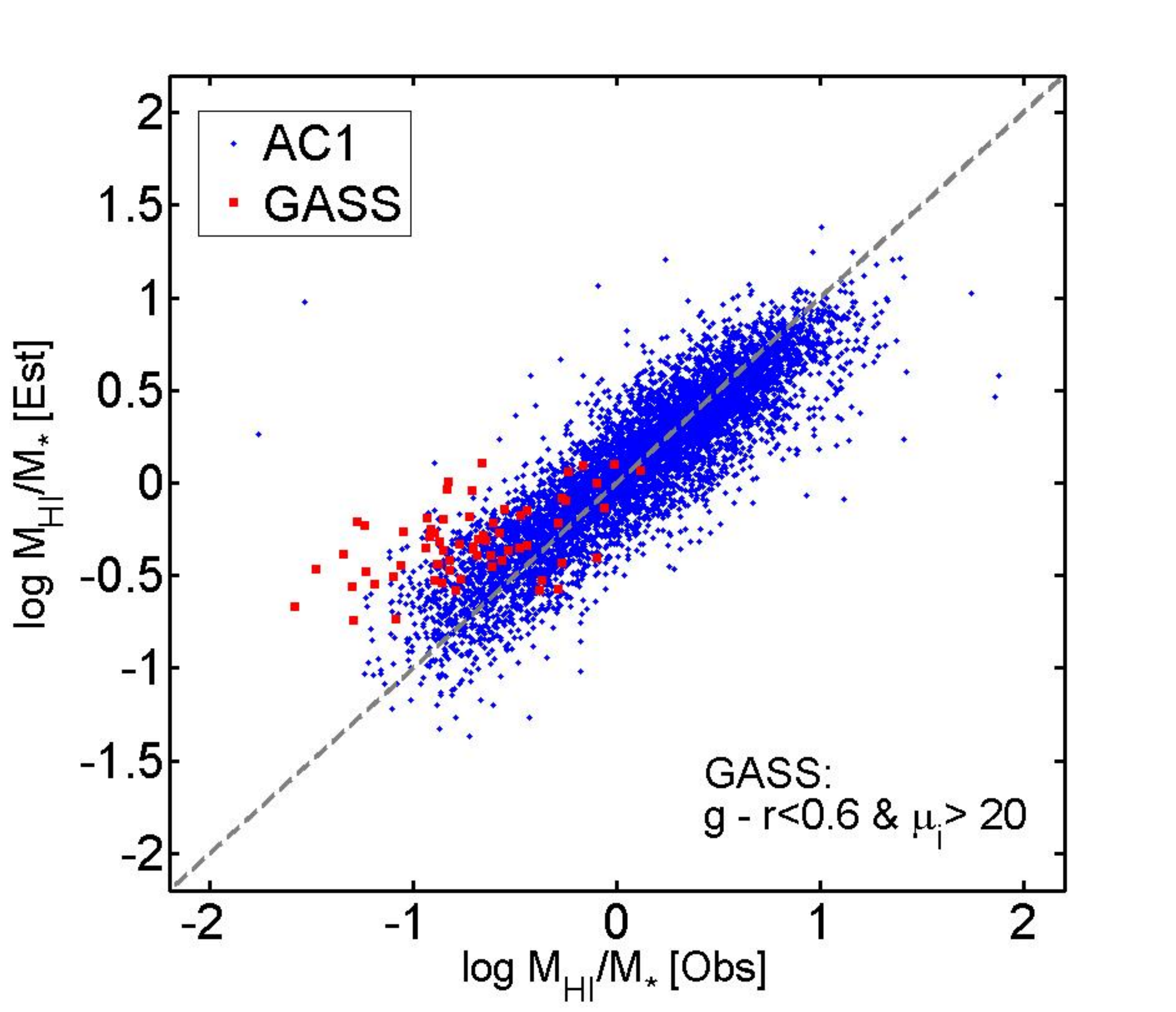}
\caption{Top panel: $\mu_i$ vs. $g-r$  for AC1 (blue) and GASS (red); 
the two samples are preferentially located in different regions of parameter space.
Lower panel: As for Fig. \ref{fig-alfa-gass-4-par}, but restricting GASS data points
to the region  $\mu_i>20$ and $g-r<0.6$, where the AC1 sample is mostly located.}
\label{fig-gr-mu}
\end{figure}

Despite the significant improvement in AC1, Fig. \ref{fig-alfa-gass-4-par} shows that 
no improvement in scatter or systematic offset is seen in the GASS dataset.  The persistent
systematic offset shown in  Fig. \ref{fig-alfa-gass-4-par} for the GASS data could
be due to the two samples (AC1 and GASS) representing galaxies of different physical 
nature, so that even the matrix form of the estimator 
cannot extrapolate it in a suitable manner. This is demonstrated in Fig.  \ref{fig-gr-mu} 
where we plot $\mu_i$ vs. $g-r$  
for the two samples; whilst there is some overlap, the GASS galaxies are preferentially
located at the reddest $g-r$ colours and lowest values of $\mu_i$, that is poorly represented
in AC1. It is possible to select a subset of the GASS data that are relatively well
represented in the AC1 data, for which we might expect the calibration to perform better. 
In the lower panel of  Fig.  \ref{fig-gr-mu} we again plot the comparison of
estimated and observed \fgas\ shown in  Fig. \ref{fig-alfa-gass-4-par}, but now limiting the GASS
data to the range $\mu_i>20$ and $g-r<0.6$, which is more populated by the AC1 sample.  
The estimated \fgas\  does not remove all the offsets; however, this estimate is now closer to the observed one for this limited GASS sample.
These tests demonstrate that whilst a single ANN model could not be found
that is a good representation of all of the galaxies in the combined GASS and ALFALFA samples,
it is nonetheless possible to post-facto exclude those galaxies for which the ANN is not suitable.
For this reason, our approach for the remainder of the paper is to train our networks
only using the ALFALFA sample, and then to develop a set of criteria from which we
can assess the robustness of the gas fraction estimate.
The cuts used in   Fig.  \ref{fig-gr-mu} represent a rather crude approach, 
and practical here only because we are 
dealing with a two-dimensional input parameter space. With greater dimensionality (higher
number of input parameters for the calibration), such simple cuts are cumbersome and complex, and
ultimately subjective with no quantitative assessment of suitability.  Therefore, whilst
the general approach of limiting the data suitable for the application of a given calibration
is desirable, a more sophisticated method is required.  We return to this issue in Section
\ref{sec-pr}.

In this section, we have demonstrated that for a given set of input parameters
a non-linear approach can yield a significant improvement in the estimation
of \fgas\ over a linear fit to the data.  The matrix representation 
provides a very good option for determining the gas fraction for galaxies with only 
photometric data.  However, using a larger number of input  parameters, the accuracy of the \fgas\
estimation may be further improved. As we move to larger numbers of
input parameters, the complexity of the matrices increases, and ultimately becomes very cumbersome,
such that ANN  can provide a better framework for the extension into many
variable space.  Another important result from this section is the potential
pitfalls associated with combining different datasets for training.  Homogeneity is key 
in the training process. If the training sample is heterogeneous (either in its intrinsic properties 
or in the methods used to obtain measured or derived properties) the network will be compromised.  
Moreover, the mixed quality or depth of a combined dataset (e.g. ALFALFA+GASS) makes it very challenging to 
determine a unique set of control and quality parameters, which is a 
critical part of our analysis.  It is therefore extremely rare that different datasets are 
(or can be) combined for training.  Instead, we use a single sample for our training set
(ALFALFA) and then define the regime for which it can be robustly applied to other samples.
This might be done in linear examples using a simple criterion such as a cut  
on a physical parameter. Since the parameters can be correlated with each other in 
different and complex ways,  if we want to use  a higher dimensional space as a
training set then  some simple cuts on only some input data might not be a suitable approach.  
In Section \ref{sec-pr}
we will return to this point and describe a pattern recognition method using an ANN model 
to achieve this objective.  

\section{Input parameters, performance function and ranking}
\label{sec-performance}

An important pre-analysis step is the selection of the input parameters on
which the ANN will be trained.  Galaxy evolution involves a complex interplay
between many parameters and the manual exploration of parameter space
normally restricts investigation to a handful of variables at a time.  However,
with an ANN approach, the parameter space can be expanded straightforwardly
to enable a multi-variate analysis.

\subsection{Selection of galaxy variables for training}

In selecting the parameters used for our training set, we have attempted to incorporate
the primary physical variables that may affect the HI gas fraction. Some of the parameters are not independent; for example, we use both g- and r-band magnitudes as well as g-r colour.  This repetition is not detrimental to the network's performance, but contributes stability. Photometry
and galaxy colour represent some of the raw observables that have been shown to
correlate with gas fraction (e.g. Kannappan 2004; Denes et al. 2014; Eckert et al. 2015).  
These are in turn related to physical parameters
such as galactic stellar mass, which has been shown to exhibit a strong anti-correlation
with \fgas\ (e.g. Zhang et al. 2009; Catinella et al. 2010; Huang et al. 2012).  
However, internal galaxy structure, size and the distribution
of stellar mass appear to be even more tightly correlated with gas fraction than stellar mass itself
(Zhang et al. 2009; Catinella et al. 2010; Toribio et al. 2011; Wang et al. 2016).  
Furthermore, Brown et al. (2015) have
argued that specific star formation rate also modulates \fgas\ at fixed M$_{\star}$.  Environment
also appears to play a role in determining gas fraction with gas fractions suppressed in both
the cluster (e.g. Chung et al. 2007; Cortese et al. 2011; Denes et al. 2014) and group
(e.g. Verdes-Montenegro et al. 2001; Rasmussen et al. 2008; Kilborn et al. 2009; Catinella et al. 2013; Hess \& Wilcots 2013;  Denes et al. 2016; Odekon et al. 2016) environments. We have included two environmental
metrics in our list of training variables, halo mass and $\delta_5$, although Brown et al. (2016)
have recently shown that it is the former of these that dominates the environmental
dependence of \fgas.

The full list of 15 parameters used in our work, based on the SDSS imaging and 
spectroscopic data, is presented in Table \ref{tab1}, which includes photometry, 
metrics of internal size and structure, star formation and environment.
We note that not all the 15 parameters are available for
all galaxies. This requirement reduces the number of galaxies in both the GASS and NIBLES samples from their complete data release.  In the analysis that follows,
the full AC1 and AC3 samples are therefore sometimes restricted
further by the lack of input parameter data.  If a certain input variable
(such as a magnitude in a band) is required for a given training run,
galaxies without a robust measurement of that input variable are excluded.  Amongst the
parameters in Table \ref{tab1},  there are
two notable omissions of parameters that may significantly dictate \fgas. The first is
angular momentum, which has recently be proposed by Obreschkow et al. (2016) to dictate
HI gas fractions based on arguments of gas instability in the interstellar medium (see
also Huang et al. 2012; Maddox et al. 2015).
Unfortunately, we do not have metrics of angular momentum available for our sample.
We have also not included a NUV colour, proposed by several works (e.g. Cortese et al. 2011;
Catinella et al. 2013, Brown et al. 2015) to be the single most important variable in their samples.
Requiring NUV photometry reduces our sample size by a factor of more than 4, to only $\sim$
1400 galaxies, which was found to be inadequate for ANN training.

\begin{table}
\begin{center}
\caption{Galaxy variables used in training sets.}
\begin{tabular}{ll}
\hline\hline
Input data & Description\\
\hline
 M$_*$ & stellar mass    \\
 M$_u$ &  $u$ band absolute magnitude  \\
 M$_g$ &  $g$ band absolute magnitude   \\
 M$_r$ &  $r$ band absolute magnitude   \\
 M$_i$ &  $i$ band absolute magnitude   \\
 M$_z$ &  $z$ band absolute magnitude   \\
 $g-r$ & observed colour  \\
$\mu_{*i}$ & $i$-band stellar mass density  \\
SFR  &star formation rate\\
sSFR & specific star formation rate   \\
 M$_{\rm{Halo}}$ &halo mass  \\
 $\delta_{5}$  & local galaxy density\\
 $\rm{rhalf_r}$  &half light radius (kpc) in the $r$ band\\
 B/T & bulge-to-total fraction in the $r$ band\\
  $\rm{rd_{disk}}$ &disk radius (kpc) in the $r$ band\\
\hline
\end{tabular}
\label{tab1}
\end{center}
\end{table}

\subsection{The ANN performance metric, R}

In this section, we introduce the use of a performance function, as a metric of the
quality of the fit between gas fraction and a given variable.  This performance metric
can be a simple linear regression of  the estimated and the observed values or a Spearman's 
rank correlation number (see Huang et al. 2012 for more details), but it may also be a more
complicated figure of merit (e.g. Teimoorina et al. 2016).   We use the
coefficient of determination, $R^2$, which is a measure of goodness of fit  and is defined as:

\begin{equation}
R\rm{^2=1-\frac{\sum\limits_{i=1}^n (target-y_{fit})^2}{(n-1)\times Var(target)}}.
\label{eq-R}
 \end{equation}

In this equation $y_{\rm{fit}}$ is a linear fit to the target (observed data) and the estimated 
values (obtained by ANN).  Var is the variance and $n$ is the number of objects in the sample. 
$R$ ranges from 0 to 1 in which $R\sim0$ indicates that the fit is not significantly better
than a model in which $y_{\rm{fit}}$ = constant. A value of $R$=1 indicates that the linear 
equation ($y_{\rm{fit}}=aX+b$), where X is the observed \fgas, predicts 100\%  of the 
variance in the target ($y$, in this case the predicted gas fraction).  Each parameter 
listed in Table \ref{tab1} has a certain contribution to estimating \mhi\ which can be 
considered as a weight in the fitting procedure, such that parameters with higher values of R
contribute more significantly to the combined estimate of \fgas.

\begin{figure}
\centering
\includegraphics[width=9cm,height=7.5cm,angle=0]{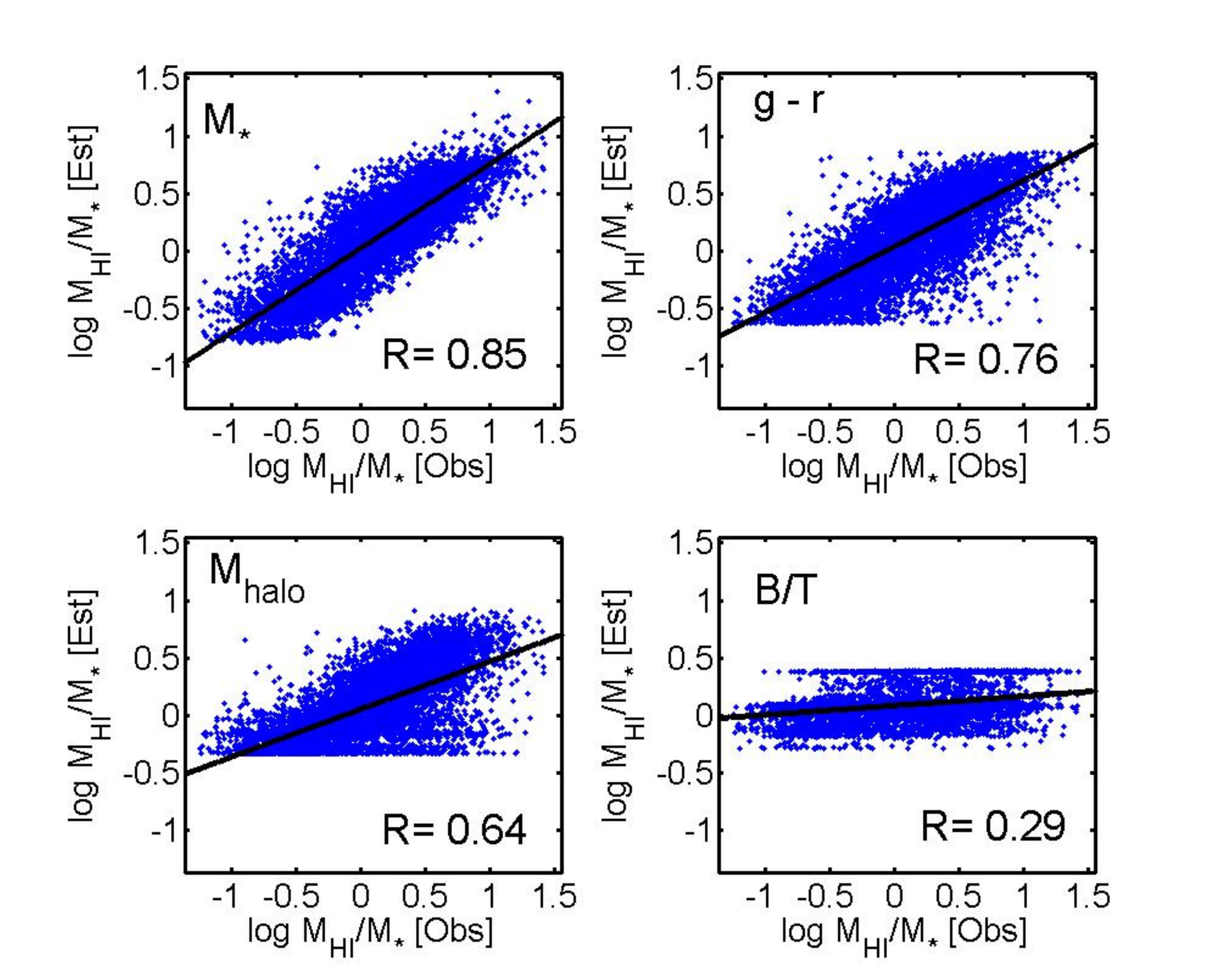}
\caption{Comparison of predicted and observed \fgas\ for four individual parameters taken
from our complete list of 15 (Table 1).  These examples are selected because
they span a wide range of performance, as parametrized by the R value given in the
lower right corner of each panel. }
\label{fig-perform}
\end{figure}

In Figure \ref{fig-perform} we show four different parameters taken from 
Table \ref{tab1} as an example of the performance metric functionality. The value
of R for each variable is given in the lower right of each panel.  Amongst these four
examples, it can be seen that stellar mass has the highest value of R=0.85, and indeed the
scatter between the observed and predicted \fgas\ is relatively small.  Although parameters
with low R, such as B/T (R=0.29) provide little improvement in our predictions of \fgas\
their inclusion with the ensemble of parameters can still provide stability to the network.
We also note that in some regimes, some of our parameters may become unreliable, such as
M$_{halo}$ in the low mass regime.  In these cases, variables act simply as random numbers,
and in the limit of a large training sample such as ours, such random variables do not
decrease the performance of the network.  For these reasons, all 15 parameters are used in
our final ANN.

\begin{figure}
\centering
\includegraphics[width=9cm,height=4.5cm,angle=0]{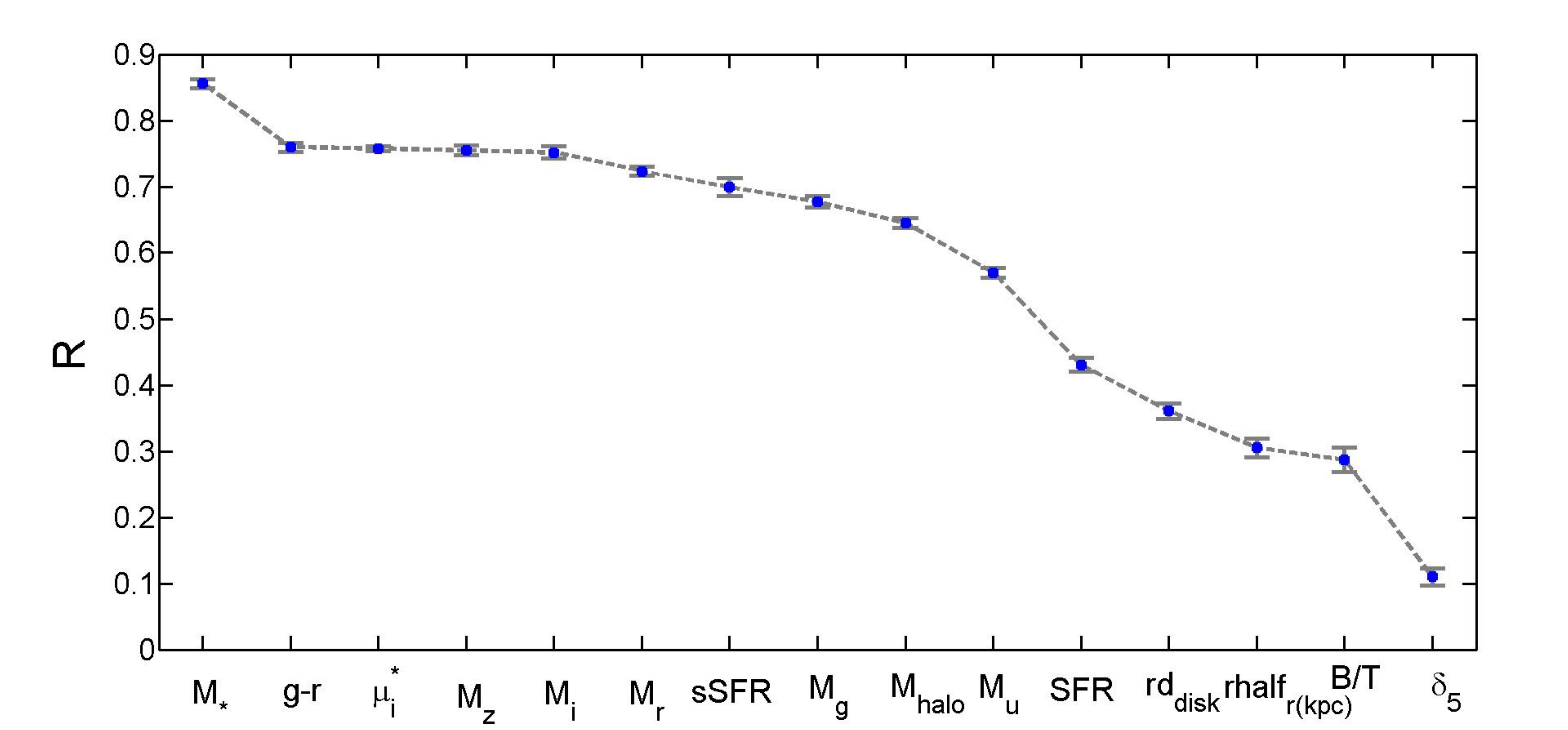}
\caption{The performance number R for all 15 parameters listed in Table \ref{tab1}.  The R parameter reflects the amount of scatter present between the observed and predicted gas fraction and hence the relative weight of a given parameter in a fit. } 
\label{fig-perform15}
\end{figure}

In Figure \ref{fig-perform15} we show the performance numbers for all the 15 input parameters.  It should be noted that these performance numbers are only the weights that  show the contribution of each single parameter in the fitting procedure (i.e., when we use only a parameter from 15 for fitting) and should not be considered as a ranking in physical importance.  In order to extract physical determism, different statistical methods such as receiver operating characteristics (ROC) (Teimoorinia et al. 2016 and reference therein) can be applied.

\subsection{The physical link between galaxy variables and gas fraction}
\label{sec-rank}
\begin{figure}
\centering
\includegraphics[width=9cm,height=4.5cm,angle=0]{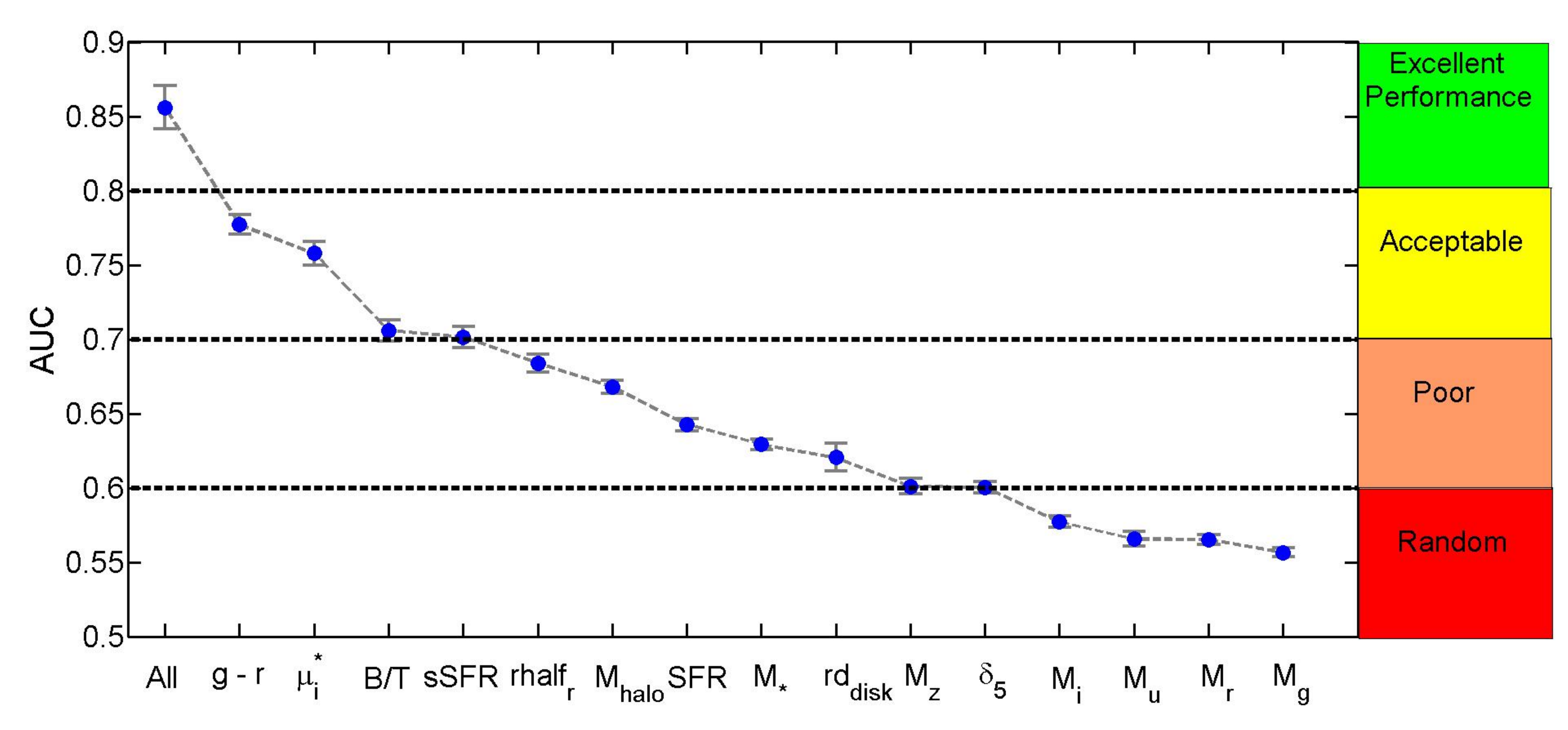}
\caption{The performance number AUC for all 15 parameters listed in Table \ref{tab1}.  In contrast to R, the AUC indicates the relative importance of a given parameter for determining \fgas.}
\label{fig-AUC}
\end{figure}

In Teimoorinia et al. (2016) we described a different performance number, namely the area under the curve (AUC) of ROC plots which can be used to link the physical significance of one variable to the value of another.  This method was applied by Teimoorinia et al. (2016) to determine which were the most important variables for determining whether or not a galaxy's SFR was quenched.  A full description and worked examples of the AUC parameter can be found in that paper.  In brief, the AUC ranges from 0.5 (a random number with no physical import) to 1 (outstanding performance governing the physical link between two variables).  In Figure \ref{fig-AUC} we show the ordered values of AUC for the 15 parameters used in this work.  The numerical values of AUC are traditionally associated with qualitative ranks, ranging from `outstanding' (AUC$>$0.9) to `random' (no physical importance, AUC$<$0.6), e.g. Hosmer \& Lameshow (2000).  Figure \ref{fig-AUC} shows that $g-r$ colour and stellar mass surface density are the two best performing indicators of \fgas, and indeed these parameters have featured widely in the literature (e.g. Zhang et al. 2009; Catinella et al. 2010, 2012; Cortese et al. 2011).   Whereas B/T has a low value of R $\sim$ 0.3, it has an AUC = 0.71, placing it as the 3rd most important parameter for determining \fgas.    Specific SFR also plays a marginally acceptable role in governing \fgas, but the 11 other parameters in our list have a formally low impact on the gas fraction.  This includes parameters that are linked to environment: halo mass and $\delta_5$, indicating that such parameters are not the \textit{prime} drivers of \fgas, although they may still contribute at a lower level once higher performing variables have been accounted for.  Moreover, although the value of R for stellar mass is high, we can see from Figure \ref{fig-AUC} that the AUC value associated with M$_{\star}$ is 0.63, and this parameter therefore performs little better than a random variable.  Although none of the individual parameters has a particularly high AUC, when all 15 are considered together, the AUC = 0.86, characterized as an `excellent' (but not `outstanding') indicator.

\section{Fitting results}
\label{sec-fitting}

Having defined our samples (Section 2.1), removed galaxies with contamination (Section 2.1), 
chosen 15 relevant variables for fitting the target data (Section 4.1 and Table 1), defined
their respective weights (Section 4.2) and ranked their relative physical importance (Section 4.3), 
we now proceed with a Bayesian neural network model 
(e.g., Ellison et al. 2016a) to fit the 6329 uncontaminated galaxies from AC1.  In practice,
we train 25 different networks with different initialization conditions
and select the 20 best performing networks. Hence,  
each galaxy has 20 estimations of \fgas, from which we adopt the mean value.  We can also quantify the
scatter (\sigf) in the estimated \fgas\ (for a given galaxy, and for a given variable)
for these best 20 performing networks, which can be used to quantify an uncertainty in the network
estimation.  That is, if a given galaxy's \fgas\ is robustly estimated by the 20 networks
(for a given variable) \sigf\ will be small.  If the \fgas\ estimation is unstable, \sigf\ will
be large.  We have previously used \sigf\ as a way of
identifying which subset of galaxies have robust ANN estimations (see Ellison et al. 2016a, for a more technical discussion).

\begin{figure}
\centering
\includegraphics[width=8.2cm,height=6.3cm,angle=0]{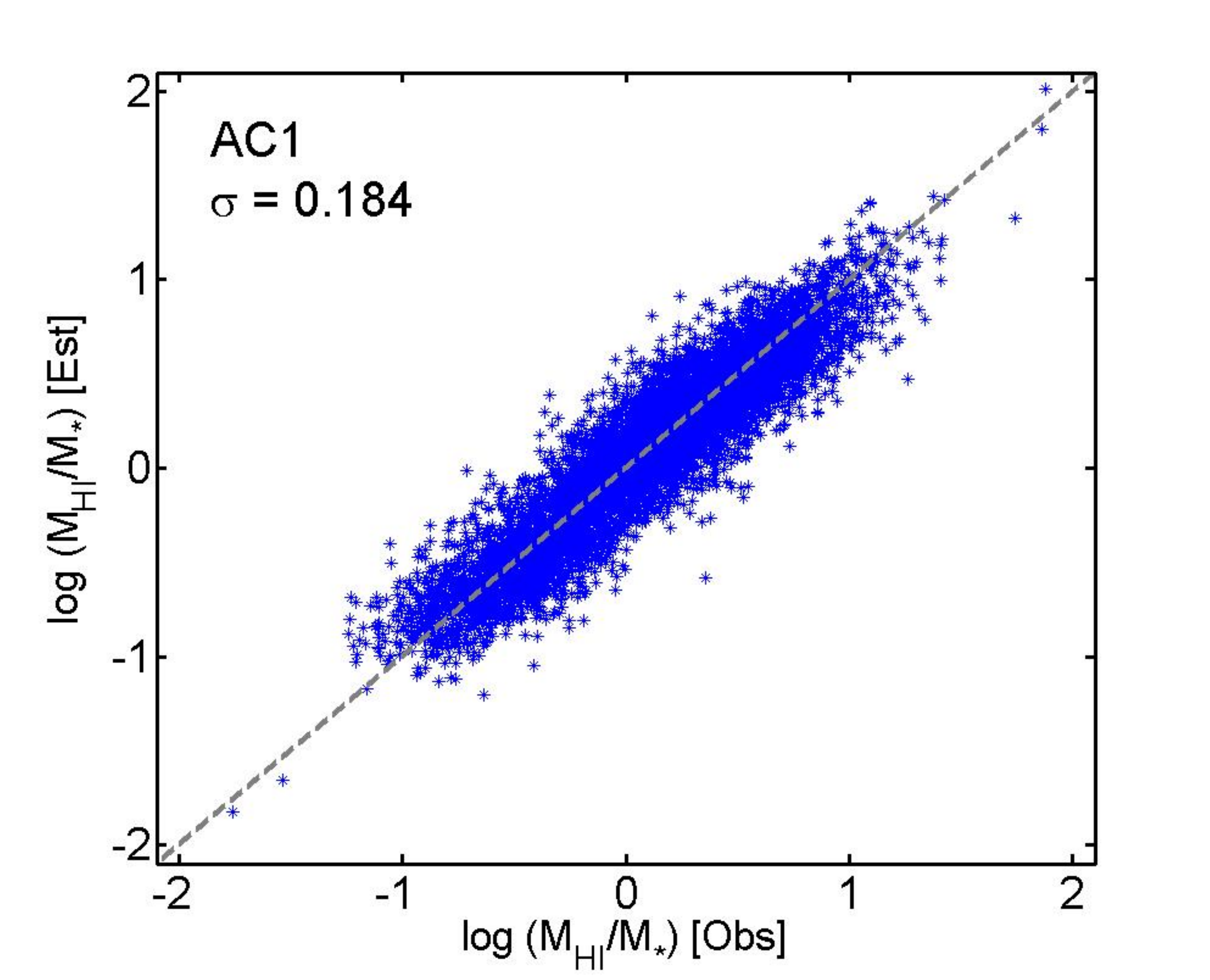}
\includegraphics[width=8.2cm,height=5.5cm,angle=0]{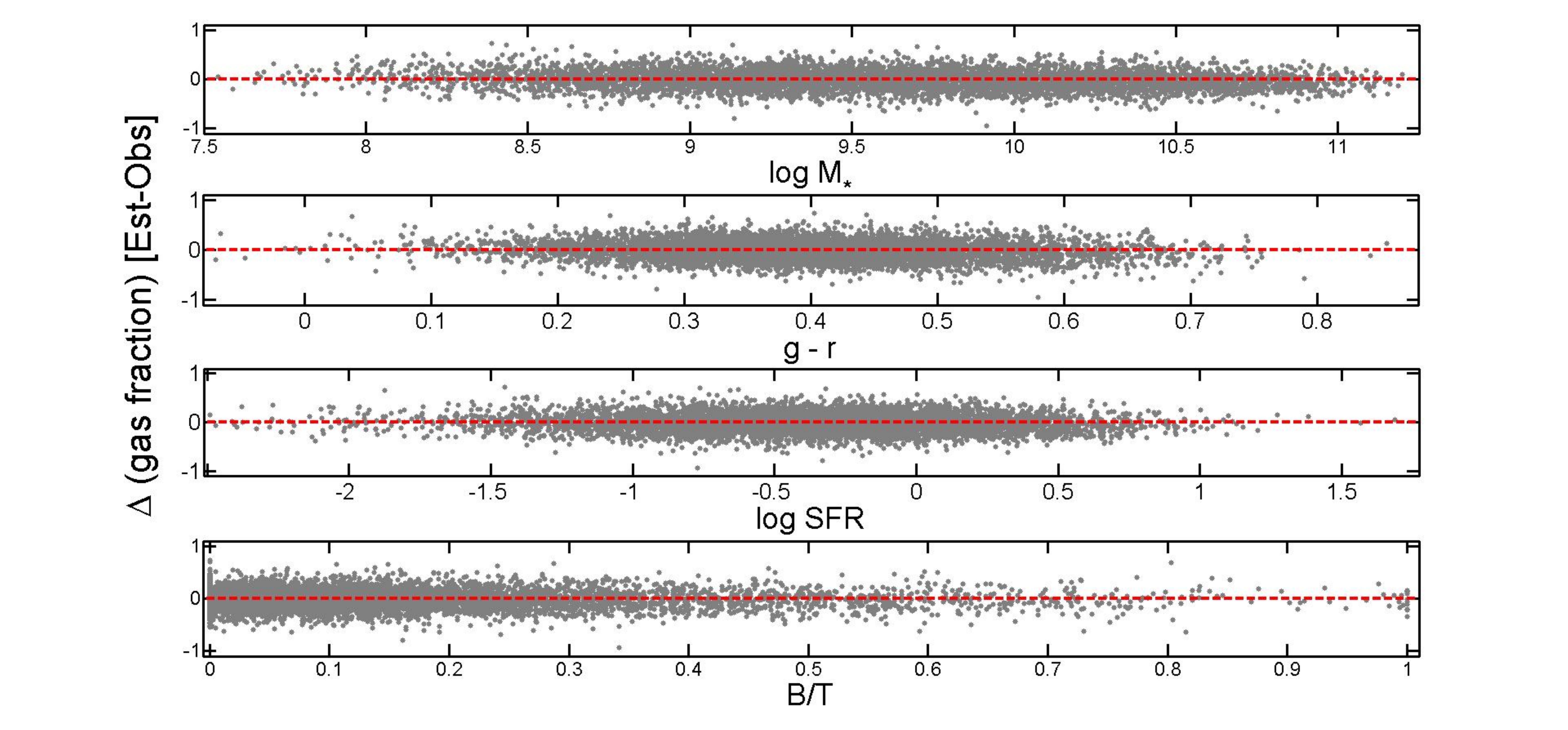}
\caption{The top panel shows the estimated vs observed gas fraction for AC1. 
The lower panels show the difference in estimated and observed gas fraction vs. four 
different physical parameters, to demonstrate that there are no systematic
trends or residuals.}
\label{fig-fit-ac1}
\end{figure}

In Figure \ref{fig-fit-ac1} we  show the fitting results for the 15 
parameters  for the training sample AC1, by comparing
the estimated and observed \fgas\ in the top panel. The scatter is low: 
$\sigma=0.184$ dex and there is no systematic offset at any value of \fgas. 
In the lower panels, we demonstrate that there is no systematic error in the
estimated \fgas\ as a function of four of the 15 variables.  These four
variables are selected as representative examples; indeed, we find
no systematic offset for any of the 15 parameters in our input list.
 
\begin{figure}
\centering
\includegraphics[width=8.2cm,height=6.3cm,angle=0]{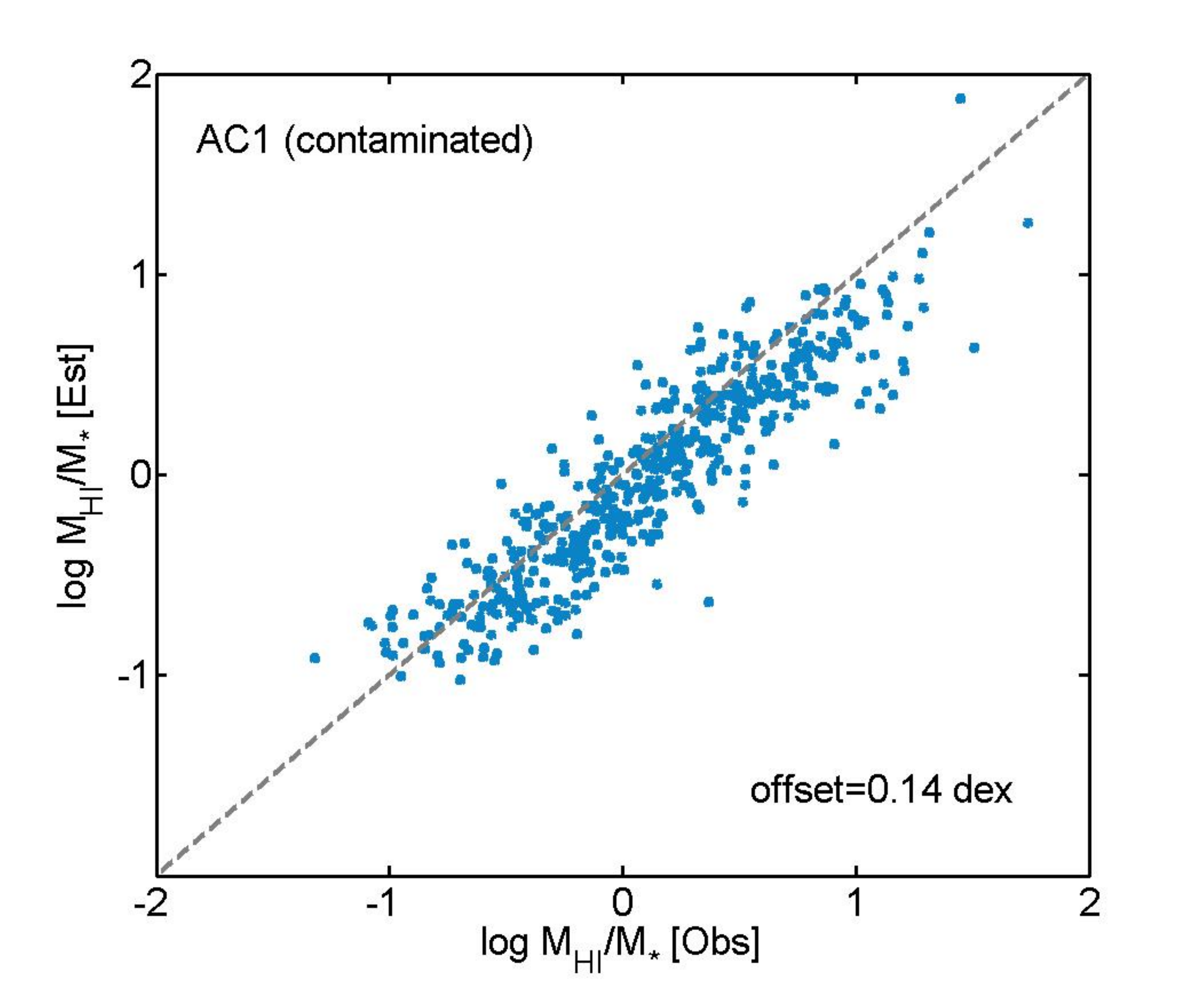}
\caption{Comparison between the observed and predicted \fgas\ for the 508 galaxies in the A70 sample suspected of being contaminated by near neighbour within the Arecibo beam.  On average, the observed \fgas\ is 0.14 dex higher than predicted by the ANN, indicating that these galaxies do indeed have contaminated 21 cm fluxes.}
\label{fig-fit-ac1-unclean}
\end{figure}

In Figure \ref{fig-fit-ac1-unclean} we show the predicted \fgas\ for the 508 galaxies that were removed from AC1 due to the possibility of contamination from a near neighbour galaxy in the Arecibo beam (Section \ref{sec-sample}).  There is now a systematic difference between the predicted \fgas\ and the observed value, with the latter being on average 0.14 dex higher than the prediction.  This result confirms that a significant fraction of the 508 galaxies that were excluded from the training set due to suspected contamination do indeed have additional 21 cm flux from a companion. 

\section{Pattern recognition and detection probabilities}
\label{sec-pr}

Before we can apply the trained networks to the SDSS, we must
develop a technique for identifying which of its galaxies are suitable
targets for the ANN to be applied to.  As we have shown
in Fig. \ref{fig-gr-mu}, applying a solution to galaxies that are not
well represented in the training set can lead to large errors in the predictions.  
In the case of ALFALFA, which is a shallow, blind survey, approximately 4/5 of the 
SDSS galaxies are undetected at 21~cm. Galaxies detected by ALFALFA
are therefore those with the highest gas masses for their stellar mass.
This well known selection effect contributes to the observed anti-correlation between
M$_{\star}$ and HI gas fraction and motivates surveys such as GASS that are designed
to detect galaxies down to a fixed gas fraction.  

The selection of gas
rich galaxies in ALFALFA  has an important impact on our approach to \fgas\ 
estimation, since we are only training our network with this biased population.
That is, if the calibration is determined for the relatively gas-rich galaxies in AC1
and then applied to all SDSS galaxies, we are essentially forcing all galaxies
to follow gas-rich behaviour.  It is a critical step of our analysis to
determine which galaxies can be legitimately calibrated using our \fgas\
estimator.

The approach adopted here is to attempt to distinguish which galaxies
are represented in AC1, as opposed to those that appear as non-detections
in AC3. Artificial neural networks are powerful tools for such
pattern recognition problems and they can also provide us with
statistical information about a data set in order to categorize the data
in a quantitative way (Teimoorinia 2012).  We recall that sample AC1 contains
6837 galaxies detected at 21~cm, all of which have measurements
of the 15 parameters  listed in Table \ref{tab1}.  In order to create a statistical balance
within the network, we randomly select 6837 galaxies out of the 23,652
non-detections in AC3. These 13674 galaxies  are used as  the
main training sample for the pattern recognition step of our
analysis. The objective of this step is to train a network that can
distinguish the 21~cm detections from the non-detections, based solely
on the 15 input SDSS parameters. Note that we use all 6837 of AC1 regardless 
of the potential contamination described in Section \ref{sec-sample}, 
since what is important here is whether or not a galaxy is detected, not its exact
21 cm flux.

We use a binary classification for our training set such that a value of
0/1 is initially assigned to all galaxies in AC3 (non-detections) and AC1
(detections), respectively.    During the training procedure,
the 13674 galaxies of the training set  are randomly separated into two
sub-samples of training (70\%) and validation sets (30\%), to avoid any
over-fitting problems.  Using these input and target values, we train
40 networks and  use the average output of the best 20.
The network output will be the estimated probability that the input
pattern (of SDSS parameters) belongs to one of the two categories.
We refer to this probability of detection and non-detection galaxies as 
the pattern recognition detection metric, \pr, which has a value between 0 and 1.

\begin{figure}
\centering
\includegraphics[width=9cm,height=4.5cm,angle=0]{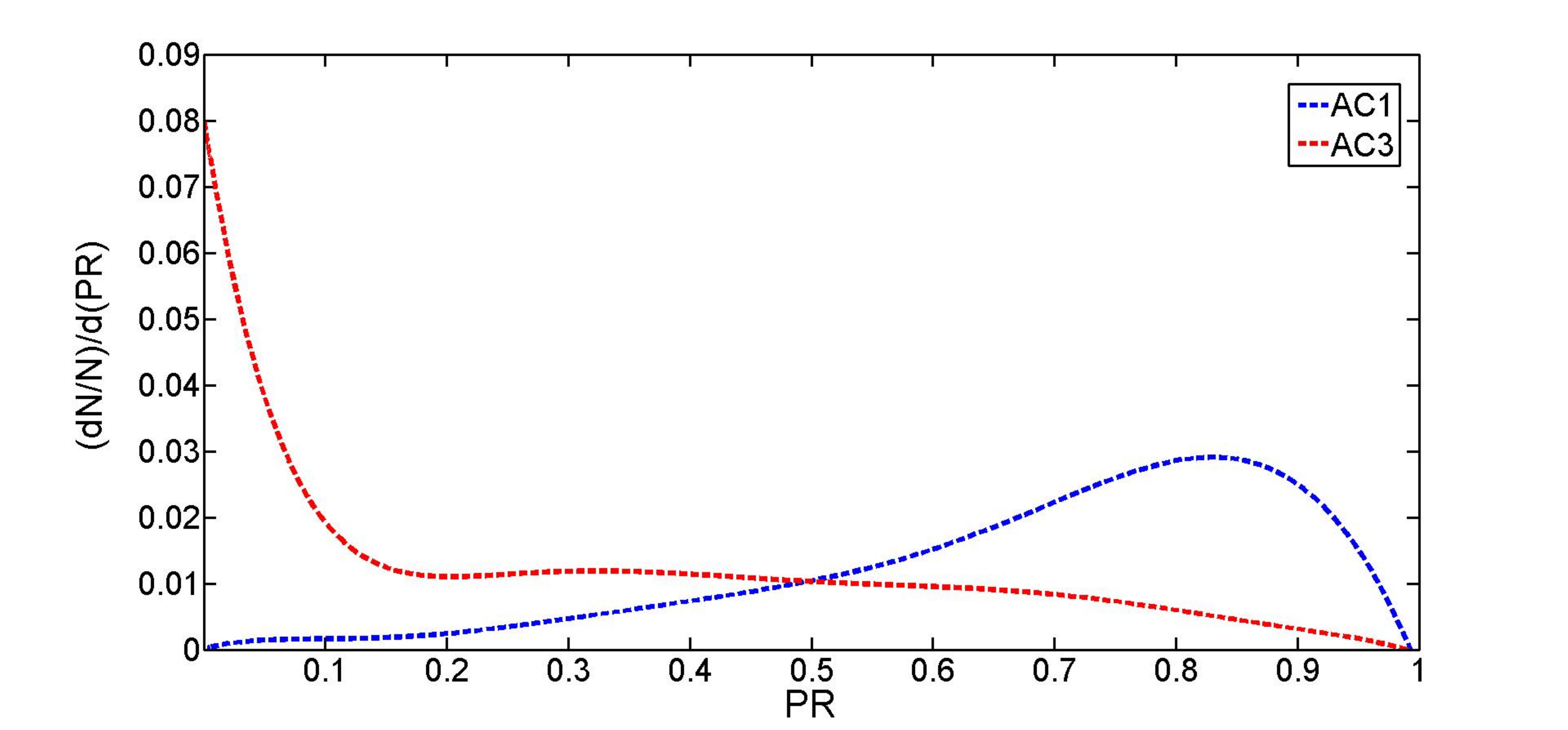}
\caption{Normalized distribution of detection pattern (\pr) for 13674 galaxies in the pattern recognition
training set.  Probabilities peak at high values for 21~cm detections (AC1,
blue line) and low values for non-detections (AC3, red line). Although
both samples have long tails,  more than  80 per cent of galaxies are correctly
classified in sample AC1 for a threshold of \pr$>$0.5.}
\label{fig-dp}
\end{figure}

In the ideal case, for the training set of 13674 galaxies, we would expect
to see two completely separated groups with \pr\ values of 0 and 1,
i.e. HI detections and non-detections that are
completely separable in terms of their SDSS properties.
However, in practice, the detection pattern
exhibits a more continuous behaviour, due to the smoothly varying properties
within the galaxy populations (see also Teimeoorinia et al. 2016).  In Figure \ref{fig-dp} we show the
actual (normalized) distribution of the detection pattern for the galaxies in
the training set, plotted separately for AC1 (blue line) and AC3
(red line).  The figure shows that the detection pattern
peaks near 1 and 0 for AC1 and AC3 respectively, demonstrating that the
network is largely successful in discriminating the two samples based
on their SDSS properties.    However,  both samples show long tails in their \pr\ distributions, indicating that
not all galaxies are correctly classified with the 15 parameters in our training set. 
In other words, this is not a perfect classification so that some detections in AC1
have low  probabilities estimated by the ANN, and some non-detections
in AC3 have high estimated values of \pr.     For a decision boundary
of \pr=0.5 such that AC1 (\pr$<$0.5) and AC3 \pr$>$0.5) represent misclassifications, we
find more than 80  per cent of galaxies of AC1 sample are correctly classified. The
exact choice of \pr\ threshold will depend on the specific application of the data
and the requisite combination of purity and accuracy.  According to binary classification methods, 
the level of separation shown in  Figure \ref{fig-dp} can be considered
as `successful'. In this method, the AUC = 0.86 and is therefore
indicative of a very good classification.  

\begin{figure*}
\centering
\includegraphics[width=8.5cm,height=4.5cm,angle=0]{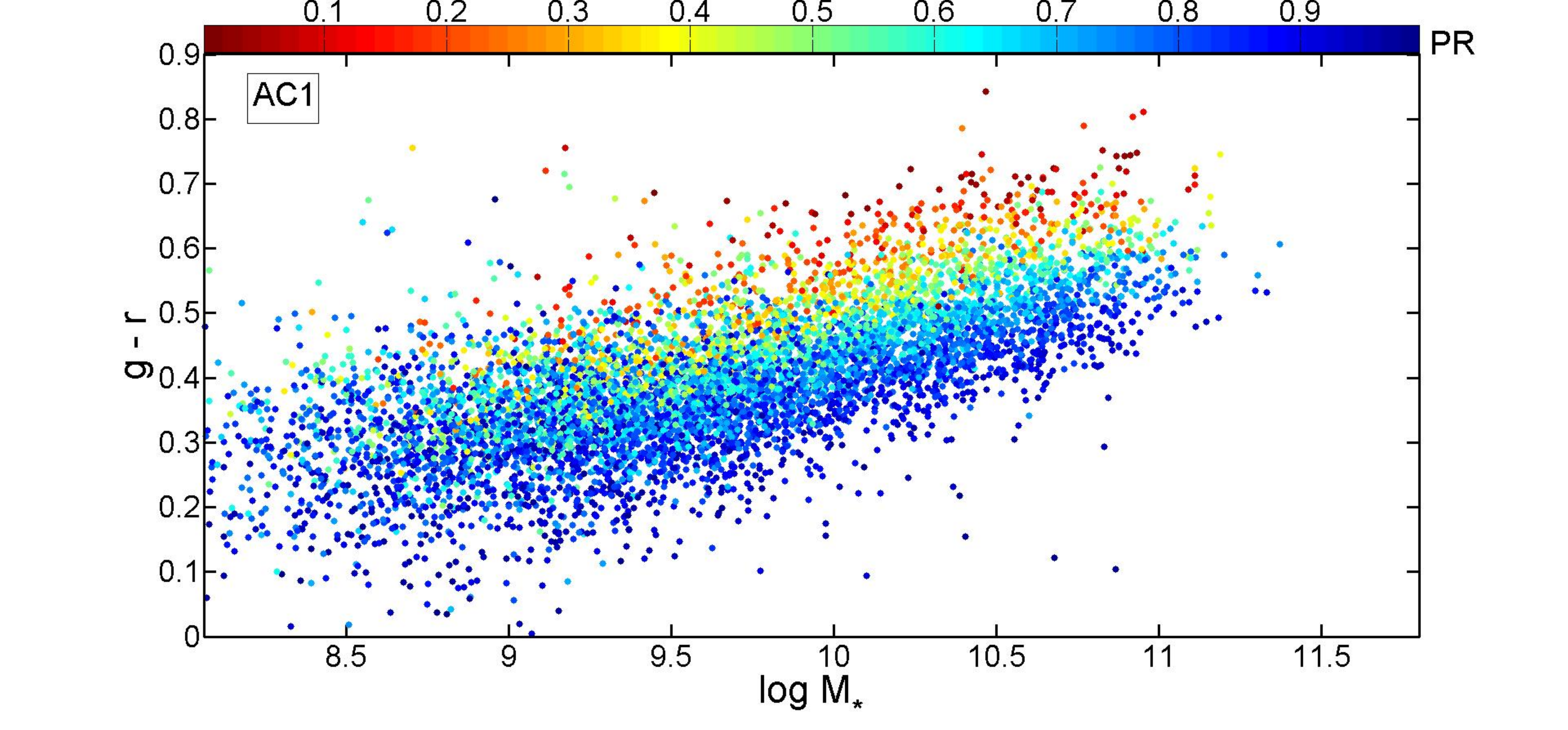}
\includegraphics[width=8.5cm,height=4.5cm,angle=0]{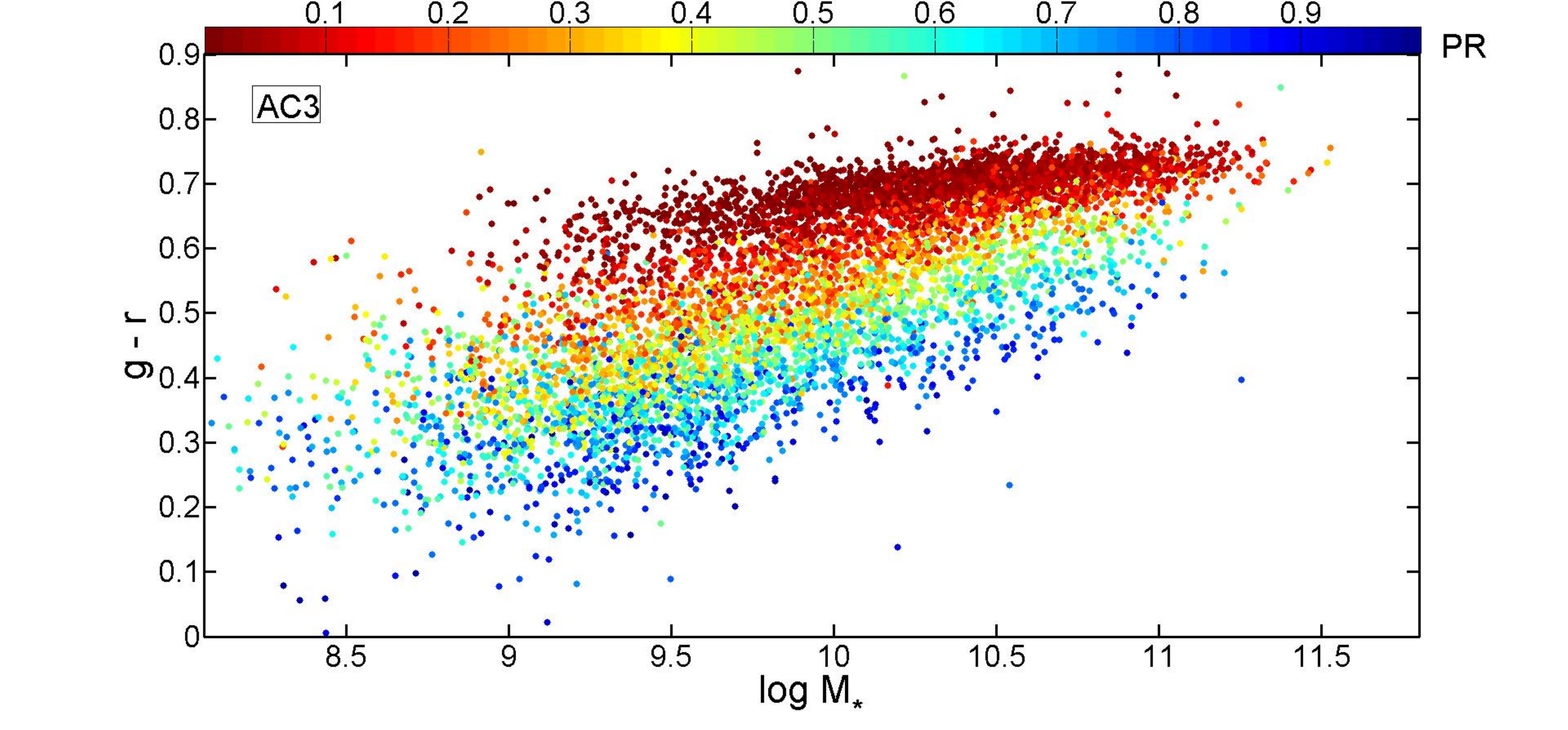}
\includegraphics[width=8.5cm,height=4.5cm,angle=0]{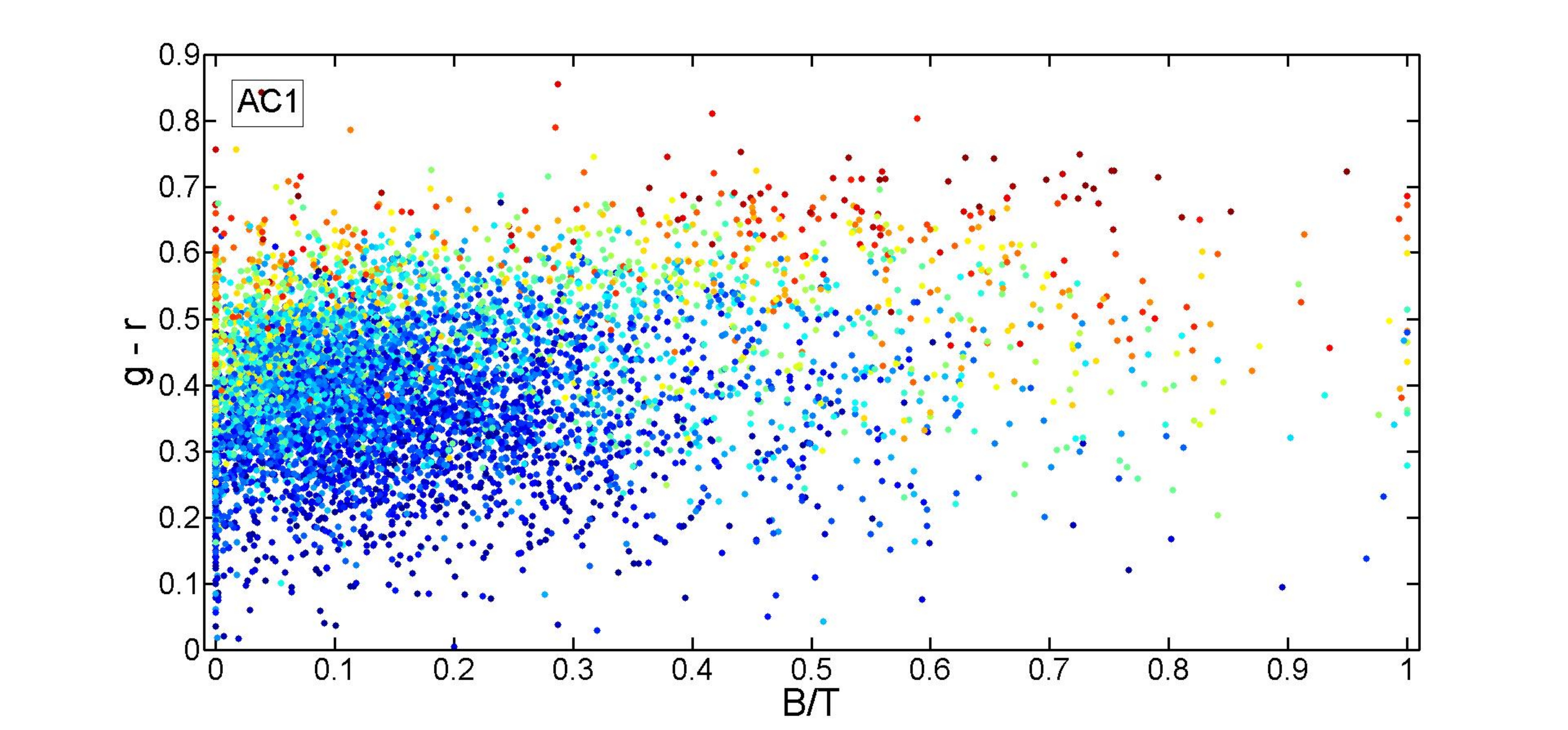}
\includegraphics[width=8.5cm,height=4.5cm,angle=0]{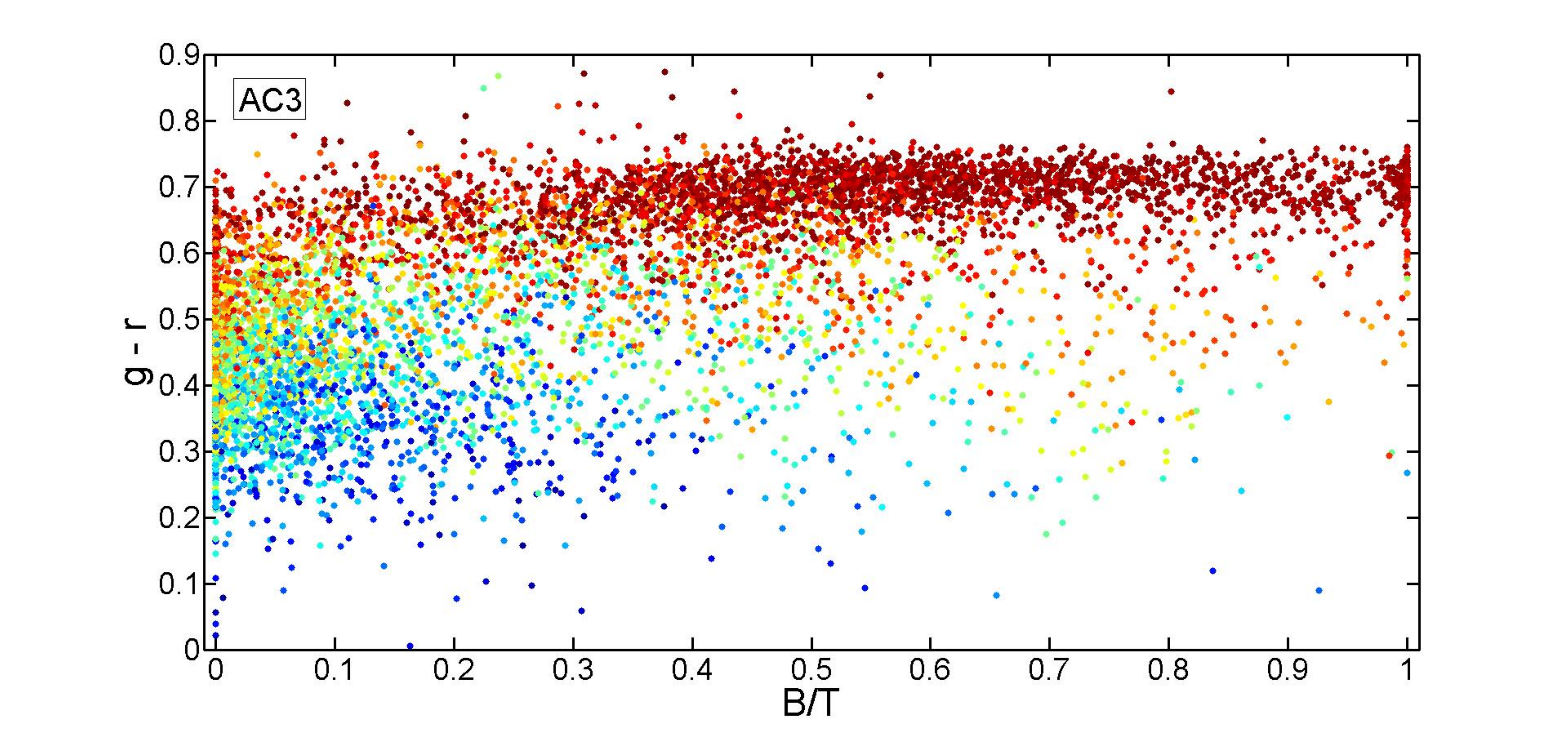}
\includegraphics[width=8.5cm,height=4.5cm,angle=0]{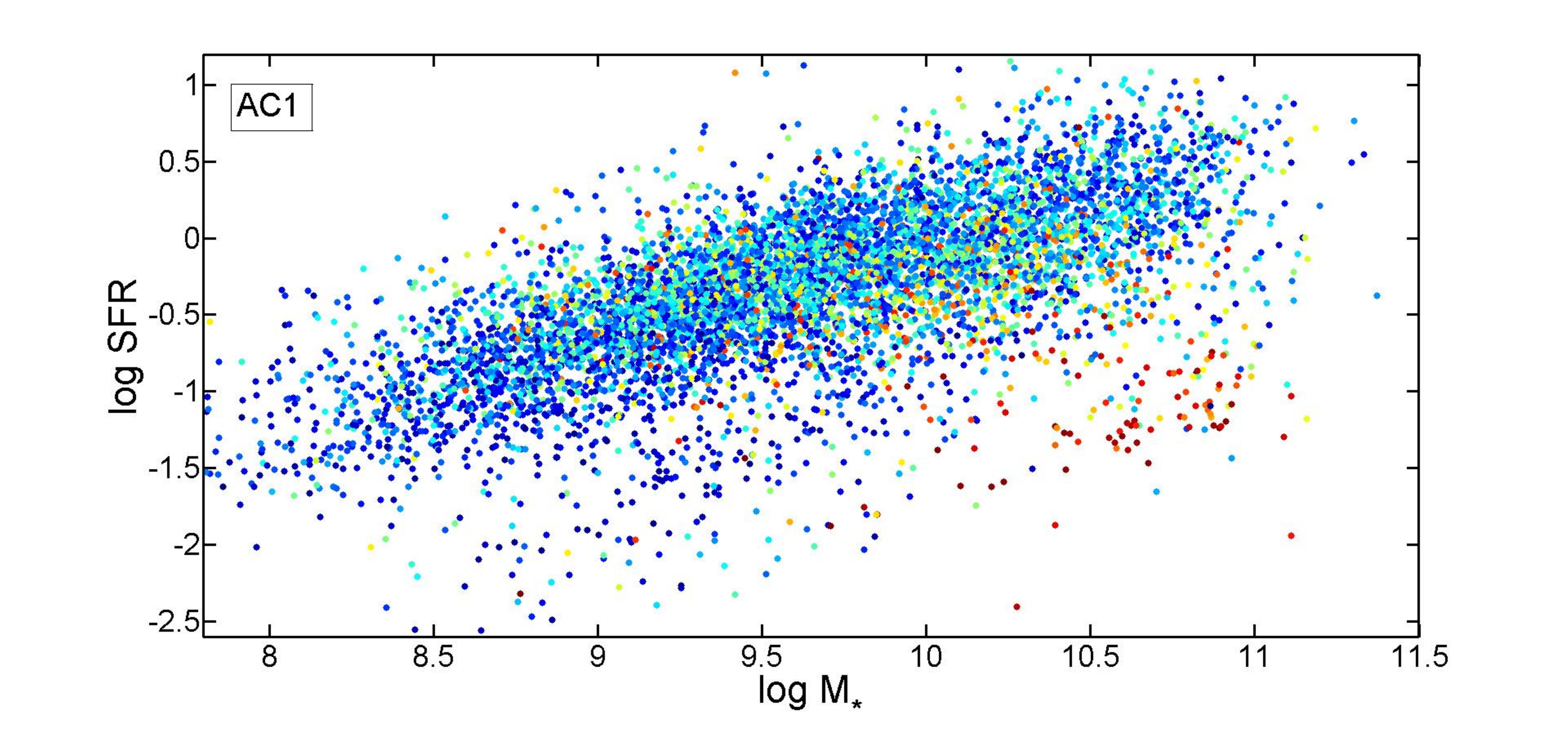}
\includegraphics[width=8.5cm,height=4.5cm,angle=0]{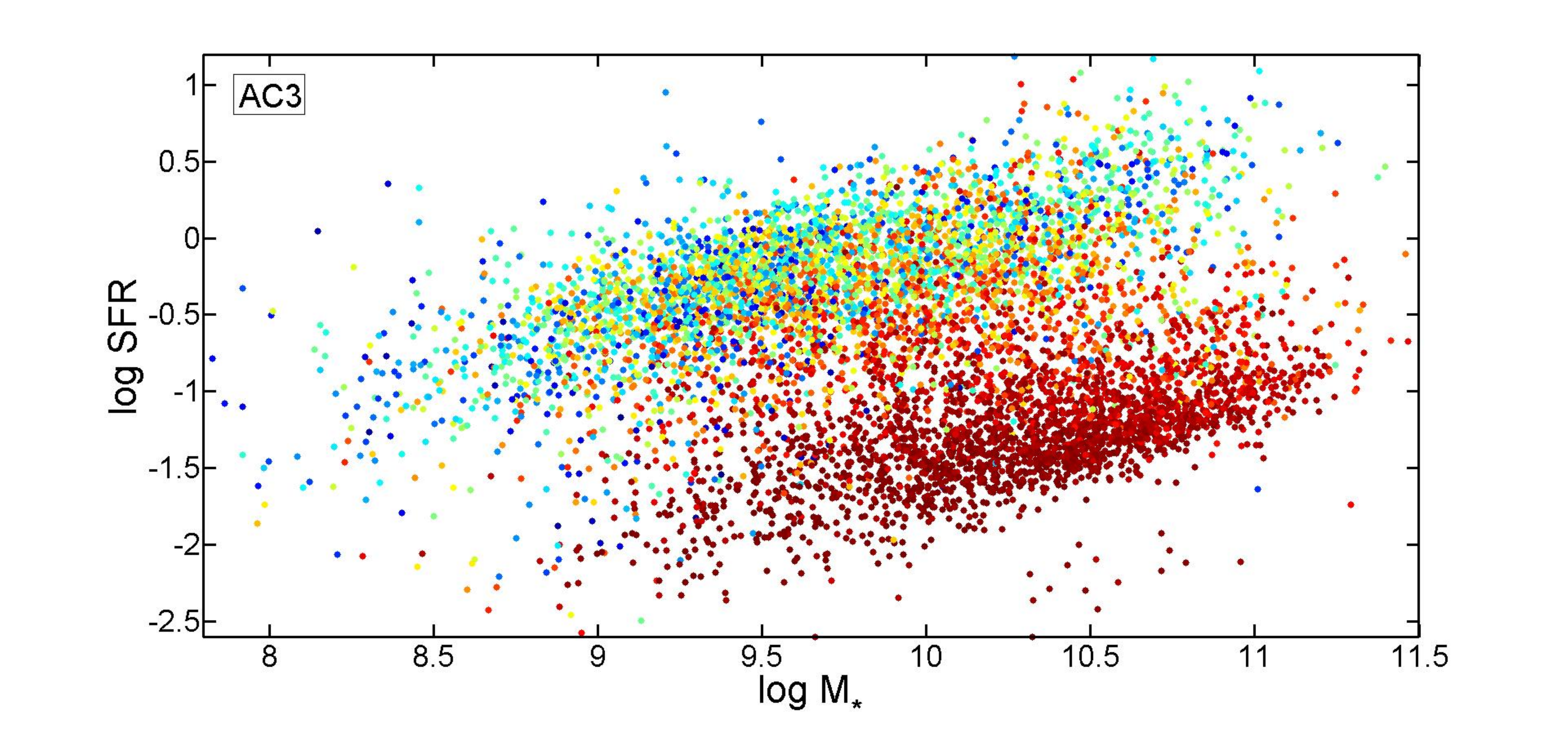}

\caption{The distribution of galaxy physical parameters colour coded by \pr. From top to bottom ,
the panels show $g-r$ vs. stellar mass, $g-r$ vs. B/T and SFR vs. stellar mass. 
The left and right panels are related to AC1 and AC3, respectively.}
\label{fig-class-pr}
\end{figure*}

In Figure \ref{fig-class-pr} we compare the physical parameter space of
galaxies in AC1 (left panels) and AC3 (right panels), where points are colour coded by \pr.
Figure \ref{fig-class-pr} shows that the distribution of $g-r$ colours, stellar
masses, B/T and SFRs are quite different between the two samples, as expected
based on the ALFALFA survey design.  For example, most of the detections
are located on the star forming main sequence and have blue colours,
whereas non-detections additionally populate the region of quiescent
galaxies with red colours and high bulge fractions. However, the distribution of \pr\ between AC1
and AC3 is qualitatively similar.  For example, the highest values of \pr\
in AC3 are seen for galaxies on the main sequence (bottom right panel)
and blue colours (top and middle right panels).  The \pr\
values are typically lower than seen for AC1, as expected given that,
in reality, the AC3 galaxies are indeed not detected.
It should be noted that in the pattern recognition procedure we do not use any direct 
connection between the observed \fgas\ values and the 15 parameters.  \pr\ is better 
interpreted as a connection between the best parameters and the nature of the survey. 
The closer the resemblance of a given galaxy with those in the original survey, the smaller 
the error in the estimation. 

\begin{figure}
\centering
\includegraphics[width=8.9cm,height=5.cm,angle=0]{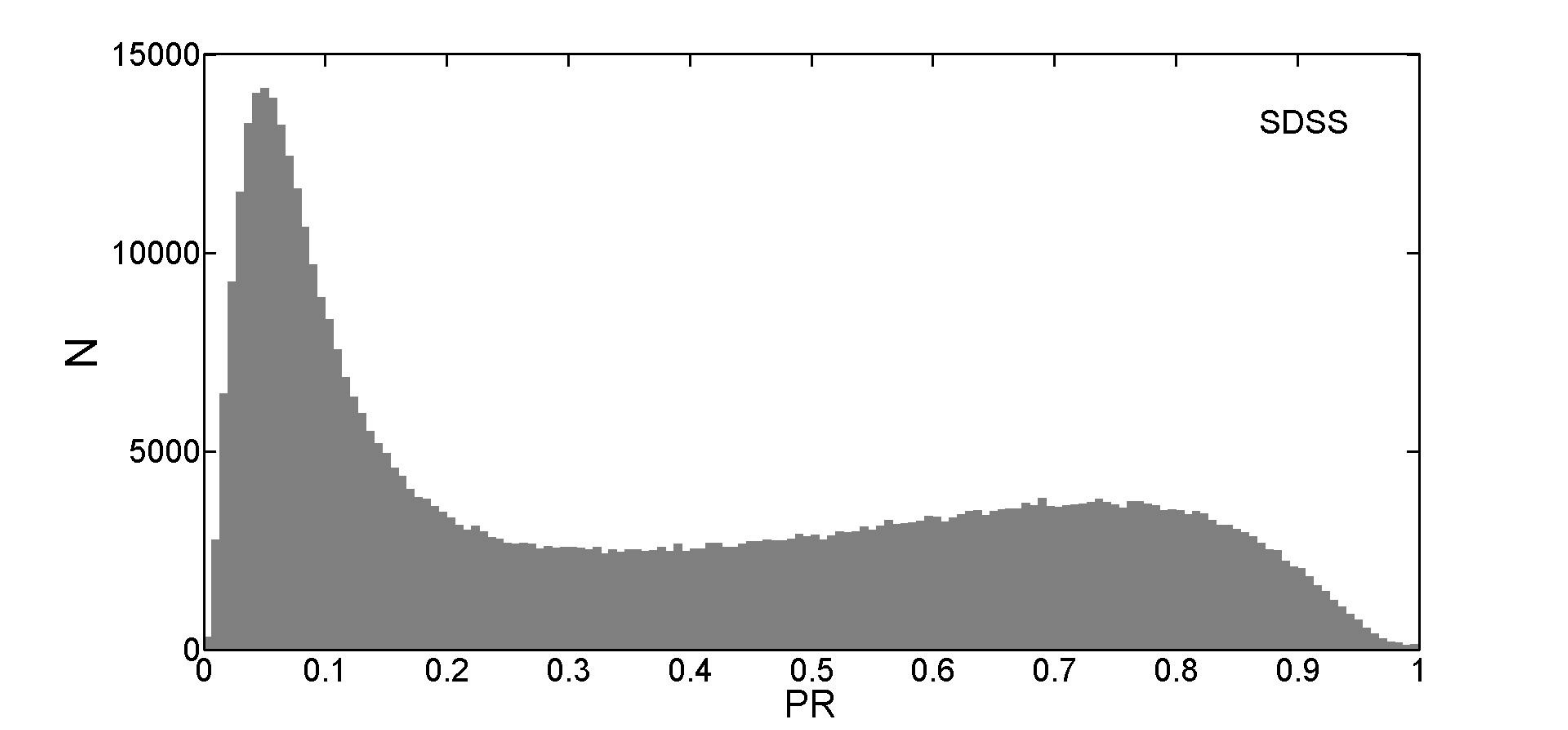}
\includegraphics[width=8.9cm,height=3.5cm,angle=0]{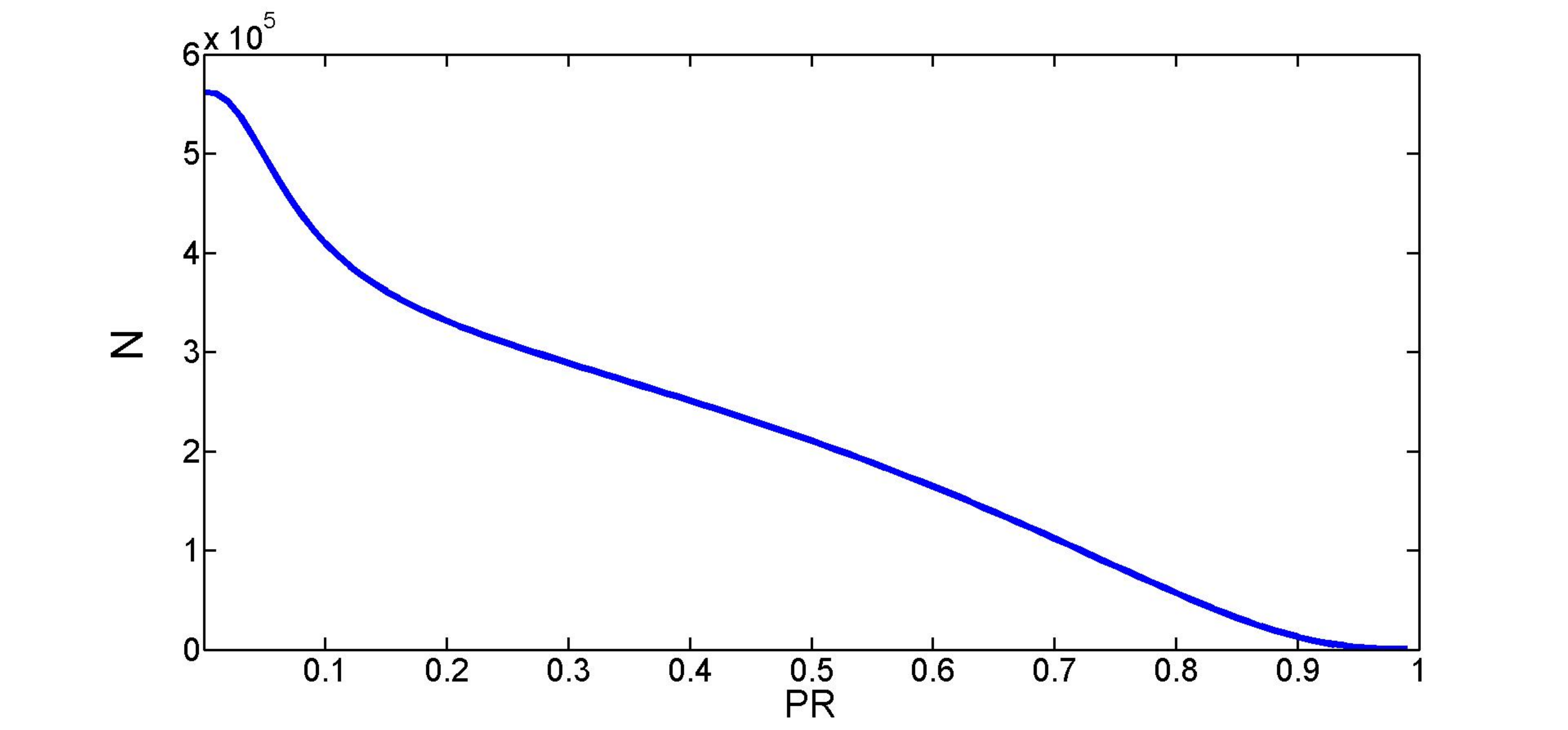}
\caption{Top panel: The distribution of \pr\ for 561,585 galaxies in the SDSS for which all 15 parameters
are available. Bottom panel:  Number of galaxies remaining in the sample as a function of \pr\ threshold.}
\label{fig-pr-det-sdss}
\end{figure}

We now apply the pattern recognition networks to all 561,585 galaxies in SDSS for which all
15 of our parameters are measured.  The resultant distribution of \pr\ values
is shown in Figure \ref{fig-pr-det-sdss}.     In the lower panel of  Figure \ref{fig-pr-det-sdss} 
we show the number of galaxies remaining in the sample as a function of progressively more stringent
\pr\ cut.

\begin{figure}
\centering
\includegraphics[width=8.8cm,height=6.5cm,angle=0]{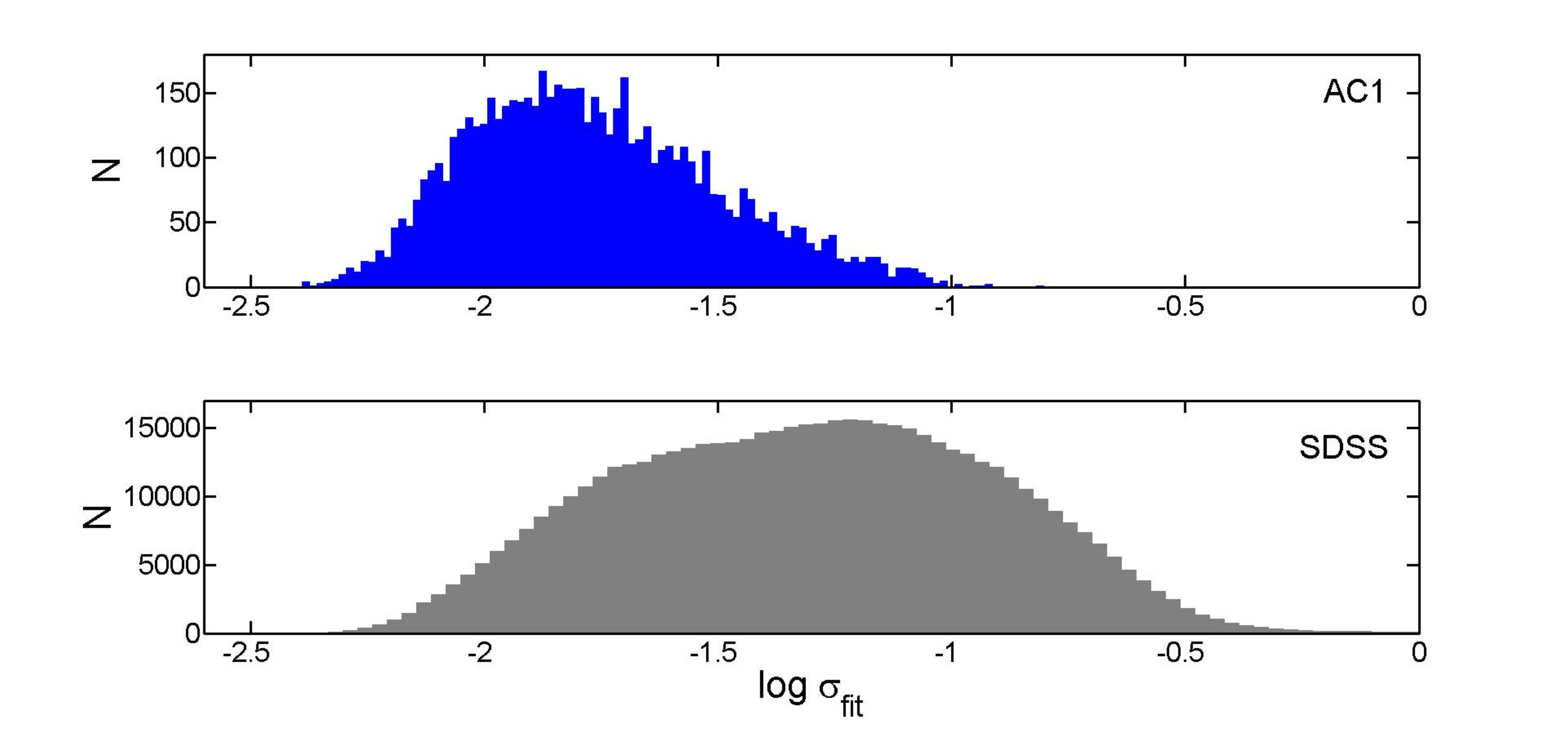}
\caption{Top panel: Distribution of log \sigf\ for the AC1 training sample. 
Lower panel:Distribution of log \sigf\ for SDSS. }
\label{fig-sig-fit}
\end{figure}

In Section \ref{sec-fitting} we described how the scatter amongst the 20 best trained
networks yields an uncertainty on the ANN estimation, \sigf; we now compute \sigf\
for the networks trained with our 15  parameters.  For reference,
in the top panel of Figure \ref{fig-sig-fit} we show the distribution of \sigf\ obtained for AC1. 
The lower panel shows \sigf\ for the 561,585 galaxies with the 15 available parameters  from SDSS.  
As can be seen from the top panel, log \sigf\ spans the interval $-2.5$ to  $-1$ for AC1. On the other hand, when 
we apply the networks to the SDSS, the distribution (in the lower panel) exhibits a more extended
distribution, notably with a tail to higher values. This indicates (not surprisingly) that
some galaxies in the SDSS have an estimation of \fgas\ that is not as stable as in the training
set.  From a pattern recognition  point of view the two distributions can be considered as 
two distinguishable  groups with  considerable overlap.  A cut at log \sigf =  $-1$ removes 
$\sim135,000$ galaxies from the SDSS sample, indicating that $\sim$1/3 of galaxies in SDSS  
have a higher uncertainty in their \fgas\ estimation than in the training set.

\begin{figure}
\centering
\includegraphics[width=4cm,height=3.8cm,angle=0]{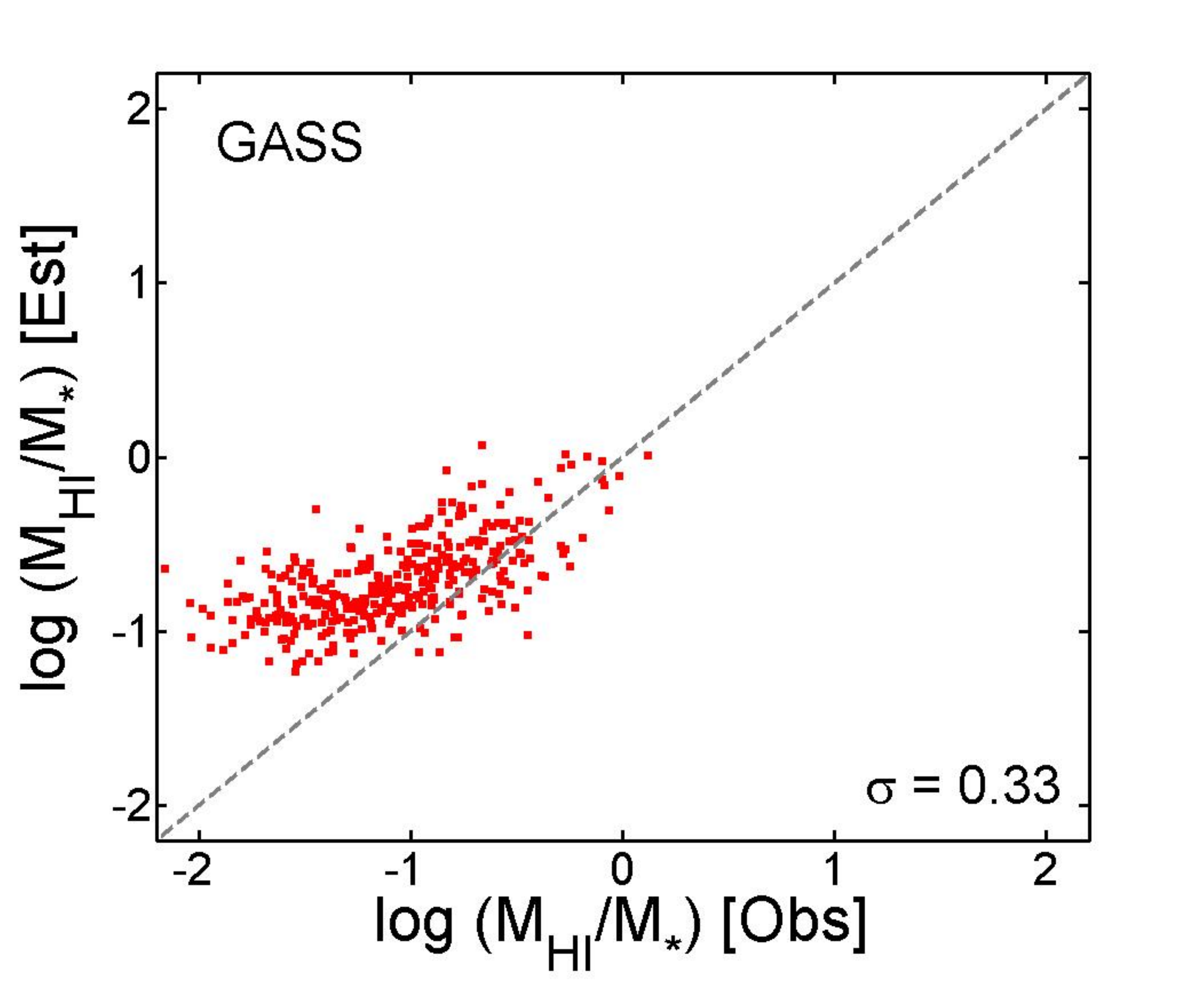}
\includegraphics[width=4cm,height=3.8cm,angle=0]{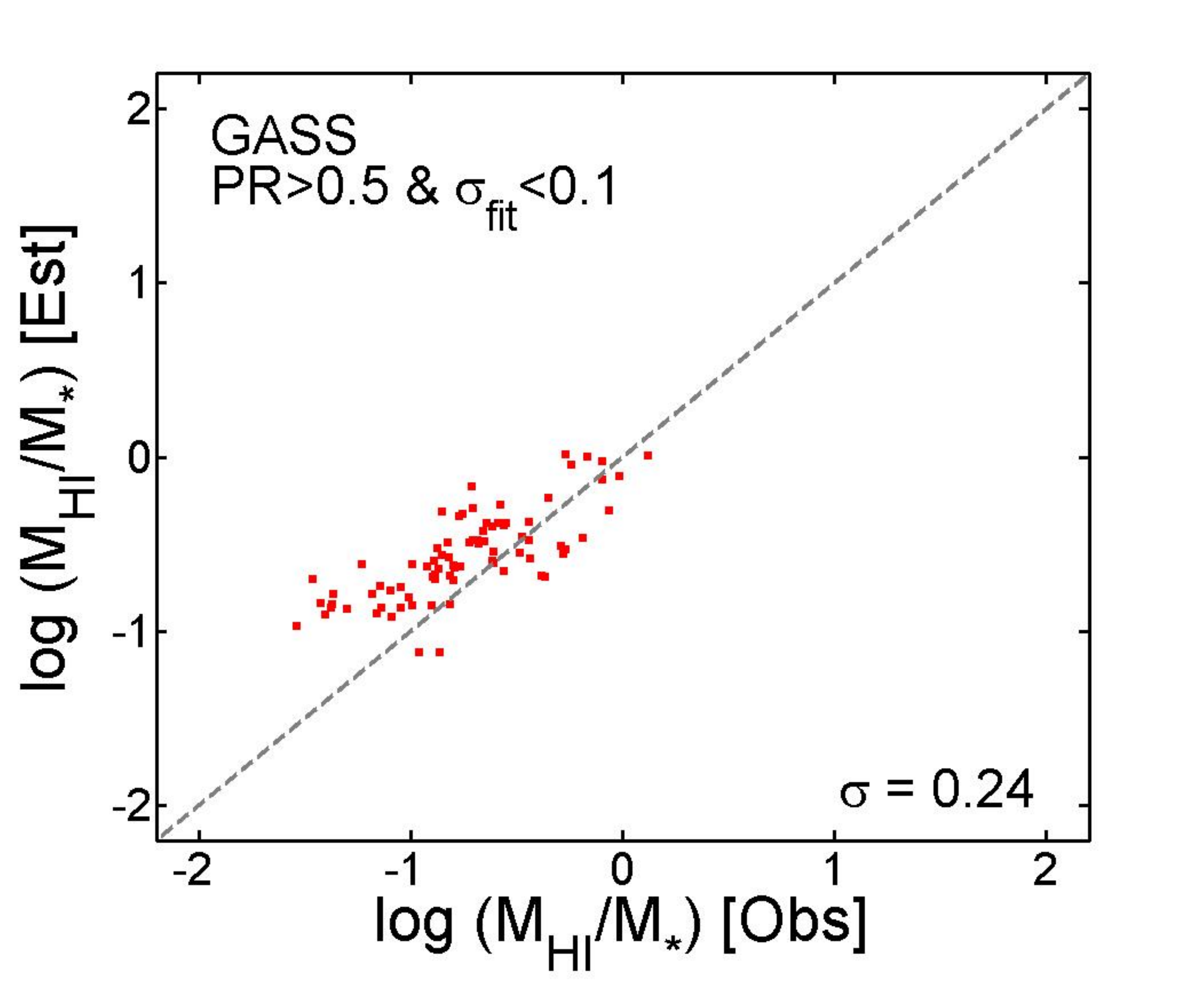}
\includegraphics[width=4cm,height=3.8cm,angle=0]{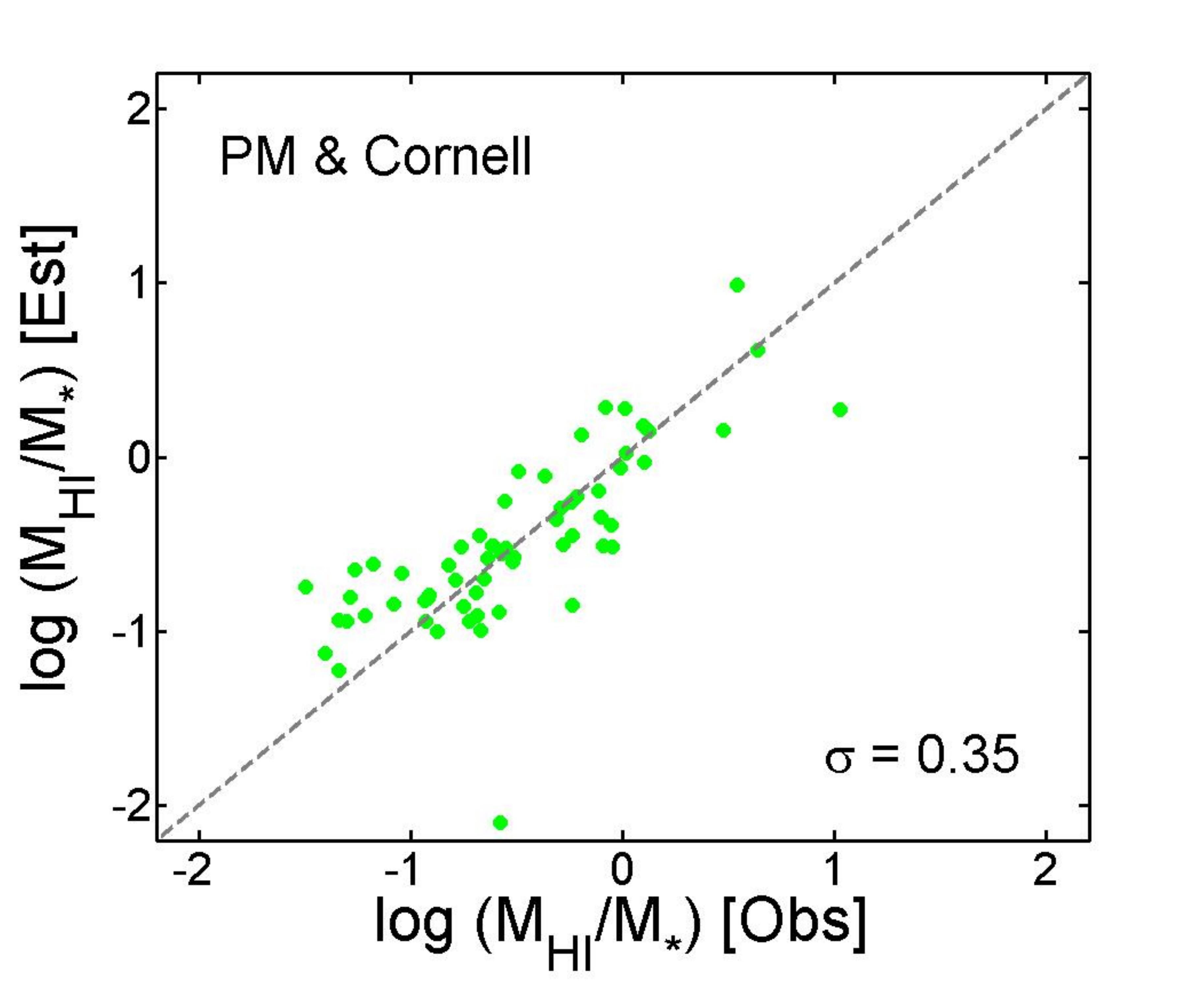}
\includegraphics[width=4cm,height=3.8cm,angle=0]{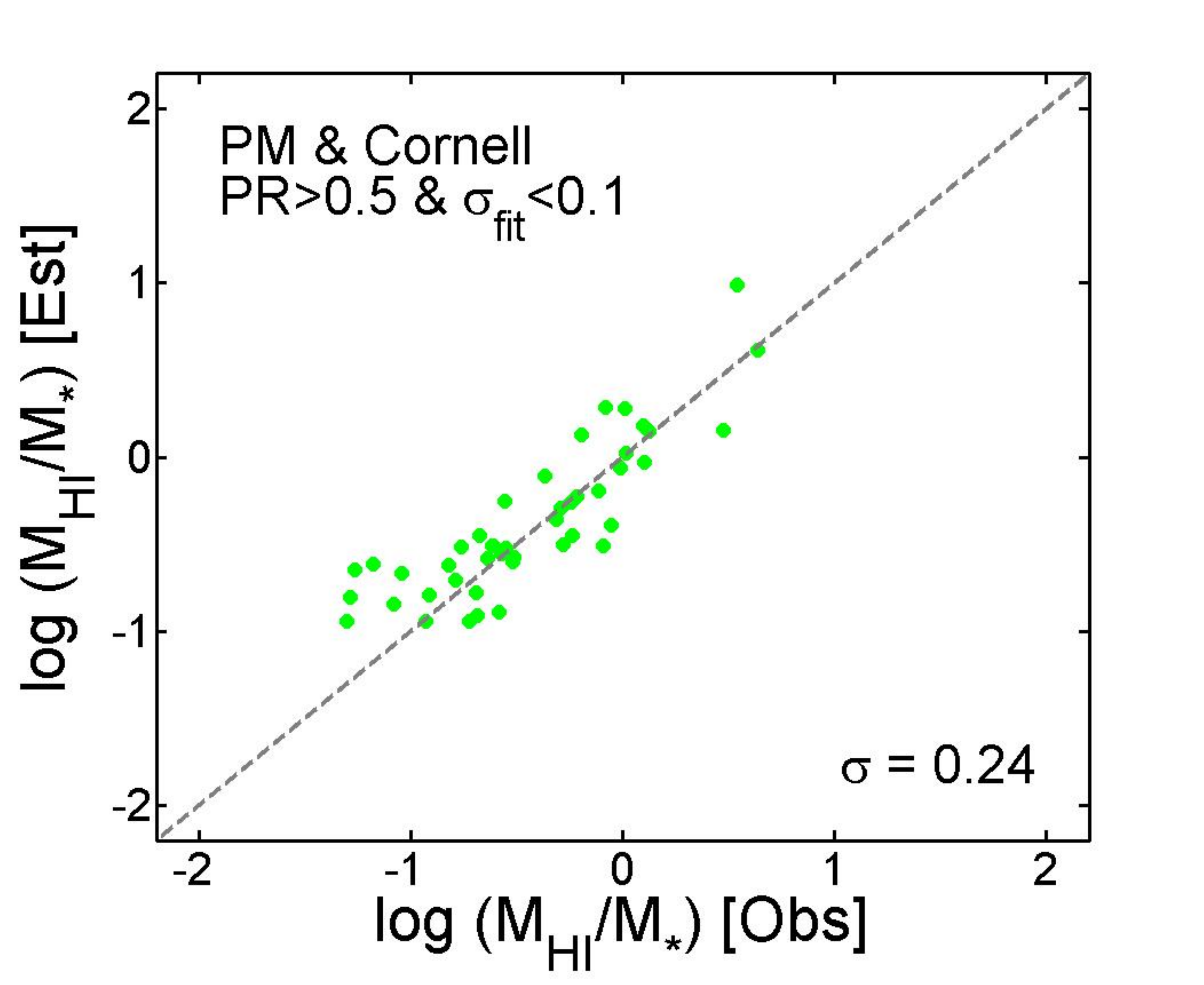}
\includegraphics[width=4cm,height=3.8cm,angle=0]{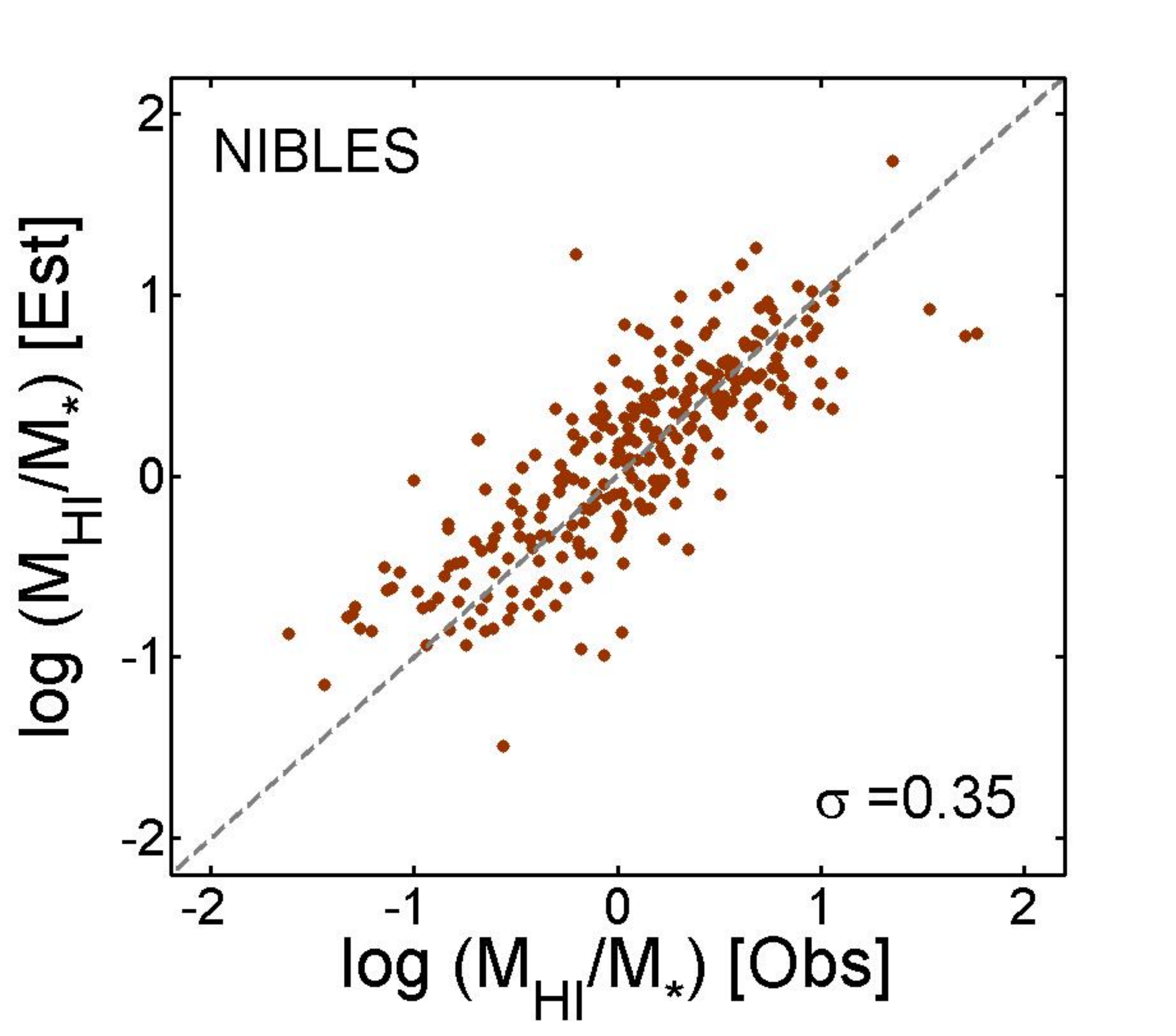}
\includegraphics[width=4cm,height=3.8cm,angle=0]{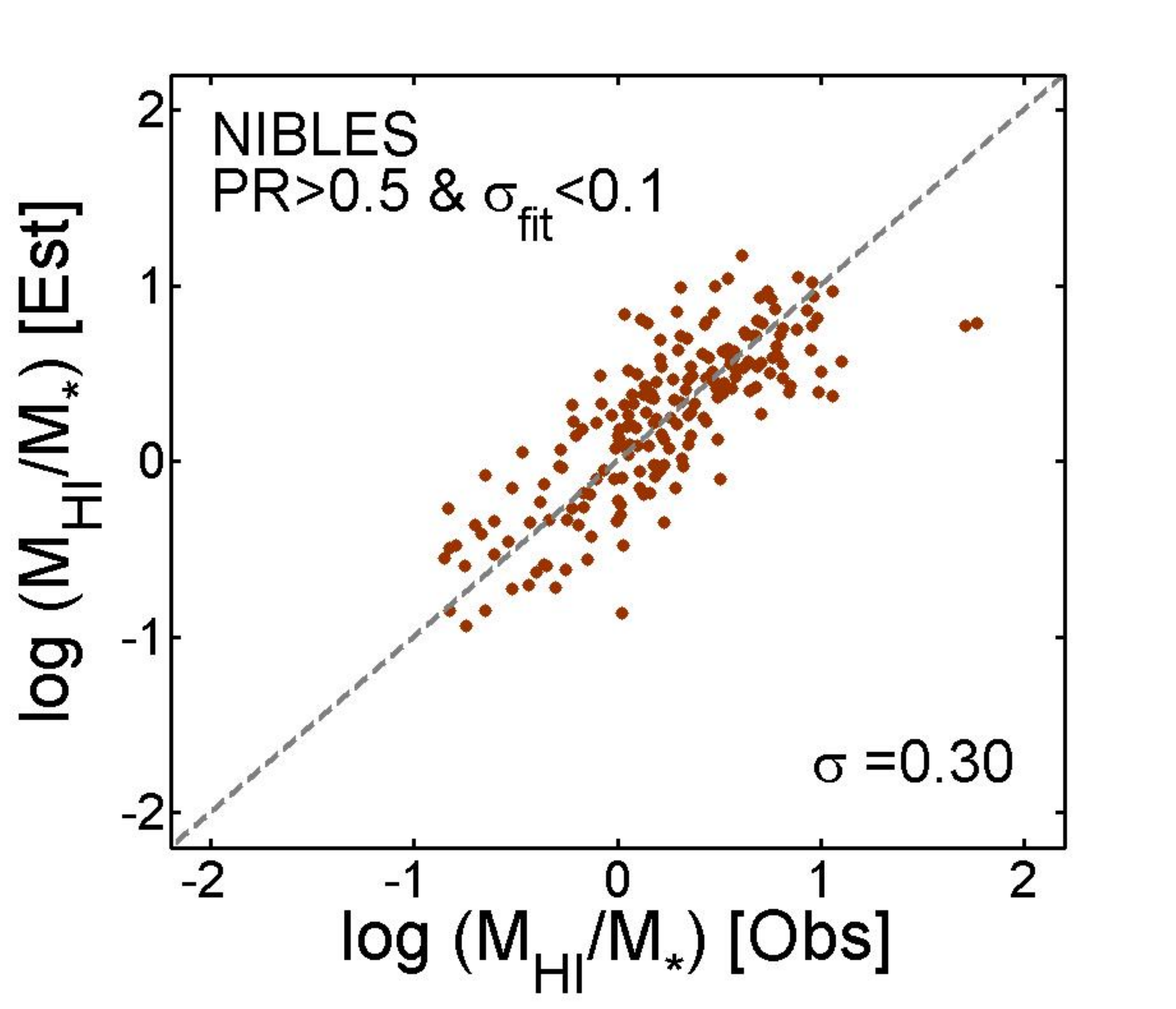}
\caption{The left panels, from top to bottom, show GASS, (PM \& Cornell) and NIBLES, respectively. They include all \pr\ and \sigf . The right panels show the same plots  when \pr\ and \sigf\ are limited, as indicated in the panel legends.}
\label{fig-fit-gass-pm-2p}
\end{figure}

We now have at our disposal two parameters that can be used to select a sample of SDSS
galaxies for which robust \fgas\ estimations can be made, based on their similarity to the
training set (\pr) and the uncertainty in the networks (\sigf).  Both of these parameters
are included in our public catalog.   As a demonstration of
how judicious cuts in these parameters can reduce the scatter between the estimated
and observed \fgas, in Fig. \ref{fig-fit-gass-pm-2p} we show results for the GASS (left panels) and
PM \& Cornell samples (right).  The top panels show all galaxies in the sample, and the
lower panels show the galaxies remaining after we restrict \pr\ $>$ 0.5 and \sigf\ $<0.1$. 
These cuts can be viewed as nominal threshold values for the SDSS catalog, given 
the natural decision  boundary at PR=0.5 and
the distribution of \sigf\ in the training set (as shown in the top panel of Figure \ref{fig-sig-fit}).
However, different cuts may be required for surveys with different properties.
For example, for the GASS sample, imposing these default cuts does reduce the scatter from 0.33 to 
0.24 dex, although a small offset still persists, indicating that even more stringent thresholds may be required.
The PM \& Cornell sample is small and shows some outliers; excluding the single
most significant outlier reduces the scatter of the full sample from 0.35 to 0.29 dex. 
Furthermore, placing a  \pr\ $>$ 0.5 and \sigf\ $<0.1$ cut on the PM \& Cornell sample further
reduction in scatter to 0.24 dex (lower right panel of Figure \ref{fig-sig-fit}).  One approach
for a practical application of \sigf\ and \pr\ is to make cuts on each of these parameters,
as we have done above, with choices that are best suited to the science application in hand.   However,
an alternative approach is to combine \sigf\ and \pr\ into a single parameter, which we
explore in the next section.

\section{A combination of \pr\ and \sigf\ and a determination of final error}

\label{sec-comb}

\begin{figure}
\centering
\includegraphics[width=8.8cm,height=6.5cm,angle=0]{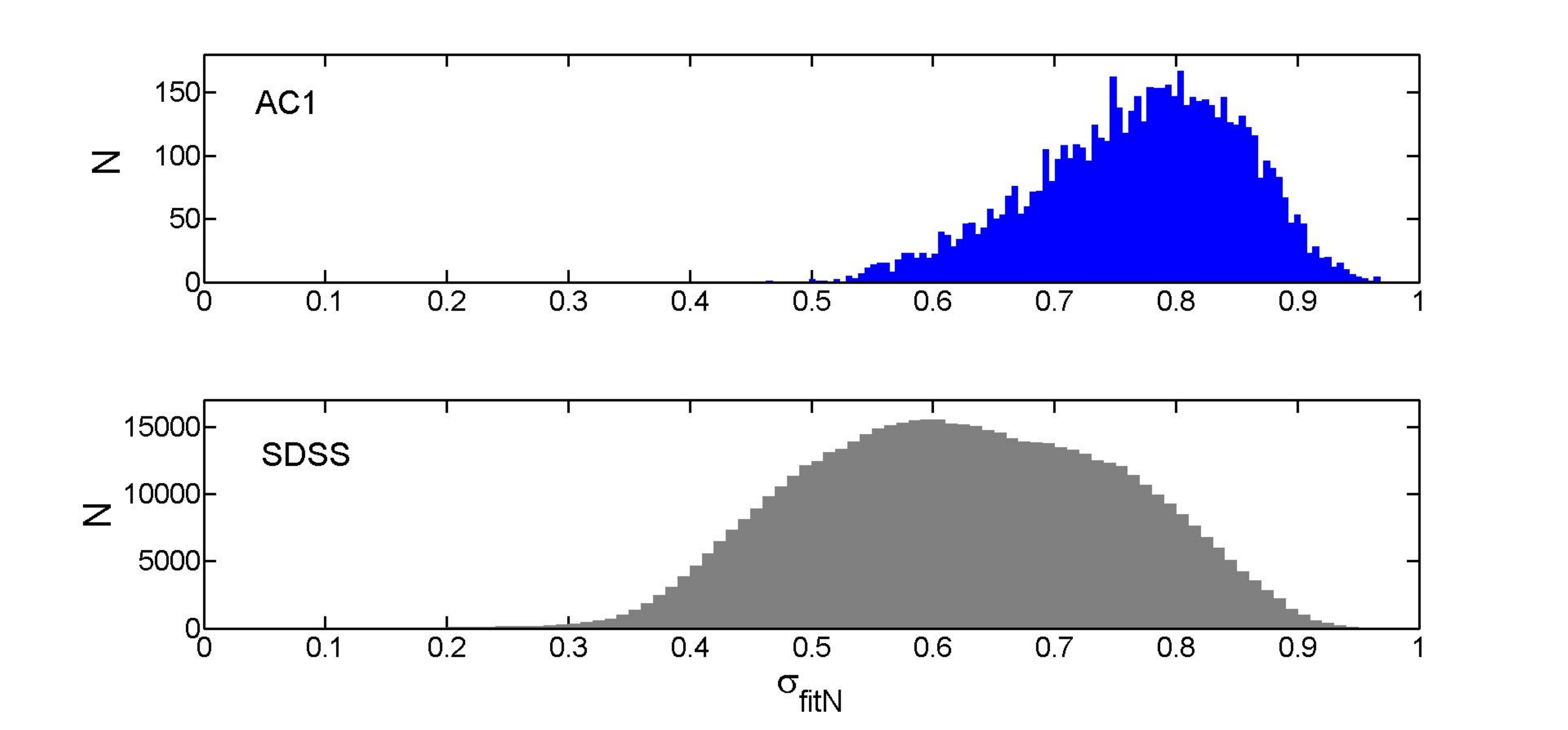}
\caption{The distribution of \sigf\ after the application of an inverse mapping process in which
the distributions of \sigfn\ are normalized the minimum and maximum values of the distribution of Figure \ref{fig-sig-fit},
with values of 0 and 1 respectively. Top panel:  AC1.  Bottom panel: SDSS.}
\label{fig-sig-fit-map}
\end{figure}

In Fig. \ref{fig-dp}, we find that  $\sim$25\% of non-detections (i.e., AC3) are predicted to be detections (i.e., false positive); so a cut only on PR is not the best way to distinguish and separate more secure estimations. On the other hand, we have introduced a control parameter  taken only from the detections in the fitting step: \sigf .  We want to go a step further to introduce a more secure control parameter: a combination  of PR and \sigf . To do this we notice that  in Figure \ref{fig-sig-fit}, log \sigf\ is extended in the interval $-2.5$ -- 0,
a range broader than AC1.  The SDSS distribution can be normalized to AC1 via a simple `inverse' mapping process,
by setting the minimum and maximum values of the two distributions to 0 and 1, where 1 is
the 'best' value, representing the minimum scatter in the distribution. We call this normalized 
distribution \sigfn.   Figure \ref{fig-sig-fit-map}  shows \sigfn\ for both SDSS (lower panel)
and AC1 (upper panel).

\begin{figure}
\centering
\includegraphics[width=8.2cm,height=7cm,angle=0]{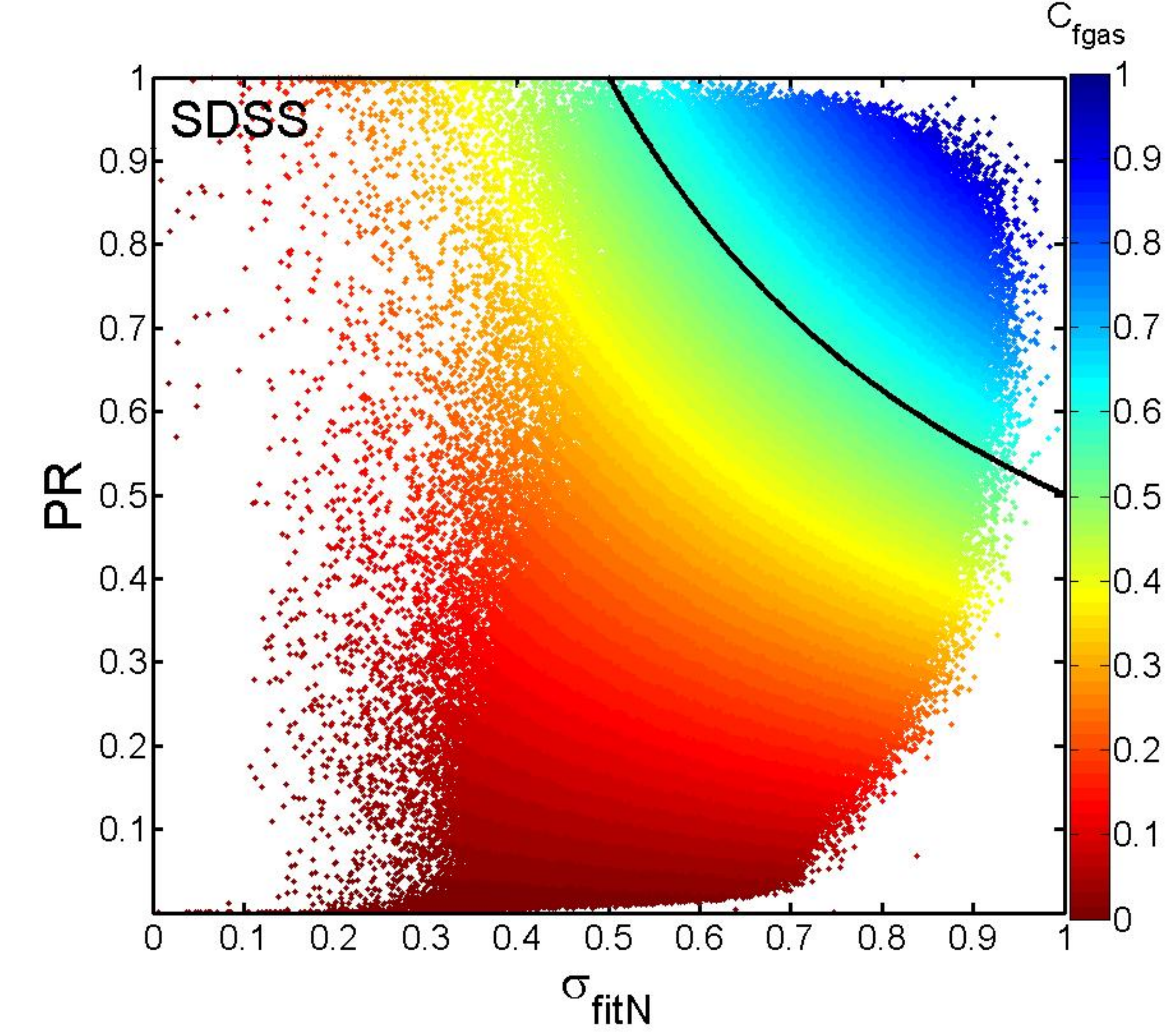}
\caption{\pr\ vs \sigfn\ colour coded by \cfgas, defined as \sigf$\times$\pr. The 
most reliable \fgas\ estimations can be made for galaxies in which both  \pr\ and \sigfn\ are large. The area above the black solid line shows $\sim150000$ galaxies with \cfgas$>$0.5.}
\label{fig-pr-total}
\end{figure}

We now have two `quality control' parameters that are distinct indicators of the \fgas\
estimation: \sigfn\ and \pr, where both have values between 0 and 1.  The best estimations
will be when both \sigfn\ and \pr\ are large (i.e. tend towards 1).  Moreover, these two parameters can now be
trivially combined (\sigfn$\times$\pr), and then normalized to define a single variable, \cfgas, that represents the confidence of
the \fgas\ estimation.  This confidence metric
also has a value between 0 and 1, where the higher the number, the more secure
the \fgas\ estimation will be.  This is demonstrated in  Figure \ref{fig-pr-total} where we  plot \sigfn\ vs \pr\ colour coded by \cfgas\ for the SDSS sample.  The area above the black solid line shows $\sim150000$ galaxies with \cfgas$>$0.5. 

\begin{figure}
\centering
\includegraphics[width=8.5cm,height=5.8cm,angle=0]{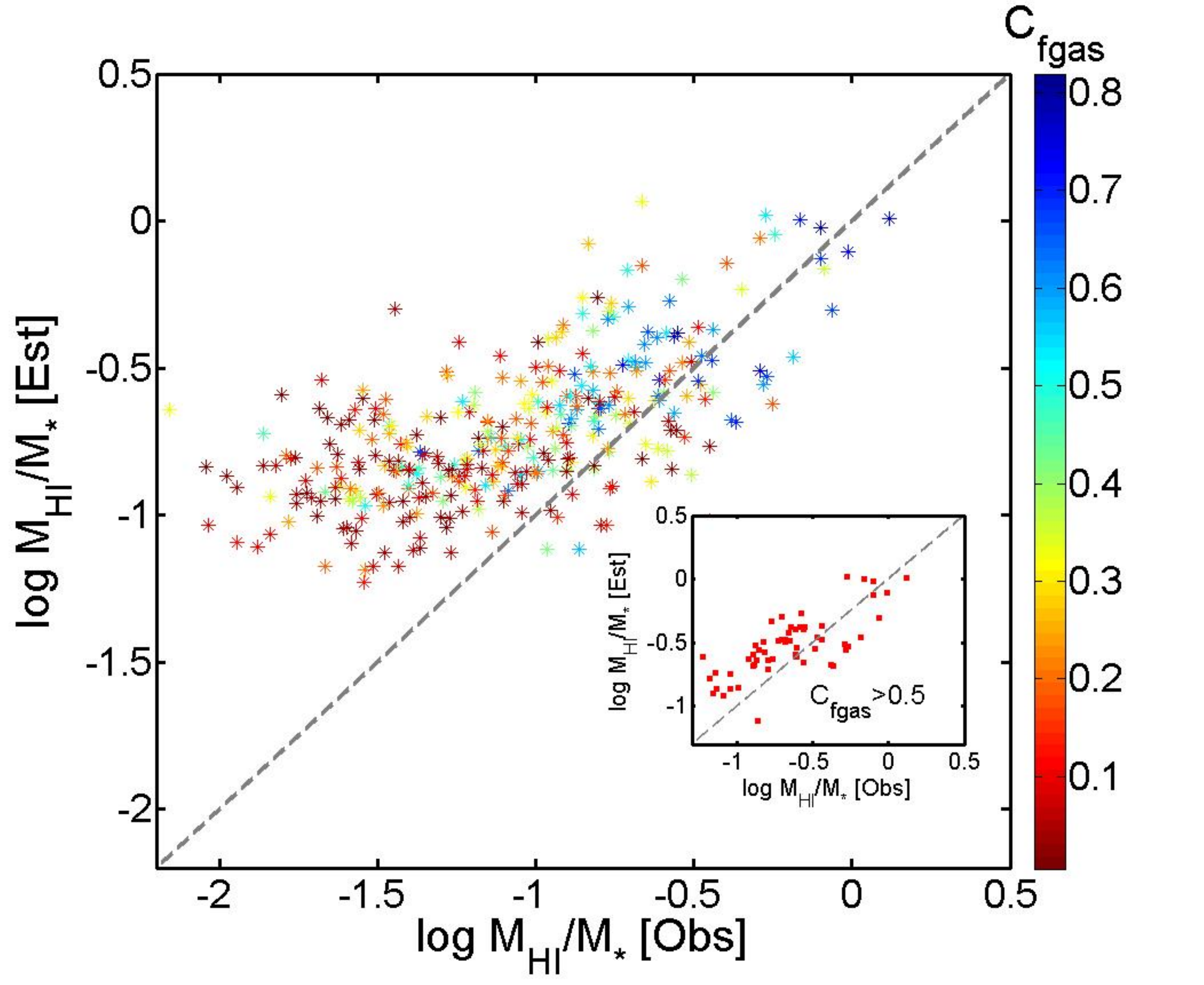}
\includegraphics[width=8.5cm,height=3.2cm,angle=0]{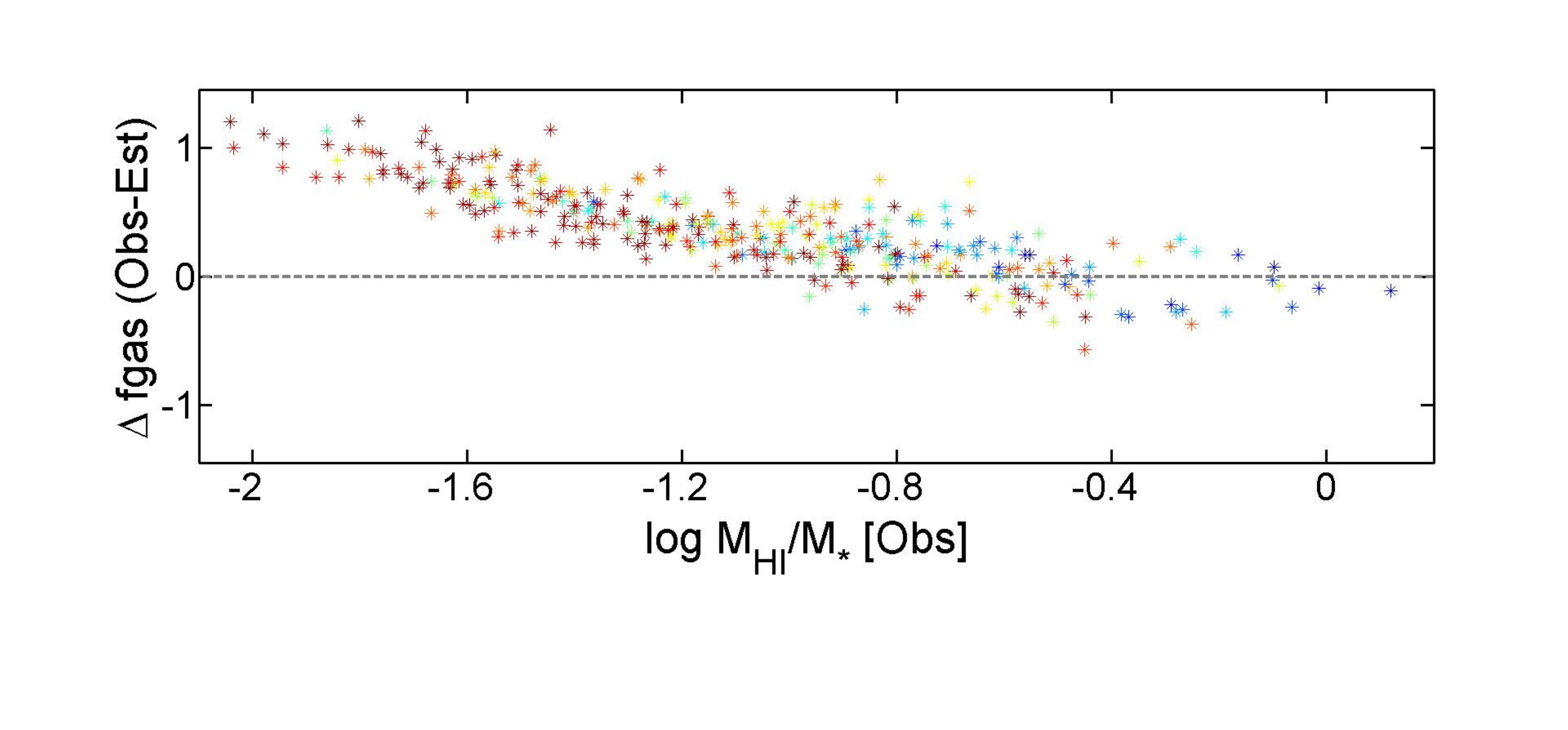}
\caption{Top panel: the estimated vs observed gas fraction colour coded by \cfgas\ for the GASS
sample. The inset show points for \cfgas$>0.5$. 
Bottom panel: The difference between estimated and observed \fgas\ vs the observed value,
again colour coded by \cfgas.}  
\label{fig-gass1-color}
\end{figure}

In Fig.  \ref{fig-gass1-color} we show how the \cfgas\ metric performs on one of our validation
samples, namely the GASS sample.  In the top panel, we show the estimated vs. observed \fgas\ colour coded
by \cfgas.  The familiar over-estimate of \fgas\ by the ANN at low gas fractions is seen.
However, the colour coding by \cfgas\ shows that the greatest offsets correspond to the
lowest values of \cfgas.  This trend is emphasized in the lower panel by plotting the 
difference between the estimated and observed \fgas\ as a function of the observed value.
In the inset panel, we show the GASS points with \cfgas $>$ 0.5;
the scatter is reduced to 0.22 dex and with a small systematic error, which can
be eliminated by increasing the cut in \cfgas\ to $\sim$ 0.7.
 For other surveys in our combined
sample, a threshold of \cfgas $>$ 0.5 provides a good nominal selection of robust
\fgas\ determinations, and is our recommended default value.

\begin{figure}
\centering
\includegraphics[width=4.1cm,height=3.9cm,angle=0]{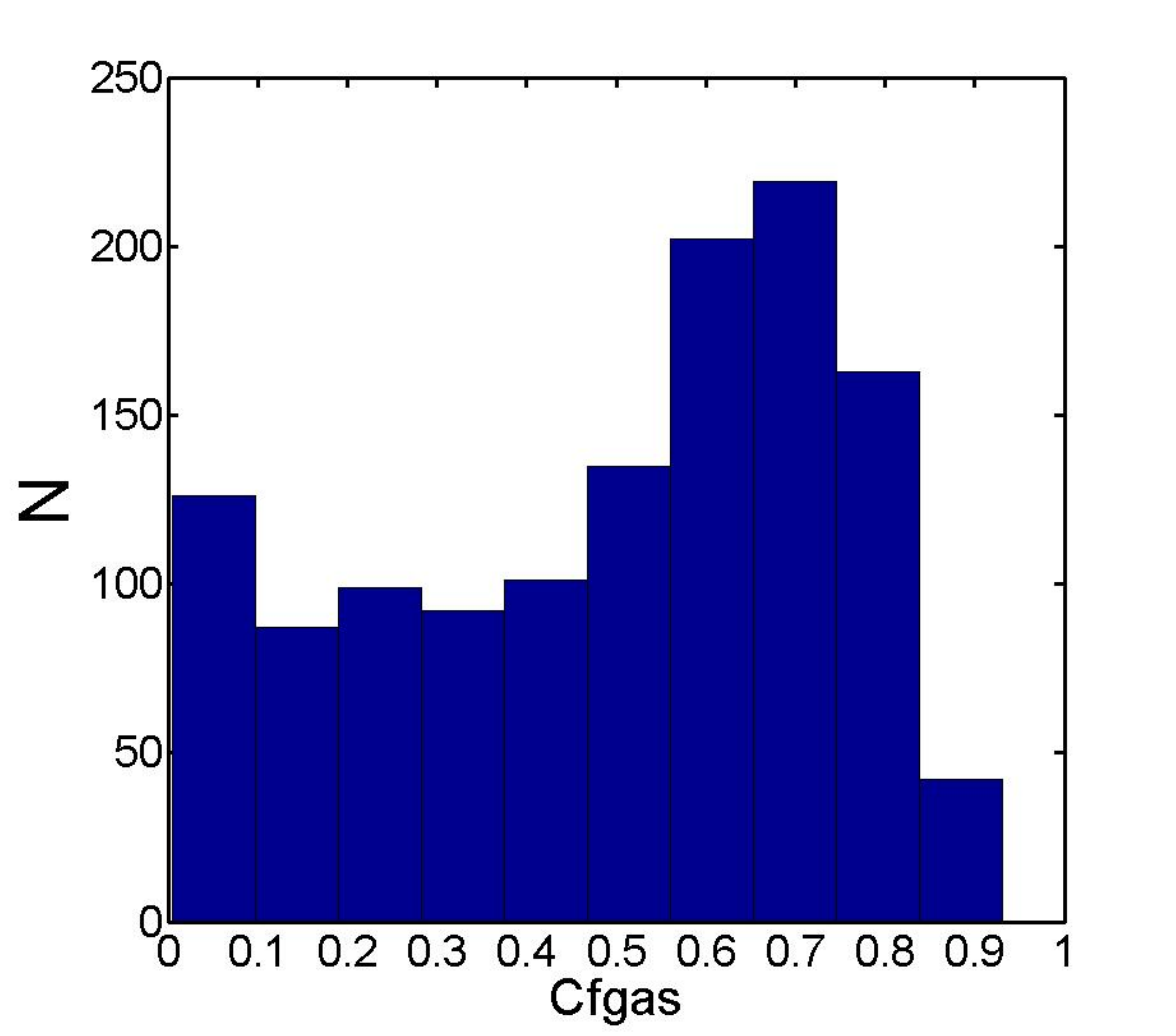}
\includegraphics[width=4.1cm,height=3.9cm,angle=0]{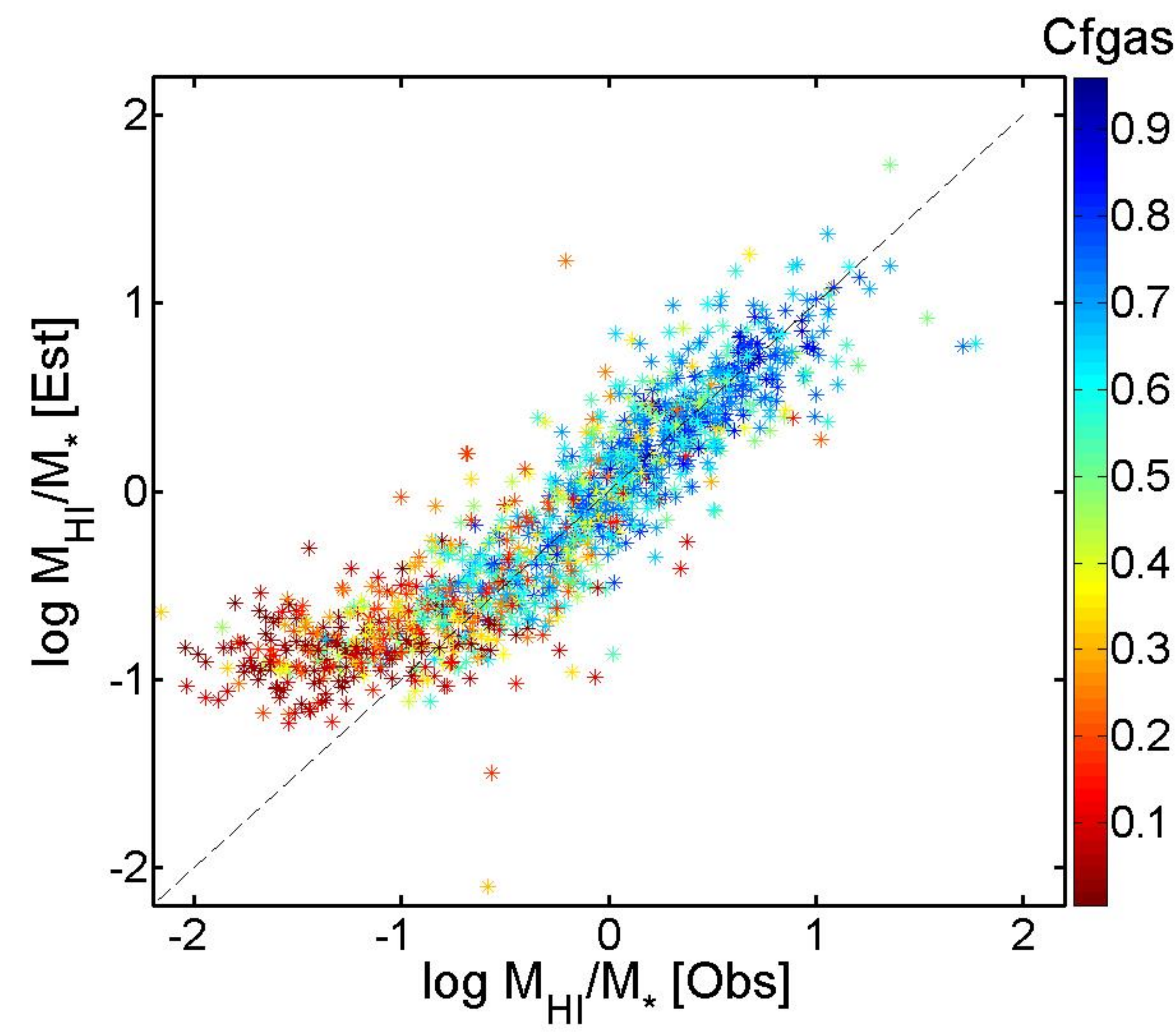}
\caption{Left panel:  The distribution of \cfgas\ for the combined sample. Right panel: 
Estimated values of \fgas\  for the sample, colour coded by \cfgas, compared with the observed values.}
\label{fig-combine}
\end{figure}

In addition to using \cfgas\ for imposing robustness thresholds on the \fgas\ estimations, we can
also use the scatter in the observed vs. estimated gas fractions as an indicator of the
likely uncertainty therein.   To do this, first  we construct  a combined  sample of all the validation sets (633 galaxies) shown in the left panels of  Fig. \ref{fig-fit-gass-pm-2p}. Then, in order to make a compromise and also to avoid any bias, we add 633 galaxies randomly selected from AC1 to our combined validation set.   In the left panel of Fig.  \ref{fig-combine}, 
we show the overall distribution of \cfgas\ in the combined sample (N=1266) and in the right panel the \fgas\ comparison. 
Again, it can be seen that points that have deviant estimations of \fgas\ have low values of \cfgas.

\begin{figure}
\centering
\includegraphics[width=9cm,height=5.8cm,angle=0]{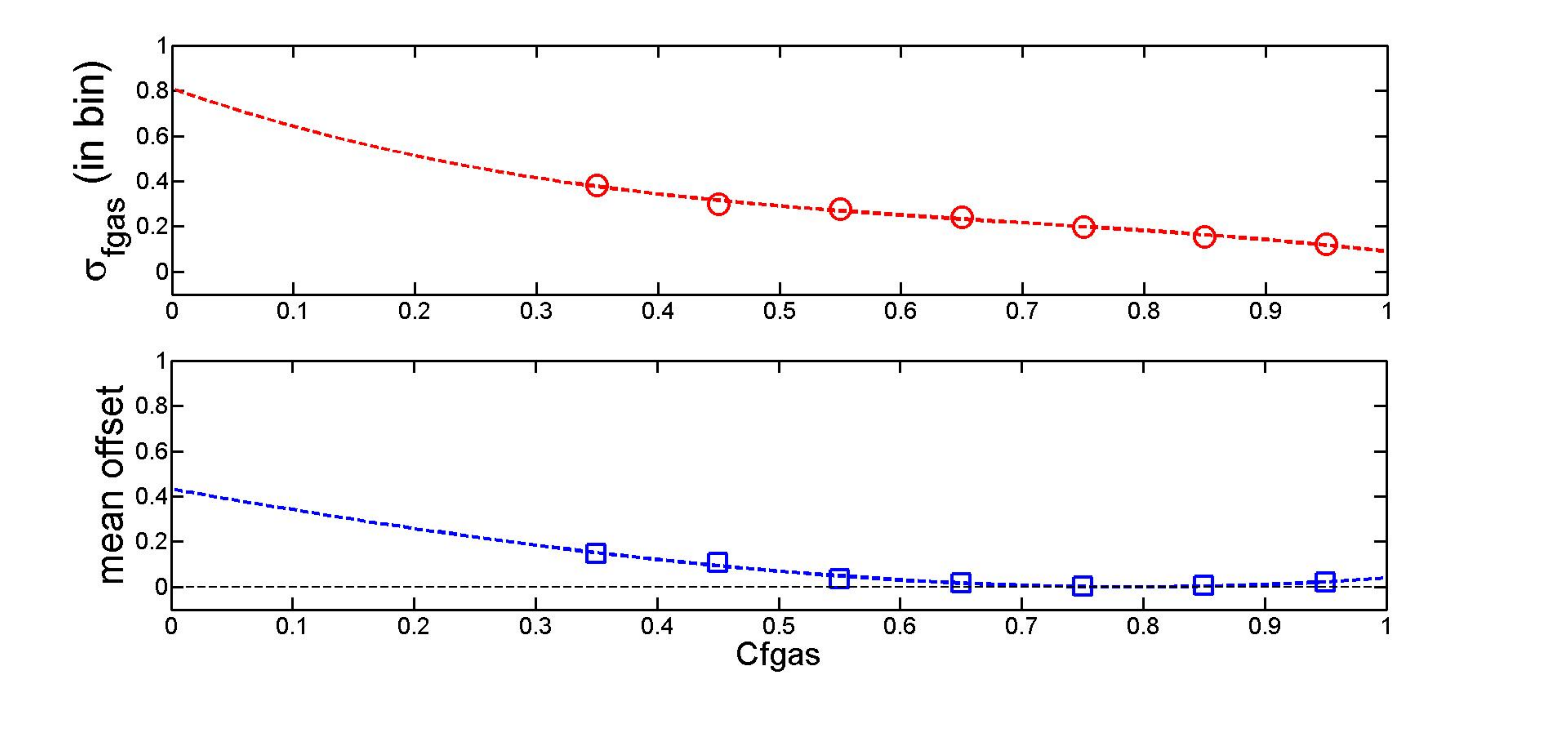}
\caption{The scatter between the estimated and observed \fgas\ as
a function of \cfgas.  The top  panel shows the error in 7 different bins of \cfgas\ ( $0.33<\rm{C_{fgas}}<1$).  The lower panel shows the mean offset between observed and predicted \fgas\. in the same area. The dashed lines are  polynomial functions fitted to the data.}
\label{fig-pr-err-color}
\end{figure}

Now, we can obtain a relationship between the scatter of the new sample and \cfgas. Since, \fgas\ estimations  are more reliable for higher \cfgas, we use data only in region $0.33 <\rm{C_{fgas}} <1$. In this way, we avoid using small \cfgas\ and also, at the same time, we have adequate data points to fit a function to them and extrapolate it to smaller \cfgas.  The upper plot of Fig. \ref{fig-pr-err-color} shows these data points (in red circles) and also a polynomial function fitted to them.  Here, we show the scatter in the \fgas\ estimation for bins in \cfgas\ (\cfgas$\pm0.1$, the value on the x-axis).  We can see that, for example,  the fitted function shows a scatter of $\sim$0.8 dex  for $\rm{C_{fgas}}=0$, which is obtained by an extrapolation. In the lower plot of this figure we use the data points (in blue rectangles) in the same regions to obtain mean offset and again fit the similar function to the points.   The average  estimated offset (by the function) for \cfgas$<$0.5 is 0.23 (median 0.22) and the scatter in this region is 0.38. This is well matched with the data from sample GASS, which is the dominant sample in this region.  There is no significant  average  offset for  \cfgas$>$0.5  and the mean scatter in this region is 0.22 dex. The recommended values to use are $0.5\leq\rm{C_{fgas}}\leq1$  or  $0.09\leq\rm{\sigma_{fgas}}\leq0.29$.

\begin{figure}
\centering
\includegraphics[width=8.8cm,height=5cm,angle=0]{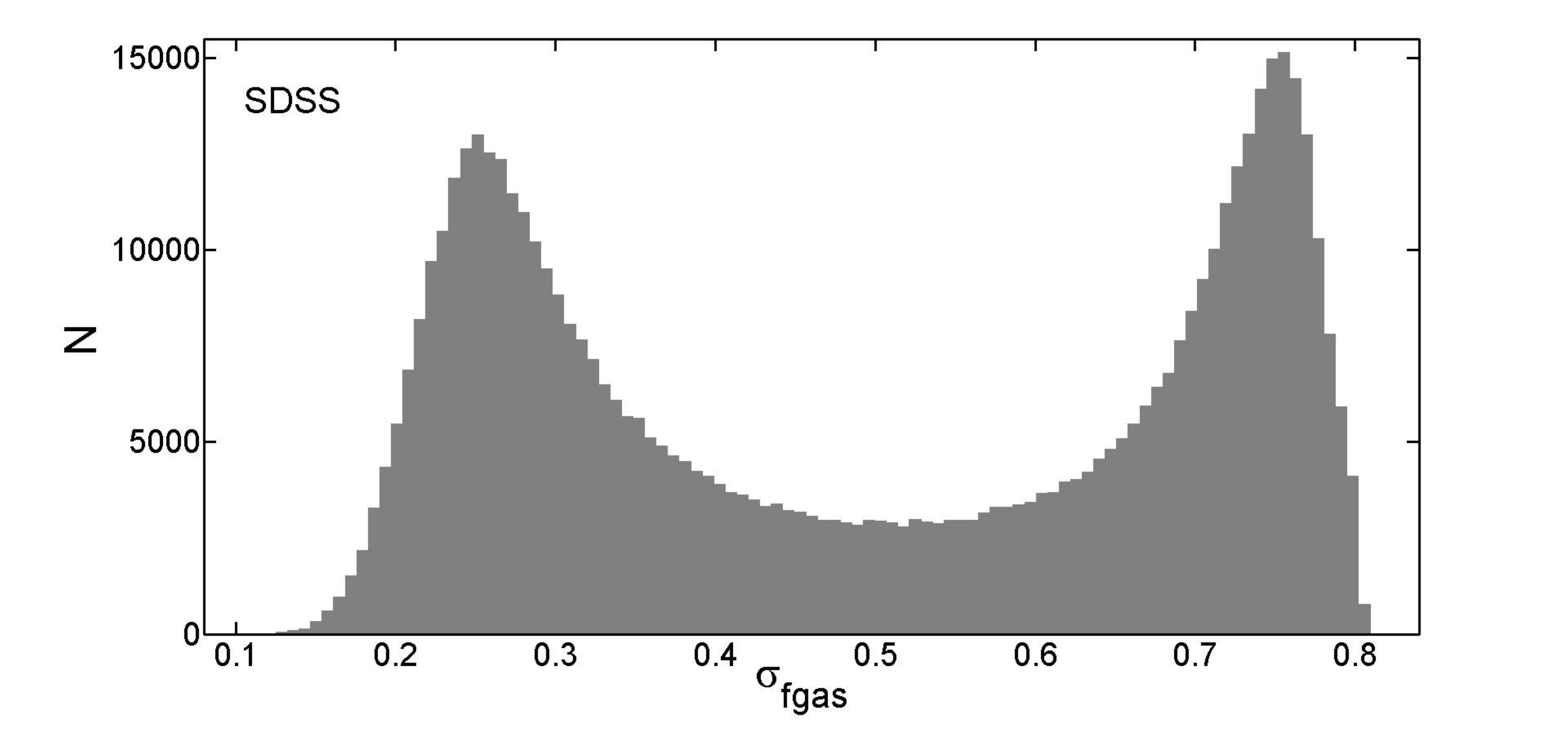}
\includegraphics[width=8.8cm,height=5cm,angle=0]{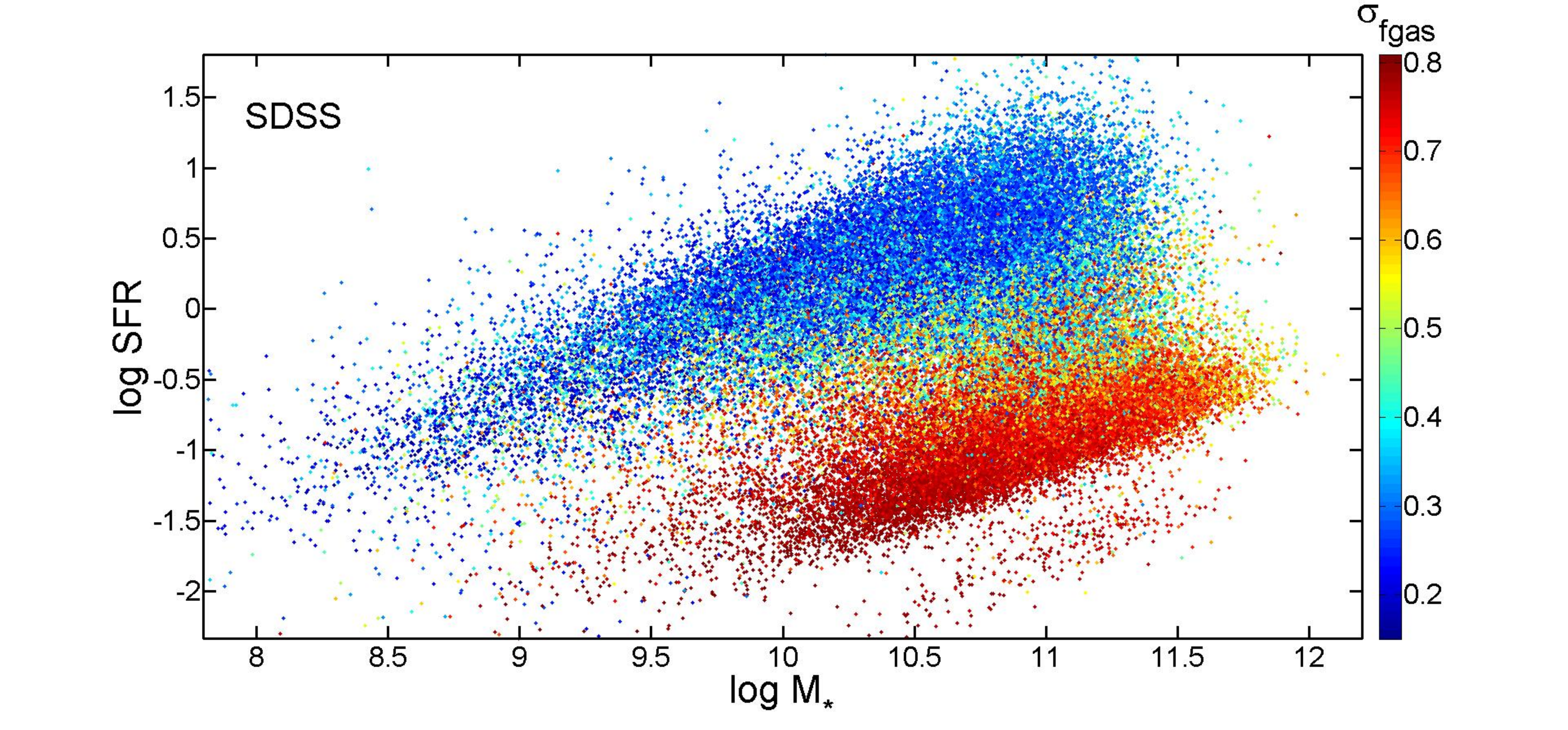}
\caption{The top panel shows the distribution of errors; \sigt.  In the bottom panel we plot the SFR vs. stellar mass for 
the SDSS sample (for presentation purposes, 60 000 galaxies have been selected at random). }
\label{fig-err-sdss}
\end{figure}

The resulting distribution of \fgas\
uncertainties for the SDSS sample is shown in the top panel of Fig.  \ref{fig-err-sdss},
and both \sigt\ and \cfgas\ are provided in our online catalog.
The distribution of uncertainties has a narrow peak at small values, corresponding to
galaxies with a high \cfgas\ value, and for which \fgas\ is expected to be well estimated.
In the lower panel, we plot the star forming main sequence, colour coded by  \sigt.
The bimodal distribution of \sigt\ seen in the top panel of  Fig.  \ref{fig-err-sdss}
is clearly present in the distribution of SFR as a function of stellar mass.  The ANN
can make robust estimations for most galaxies on the main sequence, but performs
poorly for passive galaxies, as expected from the distribution of \pr\ shown in the
lower panels of Fig. \ref{fig-class-pr}.  As we have seen repeatedly in this paper,
this is due to the nature of the AC1 training sample, whose galaxies are mostly
star-forming. 

In Table~\ref{table-samp}, we show  the first 10 entries of the ANN estimated \fgas\ along with  all the control parameters.  The full catalog is available in the online material that accompanies this paper.

\subsection{A final caveat on the derivation of \mhi\ from \fgas}\label{sec_mhi_caveat}

\begin{figure}
\centering
\includegraphics[width=8.8cm,height=7cm,angle=0]{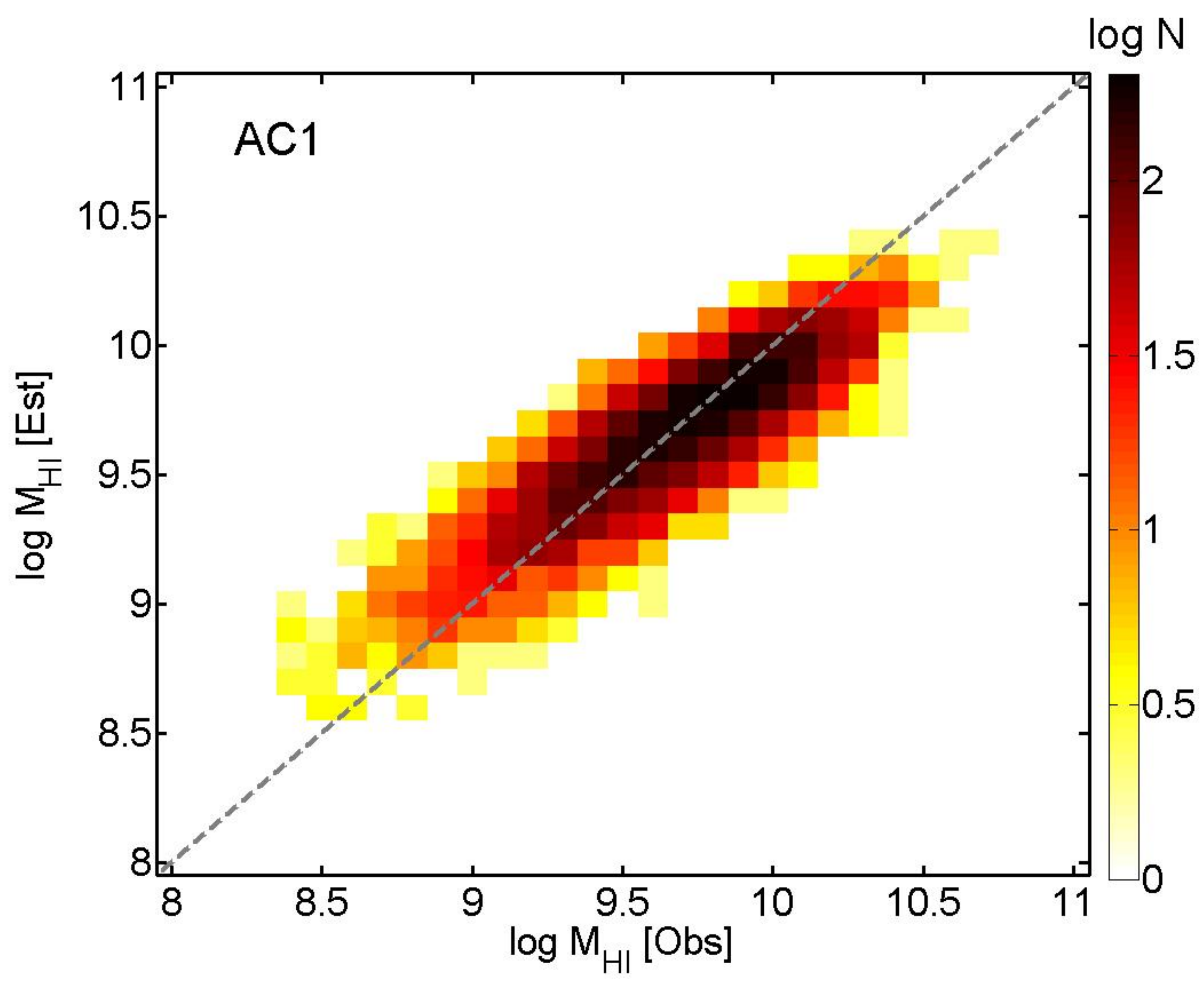}
\caption{ The estimated value of \mhi\ vs. the observed one, colour coded by number of galaxies in each point.}
\label{fig-mhi}
\end{figure}

In this paper we have concentrated on the estimation of \fgas, rather than \mhi, although the latter
can be derived from the former since we have stellar masses available.  In Figure \ref{fig-mhi} we show 
the estimated value of \mhi\ vs. the observed one, colour coded by the number of galaxies in each point. 
A mild skewness is seen in this plot, whereby low \mhi\ galaxies have their gas masses slightly over-estimated
and vice versa.  This effect persists even after the application of \cfgas\ thresholds.
Moreover, we have found that this effect persists if the network is re-trained with
\mhi\ as the target.  This check includes a complete re-assessment of the most relevant parameters for \mhi\
estimation (indeed, the parameter rankings for the prediction of \mhi\ are quite different to
those for \fgas). As was previously shown in the top panel of Figure 
\ref{fig-fit-ac1}, no such skewness exists in the estimation of \fgas.  The skewness
in the \mhi\ estimates is again due to the nature of the training sample,
the large observed scatter in \mhi\ at a given stellar mass and the shallow
nature of the survey.  We therefore caution that whilst we have demonstrated
the robustness of our \fgas\ estimates, there may be errors of several tenths
of a dex in attempts to convert these to gas masses. It is worth noting that
previous efforts to calibrate the gas content of galaxies have similarly used
\fgas\ rather than \mhi.

\begin{figure}
\centering
\includegraphics[width=8.5cm,height=6.8cm,angle=0]{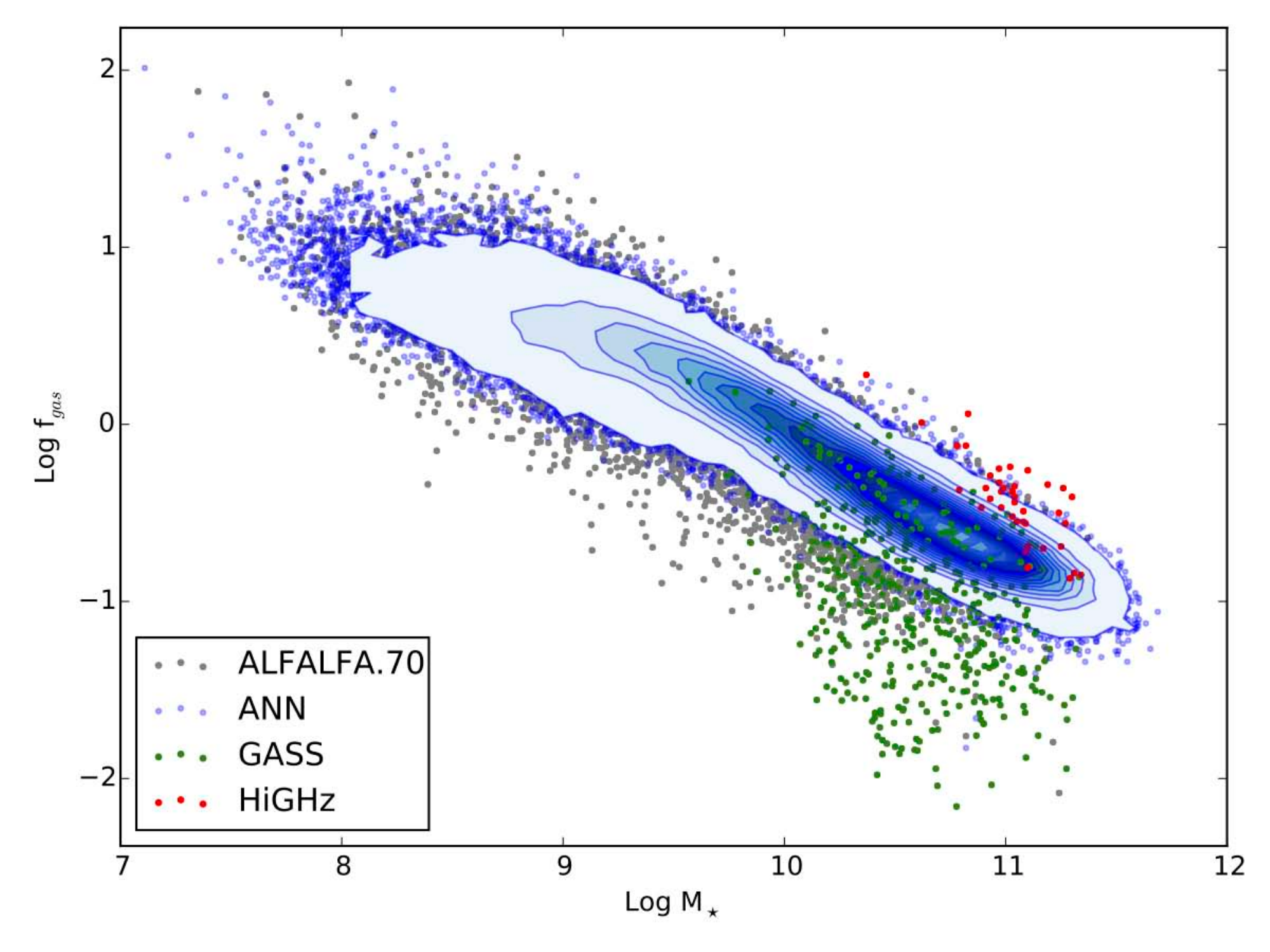}
\caption{ Distribution of galaxy gas fractions as a function of stellar mass for ALFALFA.70 (grey points), GASS (green points), HiGHz (red points) and the ANN predictions for the SDSS (blue contours and points).  The ANN predicts a significant population of high mass (log M$_{\star} >11$) galaxies with gas fractions log \fgas\ $> -1$ that are rare in current samples.  For clarity, only detections (no upper limits) are shown.}
\label{fig-25}
\end{figure}

\begin{table*}
\begin{center}
\caption{Catalog of ANN estimated \fgas\ along with  all the control parameters described in this paper. The first four columns (ID, RA, Dec and $z$) are taken from the SDSS database. The first 10 entries are shown here; the full catalog is available in the online material that accompanies this paper.}
\begin{tabular}{c|clclclclclclclclc|c}
\hline\hline

ID (SDSS)  & RA & ~~~~~~Dec & $z$ & log M$_*$& log \fgas  & \cfgas & PR&\sigfn &\sigf &\sigt \\
\hline
587736781457719496	&	237.4554943	&	34.90548896	&	0.1670	&	11.4110	&	-0.9456	&	0.1178	&	0.1972	&	0.5249	&	0.1006	&	0.6169\\
588018091616174223	&	237.5470295	&	34.70075261	&	0.0701	&	9.9120	&	-0.2466	&	0.3310	&	0.4111	&	0.7075	&	0.0269	&	0.3913\\
587736781457850591	&	237.6627645	&	34.78976388	&	0.1511	&	10.7487	&	-0.9023	&	0.3493	&	0.4945	&	0.6207	&	0.0503	&	0.3778\\
587736781457850627	&	237.691242	&	34.70155096	&	0.0751	&	10.5140	&	-0.5599	&	0.4970	&	0.6302	&	0.6929	&	0.0298	&	0.2922\\
587739130812694794	&	225.2513585	&	29.37813267	&	0.0793	&	10.6881	&	-0.8689	&	0.3917	&	0.4506	&	0.7637	&	0.0179	&	0.3491\\
587742773490155647	&	192.5694372	&	16.66269477	&	0.0252	&	9.2378	&	-0.0193	&	0.6708	&	0.7546	&	0.7809	&	0.0158	&	0.2259\\
587739630095040681	&	223.1457785	&	26.26263514	&	0.0759	&	10.0936	&	-0.1939	&	0.3101	&	0.3877	&	0.7027	&	0.0278	&	0.4078\\
587736781457916109	&	237.8366547	&	34.56556243	&	0.0488	&	10.5449	&	-0.5856	&	0.6167	&	0.6857	&	0.7901	&	0.0148	&	0.2443\\
587730816291569798	&	344.8364624	&	-10.13584859&	0.0600	&	9.9198	&	-0.0108	&	0.5532	&	0.5998	&	0.8103	&0.0128	&	0.2680\\
587742773490221138	&	192.5891741	&	16.65422642	&	0.0680	&	10.1461	&	-0.0377	&	0.7943	&	0.9399	&	0.7424	&	0.0209	&	0.1842\\

\hline
\end{tabular}
\label{table-samp}
\end{center}
\end{table*}

\section{Summary}\label{sec-sum}

In this paper we have presented a novel method to estimate HI gas mass fraction and the associated 
uncertainties based on the patterns found in our data sets, using machine learning methods.
The ALFALFA survey is used as our main training sample, and we check our model estimations
with a range of validation sets, comprised of the GASS and Cornell surveys and a small sample
of post-merger galaxies.
We have shown that, for a given set of input parameters, non-linear methods can significantly
reduce the scatter in the estimation of \fgas, compared with traditional linear fits.  
Specifically, we demonstrate that using only
the $g-r$ colour and $i$-band surface brightness, a matrix-based model can reduce the
scatter in a linear model from  0.32 dex to 0.22 dex.

In order to extend our models to include more galactic parameters, we assess the correlations
between \fgas\ and 15 parameters derived from the SDSS imaging and spectroscopic datasets.
Two performance parameters are presented:  R, which represents the relative weights
of the fit in minimizing scatter and AUC, which quantifies the physical relevance of a
given galaxy parameter to its gas fraction.  The AUC ranking shows that $g-r$ colour
and stellar mass surface density both strongly govern a galaxy's gas fraction, with specific
SFR and B/T having a further marginal relevance. 

We introduce several parameters that permit the assessment of how accurately \fgas\
is estimated for a given galaxy in the SDSS.  The scatter in the estimation of \fgas\
is determined from 20 individually trained networks (\sigf) and provides an indication of
variation in the fitting process.  The inverse normalized version of \sigf\ (\sigfn)
has a value from 0 to 1, where higher values indicate a smaller scatter.  We also
use a pattern recognition technique to identify the similarity of a given galaxy
to the ALFALFA detections used in our ANN training set, \pr.  Again, this
parameter has a value between 0 and 1, where higher values indicate a greater
similarity to ALFALFA detections, and hence, a higher likelihood that the network
can provide an appropriate estimation of \fgas.  \pr\ and \sigfn\ can by multiplied
together to yield a single parameter, \cfgas, whose value again ranges from 0 to 1,
where higher values indicate more robust estimations.  \cfgas\ can also be used
to determine an uncertainty associated with a given \fgas\ estimation.  We demonstrate how the
application of these various parameters to the validation sets effectively
removes outliers from the ANN estimations.  A \cfgas\ cut of 0.5 yields $\sim$ 150 000 
galaxies with \fgas\ estimations from our ANN approach, a factor of $\sim$ 20 greater than 
the number of firm 21~cm detections in the 70 per cent data release of ALFALFA.
All of the quality control parameters accompany the \fgas\ estimations in the online table.

Our catalog of predicted gas fractions  offer several advantages
over previous attempts to calibrate gas fractions.  For example, we have shown that there
is no systematic error in our \fgas\ estimates with various galaxy properties, and the
scatter in our estimates is lower than has been previously achieved.  Perhaps most 
importantly, we have provided a quantitative prescription for assessing the robustness of
our estimates, on a galaxy-by-galaxy basis.  Our ANN \fgas\ estimates also potentially
offer advantages over even direct Arecibo observations.  The large size of the Arecibo 
beam (3.3$\times$3.8 arcminutes) is not able to distinguish contributions from multiple
galaxies with close separations.  Indeed, as shown in Figs. \ref{fig-fit-ac1} and \ref{fig-fit-ac1-unclean}, whilst
our predictions reproduce well the gas fractions in the `clean' training sample, galaxies
with close companions have observed gas fractions that are typically 0.14 dex above the
ANN predictions. Finally,
our ANN-based gas fractions extend the number of galaxies in the nearby universe with
robust \fgas\ estimates by more than an order of magnitude.  The growth in sample
size yields predictions for large numbers of galaxies in parameter space beyond
current samples.  For example, most HI surveys are currently limited to $z < 0.06$,
whereas our \fgas\ predictions extend to $z \sim 0.2$.  Indeed, our catalog contains
$\sim$ 61,000 robust (\cfgas\ $>$ 0.5) \fgas\ predictions at $z>0.1$.  Figure \ref{fig-25}
shows the distribution of gas fractions for the GASS, HiGHz (Catinella \& Cortese 2015) and ALFALFA.70 samples (green, red
and grey points respectively), compared with the ANN predictions (blue contours and points, 
where the latter show individual galaxies with counts below the lowest contour).  It can be
seen that the ANN predicts a significant population of high stellar mass (log M$_{\star} > 11$)
galaxies with gas fractions log \fgas\ $> -1$.  Such galaxies are largely absent from current
samples and are predicted to contain some of the highest HI masses in the local universe.
Follow-up observations of these high stellar mass, high HI mass galaxies to confirm our
ANN predictions would be of great interest.  

\section*{Acknowledgments}

We are grateful to Barbara Catinella and Luca Cortese for comments
on an earlier draft of this paper and for help and advice with the
ALFALFA.70 sample. We thank Wim van Driel for providing us with his table of HI measurements in electronic format.  We also thank the anonymous referee, whose comments
led to an improved paper.  SLE and DRP gratefully acknowledge the receipt
of NSERC Discovery Grants.

Funding for the SDSS and SDSS-II has been provided by the Alfred
P. Sloan Foundation, the Participating Institutions, the National
Science Foundation, the U.S. Department of Energy, the National
Aeronautics and Space Administration, the Japanese Monbukagakusho, the
Max Planck Society, and the Higher Education Funding Council for
England. The SDSS Web Site is http://www.sdss.org/.

The SDSS is managed by the Astrophysical Research Consortium for the
Participating Institutions. The Participating Institutions are the
American Museum of Natural History, Astrophysical Institute Potsdam,
University of Basel, University of Cambridge, Case Western Reserve
University, University of Chicago, Drexel University, Fermilab, the
Institute for Advanced Study, the Japan Participation Group, Johns
Hopkins University, the Joint Institute for Nuclear Astrophysics, the
Kavli Institute for Particle Astrophysics and Cosmology, the Korean
Scientist Group, the Chinese Academy of Sciences (LAMOST), Los Alamos
National Laboratory, the Max-Planck-Institute for Astronomy (MPIA),
the Max-Planck-Institute for Astrophysics (MPA), New Mexico State
University, Ohio State University, University of Pittsburgh,
University of Portsmouth, Princeton University, the United States
Naval Observatory, and the University of Washington.


\end{document}